Short Communication

# On problematic practice of using normalization in Self-modeling/Multivariate Curve Resolution (S/MCR)

Róbert Rajkó [a,*]

[a] E-Dimension Ltd., H-6724, Szeged, Vásárhelyi P. út 13. https://sites.google.com/view/prof-robert-rajko/bio

**Abstract**

This short communication paper is briefly dealing with greater or lesser misused normalization in self-modeling/multivariate curve resolution (S/MCR) practice. The importance of the correct use of the ODE (ordinary differential equation) solvers and apt kinetic illustrations are elucidated. The new terms, external and internal normalizations are defined and interpreted. The problem of reducibility of a matrix is touched. Improper generalization/development of normalization-based methods are cited as examples. The position of the extreme values of the signal contribution function is clarified. An Executable Notebook with Matlab Live Editor was created for algorithmic explanations and depictions.

* Corresponding author, e-mail address: Prof.Rajko@gmail.com (R. Rajkó)

## 1. Introduction

The definition of the term of normalization was published in the early chemometric literature. For example, in the Handbook of Chemometrics and Qualimetrics [1] on page 12: "Note that *normalization* of an arbitrary vector **x** is obtained by dividing each of its elements by the norm $\|\mathbf{x}\|$ of the vector." Or in Malinowski's classic [2] on pages 66-67: "Since only the vector directions are of interest and not their magnitudes, each row of the abstract rowfactor matrix is normalized to unit length. After normalization, the *i*th vector is represented as $\mathbf{v}(i) = (v_{i1}, v_{i2}, \ldots, v_{in})'$ where $v_{ik} = \frac{r_{ik}}{\left(\sum_{j=1}^{n} r_{ij}^2\right)^{1/2}}$." However, Hibbert [3] remarked that "'Normalization' has many meanings in statistics (see https://en.wikipedia.org/wiki/Normalization_(statistics)). To resolve any ambiguity the nature of the scaling constant should be explained." And he also defined the term 'scaling' as "Element-wise division of a data matrix by a scaling matrix". Reducing the ambiguity of the meaning of the term 'normalization' in Self-modeling/Multivariate Curve Resolution (S/MCR), we can accept that the normalization is a kind of scaling, but here we use the same scaling constant (calculated by using a chosen norm for the vector) for a row (or a column) of a matrix resulting in the row (or the column) which has generalized length equals to 1, i.e., the chosen norm of the normalized row (or column) is 1. $\ell$p is the most known and used kind of norms, they are also known as Hölder norms: $\|\mathbf{z}\|_p = \sqrt[1/p]{\sum |z_i|^p}$, where $p \geq 1$.

In this sense, normalization keeps the shape of the graph of the vector with unit (generalized) length, but scaling does not necessarily keep the shape and/or unit length, indeed it can commonly distort them. More general term is the 'transformation' which uses more complicated function(s) on the elements of a matrix (or a vector) instead of the simple element-wise division.

The organization of the paper is as follows: firstly, some critical commentaries are given based on some recently published papers, forwarding the new types of the normalization (enlightening the problematic use of the closure property (affine combination vs. linear combination: the latter causes the rank deficiency and not the former); secondly, the two new types of normalization are introduced and their applications in SMCR and MCR are shortly provided (enlightening, e.g., the relationship between closure and $\ell$1-norm); finally, the proper positions of extreme values of the signal contribution function are given using some dominant properties of the Euclidean norm/distance.

## 2. Is rank deficiency of a data matrix caused by closure?

After the short introduction to normalization in S/MCR, a recently published paper [4] is investigated, in which band boundaries were determined for rank deficient data matrices. We can analyze here chemists' and mathematicians' approach to a problem. The proposed algorithm is scientifically correct, however the authors chose such kind of chemometric data matrices as illustrations, for which, the proposed algorithm is unnecessary. The closured data matrices are not necessarily rank deficient. And there is no partial closure at all. The mass conversations among only the parts of the components are different concept from the term closure. Their first example was inserted at a later date (confer the manuscripts dated by 01.12.2021 and 29.04.2022, and posted as 2021rankdef.pdf [5] and 2022rankdef.pdf [6], resp.), i.e., the spectrokinetic data for the reaction $X + Y \rightarrow Z$. Indeed, this type of bimolecular reaction has analytical, i.e. closed form solution. However, with the default setting, the ODE (ordinary differential equation) solver(s) in Matlab (see details later) can give numerical solution with difference error of $10^{-4}$ (Fig. 1). But the matrix is really a rank deficient data matrix. However, the chemist's solution is easy for this kind of rank deficiency: measure the system with two different initial concentration sets and use augmentation [7], e.g., $c_{X,1}(0) = 1$ mol/L, $c_{Y,1}(0) = 0.7$ mol/L, and $c_{X,2}(0) = 0.7$ mol/L, $c_{Y,1}(0) = 1$ mol/L. Thus the estimation of matrix A using the two hard-model estimated C and the two measured D matrices can be calculated uniquely using the following Matlab command: Aest = [D1;D2]'/[Cest1;Cest2]'. Or just simply, after some time, spike the reaction system with component(s) with known concentration(s) till the rank of the measurement matrix D reaching the number of chemical components [7]: confer the multiple standard addition [8] or the flow injection analysis [9] as well. It seems so there is no need for the complicated rank deficient band boundaries calculations.

The authors stated that the other examples in Ref. [4] have closure, i.e., they are closed or compositional data [10]. However, the following statements can be used here from Ref. [11]: "On the other hand, closure in the profiles refers to reaction-based systems in which concentration profiles satisfied mass balance relations. For instance, spectroscopic signals obtained during the monitoring of an evolving reaction system in a closed system will not produce closed data in the samples direction, but the resolved concentration profiles will probably fulfill a mass balance condition and therefore be closed."

The closure cannot mean linear dependency among, e.g., the column-vectors of a matrix, because for that the linear combination with nonzero coefficients should result in zero-vector. If the result is different from zero-vector, it means affine combination, but affine combination cannot cause rank deficiency. Instead, it can be used to reduce the dimensionality by 1, because we can choose this affine subspace as a new linear vector space dropping the singular direction belonging to the highest singular value (projection of the original nD space onto the (n-1)D (not linear) subspace). If there exist an almost linear combination, it can cause, however, extremely small singular value(s), and it can be a problem in noisy circumstances: that is the ill-conditioned case, the ratio of the largest and smallest singular values is extremely large. Let us consider a simulated data similar to one for the Michaelis-Menten kinetic model in Ref. [4] (Example 1.1). MATLAB® SimBiology® (https://www.mathworks.com/products/simbiology.html) was used here for simplified kinetic simulations. With default settings of SimBiology, using singular value decomposition (SVD), svd(C) results in [7.054525774793125; 3.848496628860587; 0.434969142486790; 0.000000000003752], and indeed, the fourth singular value is extremely small, still not zero, because this concentration profile-matrix is noiseless! E.g., the scree plot depicts the singular values in descending order [2]. Using the scree test, the "elbow" of the curve is determined where the singular values appear to level off, and the factors or components to the left of this point are predicted significant. The elbow value is 0 here, thus the next value to the left is 0.000000000003752 which is significant. Thus the rank of matrix C is 4 and not 3. svd(D) results in [93.840778438613270; 20.309655081533617; 15.270043490066088; 0.000000000115671]. Now, the elbow value is 0.000000000000061, and the next value to the left of this elbow is 0.000000000115671 which is not significantly small (not practically zero). Seemingly, there is not rank deficiency, just ill-conditioning. And, surprisingly, there is not closure either: the minimum of the sum of the rows of matrix C is 1.015 and the maximum is 1.100; the mean is 1.057 and the standard deviation is 0.036. If we choose 'sundials' (https://www.mathworks.com/help/simbio/ug/sundials-solvers.html) as ODE solver for SimBiology, the forth singular value will be much more significant: svd(C) = [6.650693175896303; 4.230055090441462; 0.425321106032236; 0.000000862499289].

However, there are two mass conversations in the Michaelis-Menten kinetic model (provided the usual initial conditions: $c_{SK}(0) = 0$ and $c_P(0) = 0$): $c_K(t) + c_{SK}(t) = c_K(0) = 0.1$ and $c_S(t) + c_{SK}(t) + c_P(t) = c_S(0) = 1$. From these two equations, both affine and linear combinations can be derived: $c_S(t) + c_K(t) + 2c_{SK}(t) + c_P(t) = c_S(0) + c_K(0) = 1.1$, and $c_S(t)/c_S(0) -$

$c_K(t)/c_K(0) + (1/c_S(0) - 1/c_K(0))c_{SK}(t) + c_P(t)/c_S(0) = c_S(t) - 10c_K(t) - 9c_{SK}(t) + c_P(t) = 0$.

Surprisingly, the default (ode15s) and 'sundials' ODE solvers in SimBiology resulted in wrong outcomes, because none of the affine and linear combinations was fulfilled. Fortunately, the ODE solver 'ode45' in SimBiology can produce such high accuracy results as ODE solver 'ode89' (not implemented in SimBiology) could. This is the proper proof that the spectrokinetic data matrix from the Michaelis-Menten kinetic is rank deficient and the proper ODE solver should be selected and correctly parametrized. However, as was mentioned above, the chemist's solution is easy for this kind of rank deficiency as well. Let's use some amount (0.005 mole) of doses of the enzyme at the beginning of the reaction with rate of 0.005 mole/second, and the initial concentration of the enzyme should be 0.095 mole. In this situation, the simulated profiles of the kinetic curves seemingly totally coincide with the original rank deficient ones, however now the matrices have rank 4. Thus the simple Matlab command `Aest = D'/Cest'` can be used for the unique determination, even in moderate noisy situations (Fig. 2 Left panel). Discrete dose is also possible, i.e., spiking the reaction system with some amount of enzyme. For the spiked concentration profiles, the matrix will have full rank, i.e., 4 again (Fig. 2 Right panel). We can consider, that the four differential equations will be equivalent to only two differential equations, because of the mass conversations [12, 13]. Although the concentration profiles will be rank deficient, the generated spectrokinetic data matrix will have tiny singular values from the fourth position. We can accept the quasi-steady-state assumption as well. In this case, closed form solution exist [14], and only one equation (i.e., substrate concentration profile) is enough for reproducing the kinetic process. However the estimated spectra will be still distorted. The proper answer is simpler than it is expected. In analytical chemical sense, the spectra of the substrate and/or the enzyme is known in advance, or they can be measured before the kinetic measurement(s), thus the unknown spectra can be estimated correctly: `Aest_unsp = (D-Cest_knsp*A_knsp')'/Cest_unsp'` (Fig. 3).

Someone wants to consider realistic kinetic parameters and initial concentrations, e.g., in practice, even six order of magnitude may occur in concentration ratio of substrate and enzyme. In these cases, the measurement noise can hide the signals of the enzyme and the complex, and the system will become two-component one. However, at moderate noise effect, the spectra can be revealed using the substrate spectrum measured before adding the enzyme (see the details in the Supplementary material).

Later the authors used the following statement: "Due to titration dilution the closures are not constant, but decreasing straight lines" [4]. The closure must be constant, that is the essence of a closed system [11]. Furthermore, the effect of the titration dilution is not straight line (Fig. 4). Moreover, if we use special prepared titrant containing different amount of dyes (in the Executable Notebook, the value of the variable *inds_in_titrant* is not 0), the titration concentration matrix will be of full rank. Consequently, the main problem of the paper of Ref. [4] is the used unfortunate chemical examples. The proposed procedure and algorithm work for adequate rank deficient matrices [15], but some more apt illustrations should have been chosen.

**3. New types of normalization and their applications in SMCR and MCR**

Let us assume a properly closed data matrix. If the closure constant is not 1, we can divide the full matrix with the actual constant getting the desired sum-to-1 property. "Using the closure constraint means a special Borgen normalization in which the norm will be not 1 but a value that is proportional to the appropriate total concentration; thus, the closed data matrix is normalized by nature in advance." [11]. And even, we can generate a closed data matrix, if we use the $\ell$1-norm normalization (which norm is a member of the Borgen norms family [16]) for every row. Now, we can split our way into two directions. We can use SMCR [17] or MCR-ALS [18] (and its variations).

1. *Using SMCR*. From Ref. [11]: "By visualizing microstructure of closed data sets, it is shown that all of possible solution satisfies from closure constraint. Although normalization and closure constraint resulted in the abstract space with reduced dimension, they resulted in different subspaces. So, it is not possible to apply both of them simultaneously." Normalization here is other kind than $\ell$1-norm normalization, e.g., the first scores-vector-to-1 normalization [19], which is another member of the Borgen norms family [16]. In practice, the first scores-vector-to-1 normalization (*fsvt1n*) is done to modify the transformation matrix: dividing all the elements in every row by the first values. Thus the first abstract coordinate will be a one-vector, therefore it can be neglected to get the dimensional reduction. However, this process can be called "internal normalization", because we are in and use the abstract space and not the original data matrix. The normalization process for the original data matrix can be called "external normalization" which is done before creating the SVD-based abstract space. We saw, that more than one normalizations must be avoided, thus for a closed data (which is already external normalized), any further internal normalizations cannot be used. But using an internal normalization without any checking tends to be a routine practice nowadays.

Sometimes the external normalization has its easily definable internal normalization pair. E.g., the ℓ1-norm external normalization has its internal normalization pair when we use closure constraint for the rows of the transformation matrix belonging to the SVD-based abstract space of the original data matrix. That is why, closure is strong relation with normalization, and it explains the necessity of the reasoning above about the source of the rank deficiency in the second section. But what about the *fsvt1n*. For this external normalization we have to know in advance the first complementary singular vector which can cause ones for the first eigenvector. However, in general, we do not know the proper complementary singular vector and the first eigenvector is not a one-vector. We can make an iterative consecutive process: determining the first singular vector, external normalizing the data matrix with it and again determining a new first singular vector, external normalizing the new data matrix with the new first singular vector and so on until convergence. It turned out, however, that sometimes the process diverged: at least two different limiting values alternated. It may/can indicate the "unhandleable" reducibility of a matrix. The Perron-Frobenius theorem [20] regarding to real irreducible nonnegative square matrices stated that although the maximal absolute eigenvalues might not be unique, but for the strictly positive maximal eigenvalue, the corresponding eigenvector has strictly positive components. However, the opposite is not true in general, i.e., some "handleable" reducible matrices may have strictly positive maximal eigenvalue and the corresponding eigenvector may have strictly positive components [21]. Even, the matrix can contain some negative entries getting maximal eigenvalue and corresponding eigenvector with strictly positive entries [22-24]. The other problem is that in chemometrics rectangular matrices occur and not square ones. Fortunately, we have got algorithm for testing the irreducibility of nonsquare Perron–Frobenius systems [25]. It seems, however, that tightening the curve resolution investigations for only irreducible matrices is not proposable! For example, the identity matrix is "handleable" reducible. Moreover, every reducible matrix can be modified to become irreducible.

Maeder introduced a special internal transformation [26]: force to 1 the diagonal elements of the transformation matrix **T**. This is similar to the internal form of *fsvt1n*, however now it is hard to find its external pair.

2. *Using MCR*. First we can see, that development of a normalization-based chemometric method can be very hard task indeed. Wang et al. [27] gave proofs that, using $\ell$p-norm normalization ($p \geq 1$) for the two-way data, the vertex vectors (which will be the pure variables if they exist) maximize a certain quadratic form over all points on a simplex ($p = 1$), or a polyhedral hyper-"spherical" surface ($p > 1$). Based on this theorem, they developed a procedure for determining pure variables. However, Rajkó and Faber [28] could prove that the generalization for $p > 1$ unfortunately failed.

Lopes et al. [29] published a paper on chemometric characterization of counterfeit tablets unmixing near-infrared hyperspectral data by using SISAL (simplex identification via split augmented Lagrangian), MVSA (minimum volume simplex analysis), and MVES (minimum-volume enclosing simplex) methods. They concluded that the three minimum volume criterion based methods outperform the popular and widely used MCR-ALS. Rajkó [30] argued most of their practices and conclusions, however now, we are only focusing on the problem with their routine normalization: "In the nature, the data are not normalized at all in general. The normalization may be needed for the used algorithms or for easier interpretation." When one uses normalization for matrix **A** as in eq 2 of Lopes et al.'s paper, the matrix **M** does not contain the unit concentration pure spectra anymore. Thus the comparability is violated. That is why, Figure 6 and 7 in Lopes et al.'s paper show distorted MCR-ALS estimated spectra, because Lopes et al. did not apply the closure constraint for MCR-ALS, which means the same data processing as normalization of matrix **A** according to eq 2 of Lopes et al.'s paper. It is also evident, that the normalization of matrix **A** by any Borgen norms [16] causes strong dependence in the normalized matrix **M** on the original abundance matrix. Although Vajna et al. [31] stated that the normalization could not result in significance difference (but their Fig. 3 suggests the opposite), if someone zooms in on Figure 6 and 7 in Lopes et al.'s paper, even for the simulated data, the spectra estimated by SISAL/MVSA show smaller or larger distortion, especially large one for talc.

## 4. Positions of extreme values of the signal contribution function

The final illustration can be considered as a hybrid application of SMCR and MCR. The question is whether the extreme values of the signal contribution function (SCF) [32, 33] are on the boundary of the feasible regions (FRs), or the extrema of SCF can be inside the FRs [34]. Of course the answer is that the extrema of SCF always should be on the boundary, very often coinciding in some vertices of the FRs. The SCF has the form $\|\mathbf{c} \cdot \mathbf{s}^T\|_{\text{Fro}}^2 / \|\mathbf{C} \cdot \mathbf{S}^T\|_{\text{Fro}}^2$ for a given component [33]. Since $\mathbf{c} \cdot \mathbf{s}^T$ is a rank-1 matrix, thus the matrix-norm calculations can be simplified: $\|\mathbf{c} \cdot \mathbf{s}^T\|_{\text{Fro}}^2 = \|\mathbf{c} \cdot \mathbf{s}^T\|_2^2 = \|\mathbf{c}\|_2^2 \cdot \|\mathbf{s}\|_2^2$. For the data matrix: $\mathbf{D} = \mathbf{C} \cdot \mathbf{S}^T = \mathbf{X} \cdot \mathbf{V}^T = \mathbf{U} \cdot \mathbf{Y}^T$, thus $\mathbf{c} = \mathbf{U} \cdot \mathbf{y}$ and $\mathbf{s} = \mathbf{V} \cdot \mathbf{x}$. Now $\|\mathbf{c}\|_2^2 = \mathbf{c}^T \cdot \mathbf{c} = \mathbf{y}^T \cdot \mathbf{U}^T \mathbf{U} \cdot \mathbf{y} = \mathbf{y}^T \cdot \mathbf{y} = \|\mathbf{y}\|_2^2$ and $\|\mathbf{s}\|_2^2 = \mathbf{s}^T \cdot \mathbf{s} = \mathbf{x}^T \cdot \mathbf{V}^T \mathbf{V} \cdot \mathbf{x} = \mathbf{x}^T \cdot \mathbf{x} = \|\mathbf{x}\|_2^2$. It means that the square of the $\ell$2-norm of a chemical profile is equal to the square of the $\ell$2-norm of its abstract coordinates! Thus the maximum/minimum of the SCF should be at the extreme points of the FRs with maximal/minimal Euclidean distances from the origin in the abstract space. Somebody may dispute that the maximum (or minimum) normed point in one linear subspace has the dual point with minimum (or maximum) norm in the other (dual) linear subspace, thus the product of the norms may not determine the extreme value of SCF. However this reasoning is false, because the dual of a point will be an affine subspace with less than one dimension of the original linear subspace [34], and this will be the base of the pointed cone with undefined opposite vertex which can be definitely the maximum (or minimum) normed point. Why did Neymeyr et al. [35] insist that the interior point can be the extremum of SCF? Because they used internal normalization for data matrices containing zero rows and/or zero columns; and even in these cases, the algorithms (including MCR-BANDS [36]) can calculate something wrong. The situation is similar to the calculation of false interior sparse solutions illustrated in Fig. 9 by Omidikia et al. [37].

## 5. Conclusions

The main take-home message of this note is that the external normalization is preferred against internal one, because in this case we can control the calculations more circumspectly: if some calculation errors occur (e.g., division by zero) we can correct the data matrix, i.e., we can delete the zero rows and zero columns or the low-valued rows, columns, or we can choose other data pretreatment processes. If we cannot use external normalization, we have to double-check the internal normalization and all the calculations based on it; and one should prefer theoretical reasonings against numerical ones.

Definitely, if an independent researcher wants to reproduce a recently published organic reaction or an analytical process, and it fails, this invalidates the published recipes. Similar validation should be used for the published/applied mathematical, algorithmic etc. results. Finally, the consequential effect of the problems discussed above is a scientific mathematical/philosophical fact: if we accept a false proposition, we can prove anything using that [38, pp. 206-7]!

**Appendix A. Supplementary material**

An Executable Notebook was created by using Matlab Live Editor containing the corresponding theoretical and numerical details, the visualizations, and some complementary information. The reader can change the code with some sliders or directly s/he can rewrite and rerun it.

**Conflicts of interest**

There are no any conflicts to declare.

*Clearly indicated that color should NOT be used for any figures in PRINT, just only for the pdf files.*

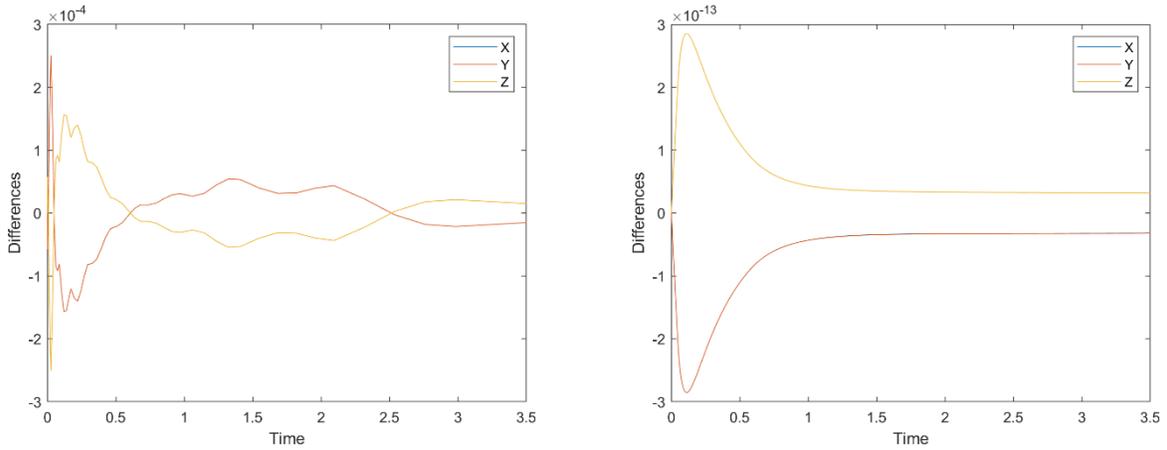

Fig. 1 Comparing the numerical and analytical solutions of the kinetic curves (concentration profiles) for the bimolecular reaction. Left panel: ode15s, AbsTol=1e-6, RelTol=1e-3; Right panel: ode45, AbsTol=1e-14, RelTol=1e-11.

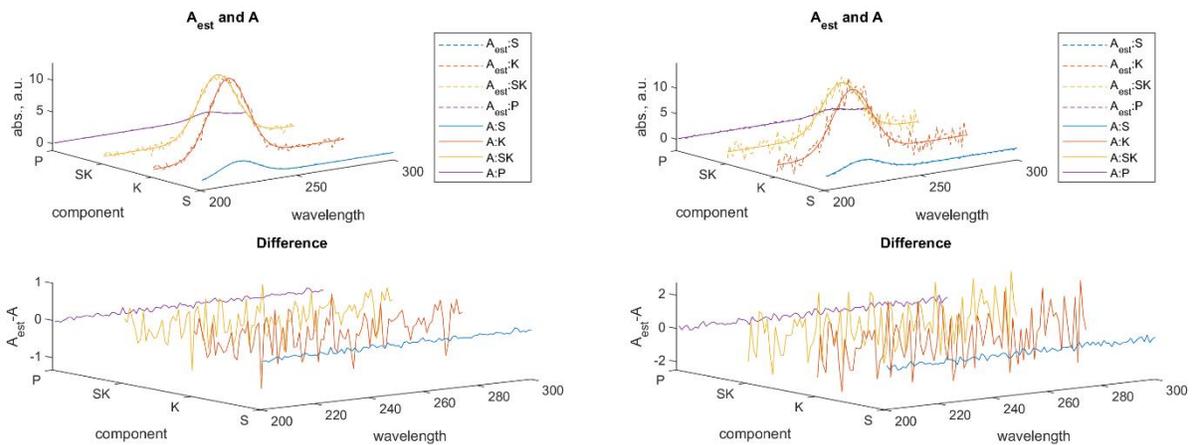

Fig. 2 Dosing for breaking rank deficiency. Left panel: continuous doses, sd=1e-4; Right panel: discrete dose (spiked reaction), sd=1e-3.

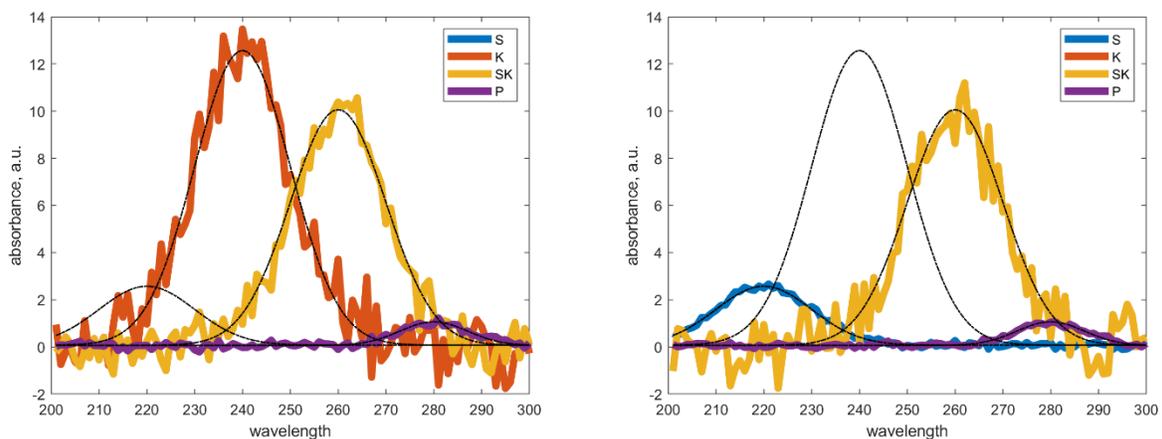

Fig. 3 Estimated spectra based on one known spectrum for Michaelis-Menten kinetics. Left panel: spectrum of substrate is known; Right panel: spectrum of enzyme is known

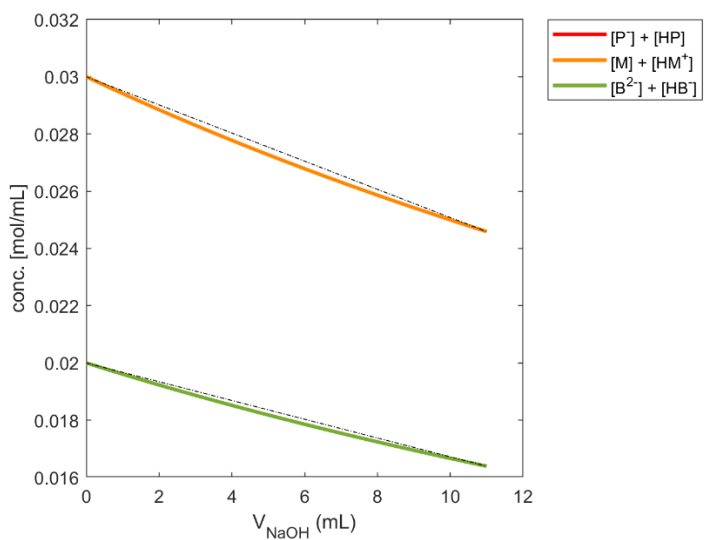

Fig. 4 The titration dilution is not straight line in case of multiple equilibria titration of monoprotic dyes

# On problematic practice of using normalization in S/MCR'

*(Supplementary Material)*

**Róbert Rajkó**

https://sites.google.com/view/prof-robert-rajko/bio

## Table of Contents

## Rank deficient spectrokinetic data for the combination reaction

"MATLAB® SimBiology® provides apps and programmatic tools for modeling, simulating, and analyzing dynamic systems. SimBiology provides a variety of techniques for analyzing ODE-based models ranging in complexity and size."

Chemical kinetic models also can be built easily. Model of the combination/addition reaction $X + Y \rightarrow Z$ can be created with the suitable Matlab code. The kinetic curves of the components are:

```
clearvars
m = sbiomodel('m');
SolvTyp = "ode15s";
AbsTol = 0.000001;
RelTol = 0.001;
configsetObj = getconfigset(m);
configsetObj.SolverType = SolvTyp;
configsetObj.SolverOptions.AbsoluteTolerance = AbsTol;
configsetObj.SolverOptions.RelativeTolerance = RelTol;
comp = addcompartment(m,'comp');
s1 = addspecies(m,'X' ,'InitialAmount',1,'InitialAmountUnits','mole');
s2 = addspecies(m,'Y' ,'InitialAmount',0.7,'InitialAmountUnits','mole');
s3 = addspecies(m,'Z','InitialAmount',0.2,'InitialAmountUnits','mole');
r1 = addreaction(m,'X + Y -> Z');
lk1 = addkineticlaw(r1,'MassAction');
p1 = addparameter(lk1,'k1','Value',12,'ValueUnits','1/(mole*second)');
lk1.ParameterVariableNames = 'k1';
csObj = getconfigset(m,'active');
set(csObj,'Stoptime',3.5);
sb = sbiosimulate(m,csObj);
```

```matlab
% [t1,C1,species1] = sbiosimulate(m,csObj);
[t1,C1,species1] = getdata(sb);
sbioplot(sb);
```

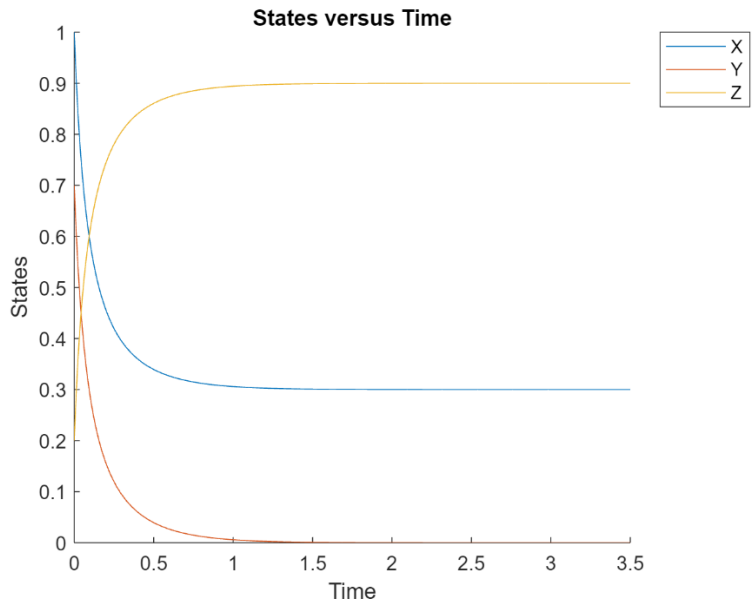

```matlab
% the diff.eq. of this type of bimolecular reaction has analytical (closed form) solution:
syms X_anal(t) k X0 Y0 Z0
cond = X_anal(0) == X0;
eqns = diff(X_anal,t) == -k*X_anal*(X_anal-X0+Y0)
```

eqns(t) =

$$\frac{\partial}{\partial t} X_{\text{anal}}(t) = -k\, X_{\text{anal}}(t)\, (Y_0 - X_0 + X_{\text{anal}}(t))$$

```matlab
X_anal = simplify(dsolve(eqns,cond))
```

X_anal =

$$\frac{X_0\, e^{X_0 k t}\, (X_0 - Y_0)}{X_0\, e^{X_0 k t} - Y_0\, e^{Y_0 k t}}$$

```matlab
Y_anal = simplify(X_anal-X0+Y0)
```

Y_anal =

$$\frac{Y_0\, e^{Y_0 k t}\, (X_0 - Y_0)}{X_0\, e^{X_0 k t} - Y_0\, e^{Y_0 k t}}$$

```
Z_anal = simplify(Z0+X0-X_anal)
```

Z_anal =

$$X_0 + Z_0 - \frac{X_0 e^{X_0 k t}(X_0 - Y_0)}{X_0 e^{X_0 k t} - Y_0 e^{Y_0 k t}}$$

```
Xansub = NaN(length(t1),1);
Yansub = Xansub;
Zansub = Xansub;
for i=1:length(t1)
    Xansub(i) = subs(X_anal,[t k X0 Y0 Z0],[t1(i) p1.Value s1.Value s2.Value s3.Value]);
    Yansub(i) = subs(Y_anal,[t k X0 Y0 Z0],[t1(i) p1.Value s1.Value s2.Value s3.Value]);
    Zansub(i) = subs(Z_anal,[t k X0 Y0 Z0],[t1(i) p1.Value s1.Value s2.Value s3.Value]);
end
Can = [Xansub Yansub Zansub];
figure
plot(t1,Can-C1); title('Comparing the numerical and analytical solutions');
xlabel('Time');
ylabel('Differences');
legend('X','Y','Z');
```

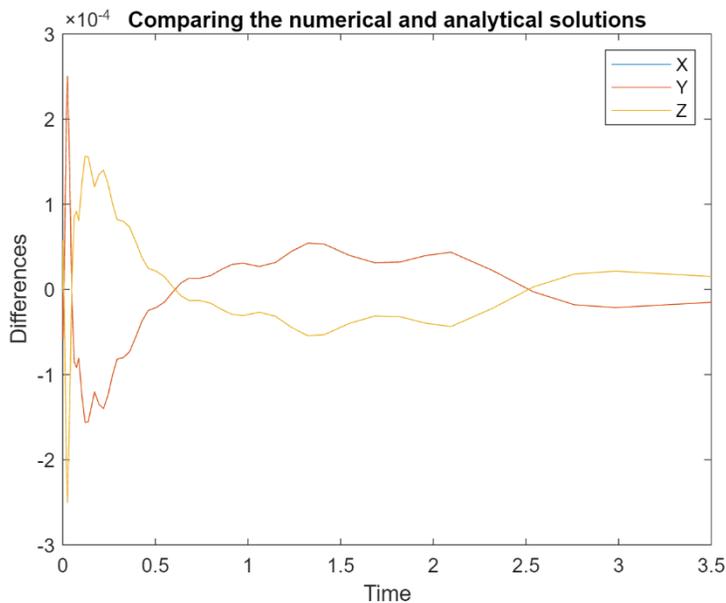

The simulated spectra are as follows:

```
x = (1:100);
A(:,1) = 2.5*exp(-(x-20).^2./200) + 0.075;
```

```
A(:,2) = 12.5*exp(-(x-40).^2./200) + 0.075;
A(:,3) = 10*exp(-(x-60).^2./200) + 0.065;

x = (201:300);
[yc,xc]=meshgrid(1:3,x);
figure
plot(xc,A); title('Simulated spectra');
xlabel('wavelength');
ylabel('absorbance, a.u.');
legend('X','Y','Z');
```

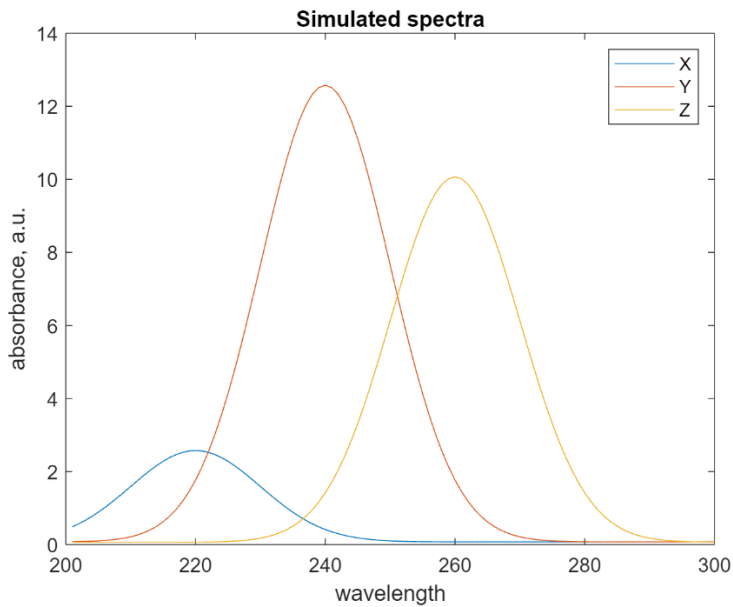

The figure of the first simulated spectrokinetic data matrix:

```
D1 = C1*A';

[xa, ya] = meshgrid(t1,x);
figure,
plot3(xa,ya,D1); hold on
[yb, xb] = meshgrid(x,t1);
plot3(xb,yb,D1); hold off
title('Spectrokinetic data matrix, D1');
xlabel('time'); ylabel('wavelength'); zlabel('absorbance, a.u.');
```

## Spectrokinetic data matrix, D1

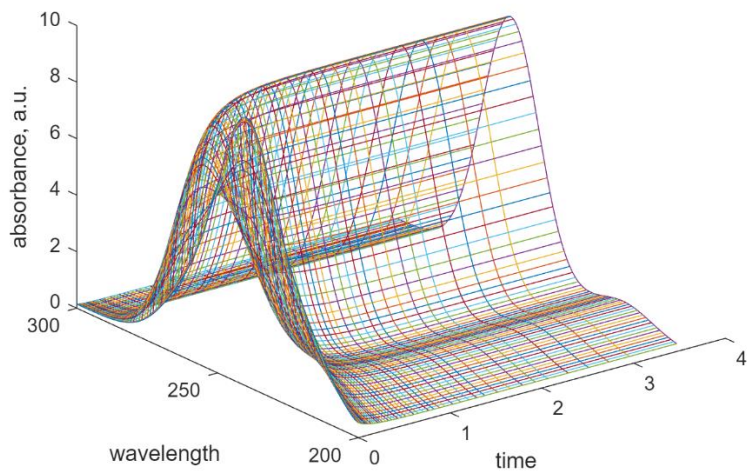

Second kinetic curves:

```
m2 = sbiomodel('m2');
configsetObj2 = getconfigset(m2);
configsetObj2.SolverType = SolvTyp;
configsetObj2.SolverOptions.AbsoluteTolerance = AbsTol;
configsetObj2.SolverOptions.RelativeTolerance = RelTol;
comp = addcompartment(m2,'comp');
s1 = addspecies(m2,'X' ,'InitialAmount',0.7,'InitialAmountUnits','mole');
s2 = addspecies(m2,'Y' ,'InitialAmount',1.0,'InitialAmountUnits','mole');
s3 = addspecies(m2,'Z','InitialAmount',0.2,'InitialAmountUnits','mole');
r1 = addreaction(m2,'X + Y -> Z');
lk1 = addkineticlaw(r1,'MassAction');
p1 = addparameter(lk1,'k1','Value',12,'ValueUnits','1/(mole*second)');
lk1.ParameterVariableNames = 'k1';
csObj2 = getconfigset(m2,'active');
set(csObj2,'Stoptime',3.5);
sb2 = sbiosimulate(m2,csObj2);
% [t2,C2,species2] = sbiosimulate(m2,csObj2);
[t2,C2,species2] = getdata(sb2);
sbioplot(sb2);
```

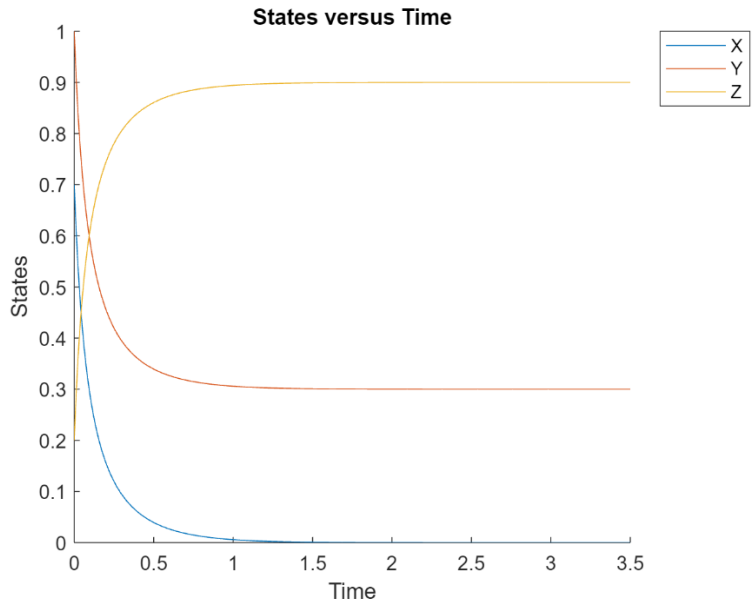

The figure of the second simulated spectrokinetic data matrix:

```
D2 = C2*A';

plot3(xa,ya,D2); hold on
plot3(xb,yb,D2); hold off
title('Spectrokinetic data matrix, D2');
xlabel('time'); ylabel('wavelength'); zlabel('absorbance, a.u.');
```

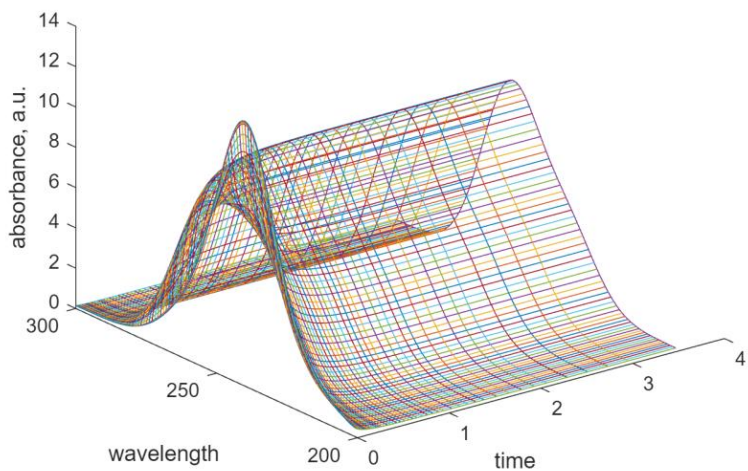

The singular values of D1, D2 and the augmented [D1;D2]:

```matlab
format shortE
sv_D1 = svd(D1); sv_D2 = svd(D2); sv_D12 = svd([D1;D2]);
array2table([sv_D1(1:5) sv_D2(1:5) sv_D12(1:5)],'VariableNames',{'D1','D2','[D1;D2]'})
```

ans = 5×3 table

|   | D1 | D2 | [D1;D2] |
|---|---|---|---|
| 1 | 2.6650e+02 | 3.4370e+02 | 4.3144e+02 |
| 2 | 9.7419e+01 | 9.6303e+01 | 1.4523e+02 |
| 3 | 3.1886e-14 | 3.4881e-14 | 2.6197e+01 |
| 4 | 2.2427e-14 | 1.7267e-14 | 2.5615e-13 |
| 5 | 5.3496e-15 | 6.8217e-15 | 2.0886e-13 |

```matlab
format default
```

Noisy simulated spectrokinetic data matrices and the estimated spectra Aest = ([Dm1;Dm2])'/[C1;C2]':

```matlab
sd =0;
Dm1=D1+sd*randn(size(D1));
Dm2=D2+sd*randn(size(D2));
% Aest = ([C1;C2]\[Dm1;Dm2])';
Aest = ([Dm1;Dm2])'/[C1;C2]';
DfA = Aest-A;
figure,
subplot(2,1,1);
plot3(xa,ya,Dm1); hold on
plot3(xb,yb,Dm1); hold off
title('Dm1');
xlabel('time'); ylabel('\lambda'); zlabel('abs., a.u.');
subplot(2,1,2);
plot3(xa,ya,Dm2); hold on
plot3(xb,yb,Dm2); hold off
title('Dm2');
xlabel('time'); ylabel('\lambda'); zlabel('abs., a.u.');
sgtitle('Noisy spectrokinetic data matrices');
```

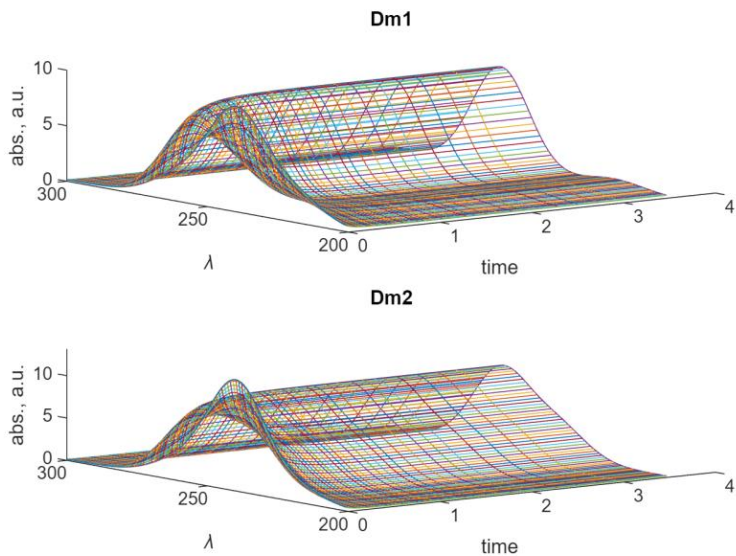

```
figure,
subplot(2,1,1);
colororder({'#0072BD','#D95319','#EDB120'})
plot3(xc,yc,Aest,'--'); hold on
colororder({'#0072BD','#D95319','#EDB120'})
plot3(xc,yc,A,'-'); hold off
legend('A_{est}:X','A_{est}:Y','A_{est}:Z','A:X','A:Y','A:Z')
title('A_{est} and A');
xlabel('wavelength'); ylabel('component'); zlabel('abs., a.u.');
yticks([1 2 3]); yticklabels({'X','Y','Z'});
subplot(2,1,2);
plot3(xc,yc,DfA);
title('Difference');
xlabel('wavelength'); ylabel('component'); zlabel('A_{est}-A');
yticks([1 2 3]); yticklabels({'X','Y','Z'});
```

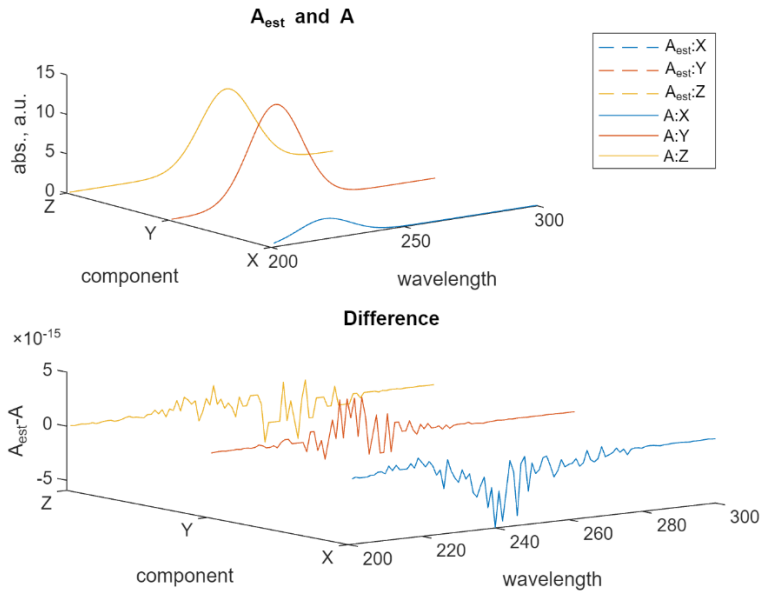

## Michaelis-Menten kinetic model (rank deficient data matrix?)

"MATLAB® SimBiology® provides apps and programmatic tools for modeling, simulating, and analyzing dynamic systems. SimBiology provides a variety of techniques for analyzing ODE-based models ranging in complexity and size."

The Michaelis-Menten kinetic model is $S + K \underset{k_{-1}}{\overset{k_1}{\rightleftharpoons}} [SK] \overset{k_2}{\to} P + K$. The kinetic curves of the components are:

```
clearvars
m = sbiomodel('m');
SolvTyp = "ode15s";
configsetObj = getconfigset(m); set(configsetObj, 'SolverType', SolvTyp);
comp = addcompartment(m,'comp');
% S_conc0 = 1; K_conc0 = 0.1; SK_conc0 = 0; P_conc0 = 0;
S_conc0 = 1;
K_conc0 = 0.1;
SK_conc0 =0;
P_conc0 =  0;
s1 = addspecies(m,'S' ,'InitialAmount',S_conc0,'InitialAmountUnits','mole');
s2 = addspecies(m,'K' ,'InitialAmount',K_conc0,'InitialAmountUnits','mole');
s3 = addspecies(m,'SK','InitialAmount',SK_conc0,'InitialAmountUnits','mole');
s4 = addspecies(m,'P' ,'InitialAmount',P_conc0,'InitialAmountUnits','mole');
r1 = addreaction(m,'S + K -> SK');
lk1 = addkineticlaw(r1,'MassAction');
p1 = addparameter(lk1,'k1','Value',20,'ValueUnits','1/(mole*second)');
lk1.ParameterVariableNames = 'k1';
```

```
r1r = addreaction(m,'SK -> S + K');
lk1r = addkineticlaw(r1r,'MassAction');
p1r = addparameter(lk1r,'k1r','Value',0.1,'ValueUnits','1/second');
lk1r.ParameterVariableNames = 'k1r';
r2 = addreaction(m,'SK -> K + P');
lk2 = addkineticlaw(r2,'MassAction');
p2 = addparameter(lk2,'k2','Value',3,'ValueUnits','1/second');
lk2.ParameterVariableNames = 'k2';
csObj = getconfigset(m,'active');
set(csObj,'Stoptime',7.5);
sb = sbiosimulate(m,csObj);
% [t,Ct,species] = sbiosimulate(m,csObj);
[t,Ct,species] =  getdata(sb);
lCt = length(Ct);
sbioplot(sb);
```

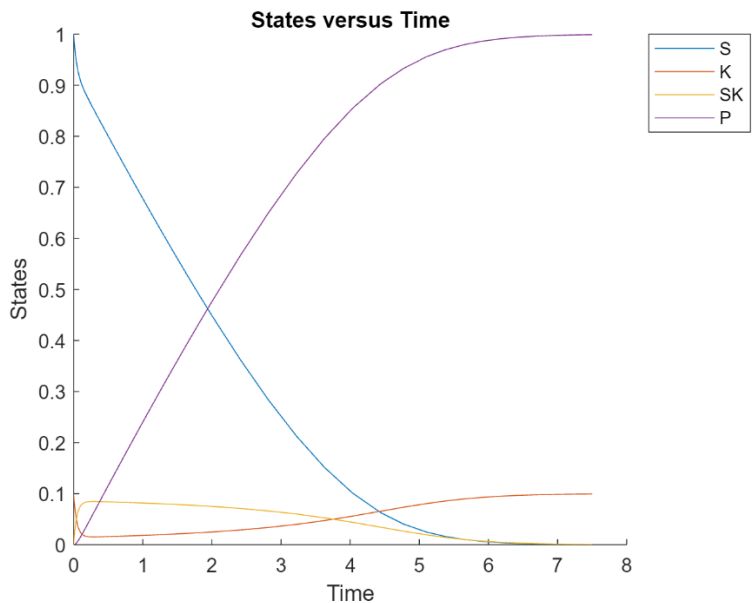

The simulated spectra are as follows:

```
x = (1:100);
A(:,1) = 2.5*exp(-(x-20).^2./200) + 0.075;
A(:,2) = 12.5*exp(-(x-40).^2./200) + 0.075;
A(:,3) = 10*exp(-(x-60).^2./200) + 0.065;
A(:,4) = exp(-(x-80).^2./100) + 0.065;

figure
x = (201:300);
[yc,xc]=meshgrid(1:4,x);
plot(xc,A); title('Simulated spectra');
```

```
xlabel('wavelength');
ylabel('absorbance, a.u.');
legend('S','K','SK','P');
```

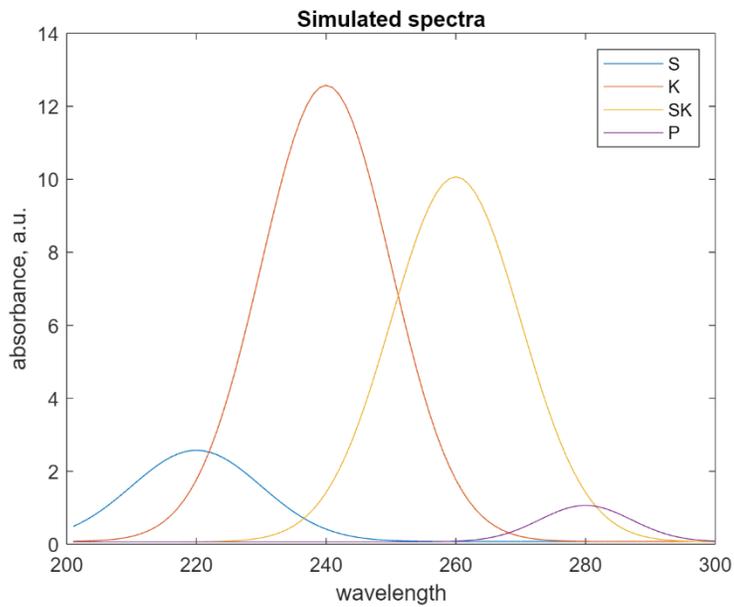

The figure of the Michaelis-Menten simulated spectrokinetic data matrix:

```
D = Ct*A';

figure
[xa, ya] = meshgrid(t,x);
plot3(xa,ya,D); hold on
[yb, xb] = meshgrid(x,t);
plot3(xb,yb,D); hold off
title('Michaelis-Menten simulated spectrokinetic data matrix');
xlabel('time'); ylabel('wavelength'); zlabel('absorbance, a.u.');
```

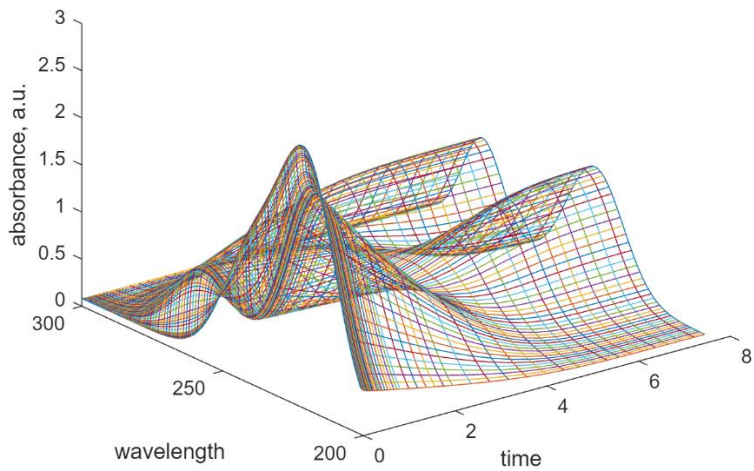

**Michaelis-Menten simulated spectrokinetic data matrix**

Singular values of matrix C by SVD:

```
format long
sv_C = svd(Ct); display(sv_C);
```

sv_C = *4×1*

```
   7.054525774793125
   3.848496628860587
   0.434969142486790
   0.000000000003752
```

```
format default
```

Singular values of matrix D by SVD:

```
format long
sv_D = svd(D); display(sv_D);
```

sv_D = *78×1*

```
  93.840778438613242
  20.309655081533617
  15.270043490066083
   0.000000000115671
   0.000000000000061
   0.000000000000049
   0.000000000000040
   0.000000000000038
   0.000000000000036
   0.000000000000034
       ⋮
```

```
format default
```

Is there closure in C?

```
sumCt = Ct(:,1)+Ct(:,2)+Ct(:,3)+Ct(:,4);
sumCt2 = Ct(:,1)+Ct(:,2)+2*Ct(:,3)+Ct(:,4);
format long
table(min(sumCt),max(sumCt),mean(sumCt),std(sumCt),'VariableNames',{'min(sum)',
'max(sum)','mean(sum)','std(sum)'})
```

ans = 1×4 table

|   | min(sum) | max(sum) | mean(sum) | std(sum) |
|---|---|---|---|---|
| 1 | 1.015381125953578 | 1.100000000000000 | 1.056640557680636 | 0.035737909005802 |

```
table(sumCt,sumCt2,'VariableNames',{'sum=c_S(t)+c_K(t)+c_{SK}(t)+c_P(t)','sum=c
_S(t)+c_K(t)+2*c_{SK}(t)+c_P(t)'})
```

ans = 78×2 table

|    | sum=c_S(t)+c_K(t)+c_{SK}(t)+c_P(t) | ... |
|----|---|---|
| 1  | 1.100000000000000 |   |
| 2  | 1.099981031930935 |   |
| 3  | 1.099962068266987 |   |
| 4  | 1.099946301581364 |   |
| 5  | 1.099930538029919 |   |
| 6  | 1.099914777600837 |   |
| 7  | 1.099813886475319 |   |
| 8  | 1.099713121313497 |   |
| 9  | 1.099612483259717 |   |
| 10 | 1.099511972605304 |   |
| 11 | 1.099254653796943 |   |
| 12 | 1.098998170104782 |   |

| | sum=c_S(t)+c_K(t)+c_{SK}(t)+c_P(t) | ... |
|---|---:|---|
| 13 | 1.098742518593655 | |
| 14 | 1.098487696240583 | |
| 15 | 1.097316052407211 | |
| 16 | 1.096161901465181 | |
| 17 | 1.095024951968377 | |
| 18 | 1.093904918746986 | |
| 19 | 1.092801522565255 | |
| 20 | 1.087900794442886 | |
| 21 | 1.083325378229122 | |
| 22 | 1.079051669841349 | |
| 23 | 1.075058005889322 | |
| 24 | 1.071324474511347 | |
| 25 | 1.067832761511925 | |
| 26 | 1.058297809014407 | |
| 27 | 1.050529205871888 | |
| 28 | 1.044185138455362 | |
| 29 | 1.038994637595994 | |
| 30 | 1.034740254898234 | |
| 31 | 1.031247506066888 | |
| 32 | 1.026821019142728 | |
| 33 | 1.023600401177538 | |
| 34 | 1.021258605060407 | |
| 35 | 1.019557129997499 | |

| | sum=c_S(t)+c_K(t)+c_{SK}(t)+c_P(t) | ... |
|---|---:|---|
| 36 | 1.018320762676041 | |
| 37 | 1.017423029151396 | |
| 38 | 1.016590242788145 | |
| 39 | 1.016059056083347 | |
| 40 | 1.015730639748041 | |
| 41 | 1.015537191354916 | |
| 42 | 1.015431826016122 | |
| 43 | 1.015384074647325 | |
| 44 | 1.015381125953578 | |
| 45 | 1.015436522125700 | |
| 46 | 1.015526393840384 | |
| 47 | 1.015635108247661 | |
| 48 | 1.015751888641345 | |
| 49 | 1.015872217307335 | |
| 50 | 1.016131208899934 | |
| 51 | 1.016405875729230 | |
| 52 | 1.016695002653713 | |
| 53 | 1.016990950312679 | |
| 54 | 1.017291848949248 | |
| 55 | 1.017601909566378 | |
| 56 | 1.018553566746067 | |
| 57 | 1.019603677723131 | |
| 58 | 1.020763743189444 | |

|    | sum=c_S(t)+c_K(t)+c_{SK}(t)+c_P(t) | ... |
|----|---|---|
| 59 | 1.022039555045714 | |
| 60 | 1.023444294758847 | |
| 61 | 1.024999474661172 | |
| 62 | 1.028911597394933 | |
| 63 | 1.033805146713057 | |
| 64 | 1.039874361753723 | |
| 65 | 1.047298426551115 | |
| 66 | 1.056022871213184 | |
| 67 | 1.065595562775193 | |
| 68 | 1.073511121916884 | |
| 69 | 1.080746807581459 | |
| 70 | 1.085608616217509 | |
| 71 | 1.089600233098746 | |
| 72 | 1.092717138343344 | |
| 73 | 1.095037153524661 | |
| 74 | 1.096692862511740 | |
| 75 | 1.097836560864105 | |
| 76 | 1.098659818250036 | |
| 77 | 1.099183275031651 | |
| 78 | 1.099457399871651 | |

```
format default
```

Mass conversations in the Michaelis-Menten kinetic model: $c_K(t) + c_{SK}(t) = c_K(0) = 0.1$ and $c_S(t) + c_{SK}(t) + c_P(t) = c_S(0) = 1$.

```
format long
table(Ct(:,2),Ct(:,3),Ct(:,2)+Ct(:,3),repelem(Ct(1,2),lCt,1),'VariableNames',{'c_K(t) +','c_{SK}(t) =','c_K(t)+c_{SK}(t)','c_K(0)'})
```

ans = 78×4 table

|    | c_K(t) +          | c_{SK}(t) =       | c_K(t)+c_{SK}(t)  | c_K(0)            |
|----|-------------------|-------------------|-------------------|-------------------|
| 1  | 0.100000000000000 | 0                 | 0.100000000000000 | 0.100000000000000 |
| 2  | 0.099981031930935 | 0.000018968069065 | 0.100000000000000 | 0.100000000000000 |
| 3  | 0.099962068266988 | 0.000037931733012 | 0.100000000000000 | 0.100000000000000 |
| 4  | 0.099946301581364 | 0.000053698418636 | 0.100000000000000 | 0.100000000000000 |
| 5  | 0.099930538029919 | 0.000069461970081 | 0.100000000000000 | 0.100000000000000 |
| 6  | 0.099914777600837 | 0.000085222399163 | 0.100000000000000 | 0.100000000000000 |
| 7  | 0.099813886475319 | 0.000186113524681 | 0.100000000000000 | 0.100000000000000 |
| 8  | 0.099713121313497 | 0.000286878686503 | 0.100000000000000 | 0.100000000000000 |
| 9  | 0.099612483259717 | 0.000387516740283 | 0.100000000000000 | 0.100000000000000 |
| 10 | 0.099511972605304 | 0.000488027394696 | 0.100000000000000 | 0.100000000000000 |
| 11 | 0.099254653796943 | 0.000745346203057 | 0.100000000000000 | 0.100000000000000 |
| 12 | 0.098998170104782 | 0.001001829895218 | 0.100000000000000 | 0.100000000000000 |
| 13 | 0.098742518593655 | 0.001257481406345 | 0.100000000000000 | 0.100000000000000 |
| 14 | 0.098487696240583 | 0.001512303759417 | 0.100000000000000 | 0.100000000000000 |
| 15 | 0.097316052407211 | 0.002683947592789 | 0.100000000000000 | 0.100000000000000 |
| 16 | 0.096161901465181 | 0.003838098534819 | 0.100000000000000 | 0.100000000000000 |
| 17 | 0.095024951968377 | 0.004975048031623 | 0.100000000000000 | 0.100000000000000 |
| 18 | 0.093904918746986 | 0.006095081253014 | 0.100000000000000 | 0.100000000000000 |
| 19 | 0.092801522565254 | 0.007198477434746 | 0.100000000000000 | 0.100000000000000 |
| 20 | 0.087900794442885 | 0.012099205557115 | 0.100000000000000 | 0.100000000000000 |

|    | c_K(t) + | c_{SK}(t) = | c_K(t)+c_{SK}(t) | c_K(0) |
|---|---|---|---|---|
| 21 | 0.083325378229123 | 0.016674621770877 | 0.100000000000000 | 0.100000000000000 |
| 22 | 0.079051669841354 | 0.020948330158646 | 0.100000000000000 | 0.100000000000000 |
| 23 | 0.075058005889332 | 0.024941994110668 | 0.100000000000000 | 0.100000000000000 |
| 24 | 0.071324474511361 | 0.028675525488639 | 0.100000000000000 | 0.100000000000000 |
| 25 | 0.067832761511941 | 0.032167238488059 | 0.100000000000000 | 0.100000000000000 |
| 26 | 0.058297809014421 | 0.041702190985579 | 0.100000000000000 | 0.100000000000000 |
| 27 | 0.050529205872377 | 0.049470794127623 | 0.100000000000000 | 0.100000000000000 |
| 28 | 0.044185138457086 | 0.055814861542914 | 0.100000000000000 | 0.100000000000000 |
| 29 | 0.038994637599309 | 0.061005362400691 | 0.100000000000000 | 0.100000000000000 |
| 30 | 0.034740254902852 | 0.065259745097148 | 0.100000000000000 | 0.100000000000000 |
| 31 | 0.031247506072265 | 0.068752493927735 | 0.100000000000000 | 0.100000000000000 |
| 32 | 0.026821019148331 | 0.073178980851669 | 0.100000000000000 | 0.100000000000000 |
| 33 | 0.023600401185528 | 0.076399598814472 | 0.100000000000000 | 0.100000000000000 |
| 34 | 0.021258605071094 | 0.078741394928906 | 0.100000000000000 | 0.100000000000000 |
| 35 | 0.019557130010974 | 0.080442869989026 | 0.100000000000000 | 0.100000000000000 |
| 36 | 0.018320762692086 | 0.081679237307914 | 0.100000000000000 | 0.100000000000000 |
| 37 | 0.017423029169287 | 0.082576970830713 | 0.100000000000000 | 0.100000000000000 |
| 38 | 0.016590242807046 | 0.083409757192954 | 0.100000000000000 | 0.100000000000000 |
| 39 | 0.016059056102240 | 0.083940943897759 | 0.100000000000000 | 0.100000000000000 |
| 40 | 0.015730639766831 | 0.084269360233169 | 0.100000000000000 | 0.100000000000000 |
| 41 | 0.015537191374869 | 0.084462808625131 | 0.100000000000000 | 0.100000000000000 |
| 42 | 0.015431826037866 | 0.084568173962134 | 0.100000000000000 | 0.100000000000000 |
| 43 | 0.015384074670290 | 0.084615925329709 | 0.100000000000000 | 0.100000000000000 |

|    | c_K(t) +         | c_{SK}(t) =      | c_K(t)+c_{SK}(t) | c_K(0)           |
|----|------------------|------------------|------------------|------------------|
| 44 | 0.015381125977459 | 0.084618874022541 | 0.100000000000000 | 0.100000000000000 |
| 45 | 0.015436522149514 | 0.084563477850486 | 0.100000000000000 | 0.100000000000000 |
| 46 | 0.015526393864196 | 0.084473606135804 | 0.100000000000000 | 0.100000000000000 |
| 47 | 0.015635108272955 | 0.084364891727045 | 0.100000000000000 | 0.100000000000000 |
| 48 | 0.015751888668428 | 0.084248111331572 | 0.100000000000000 | 0.100000000000000 |
| 49 | 0.015872217335069 | 0.084127782664930 | 0.100000000000000 | 0.100000000000000 |
| 50 | 0.016131208928229 | 0.083868791071771 | 0.100000000000000 | 0.100000000000000 |
| 51 | 0.016405875757903 | 0.083594124242097 | 0.100000000000000 | 0.100000000000000 |
| 52 | 0.016695002683833 | 0.083304997316167 | 0.100000000000000 | 0.100000000000000 |
| 53 | 0.016990950344263 | 0.083009049655737 | 0.100000000000000 | 0.100000000000000 |
| 54 | 0.017291848981132 | 0.082708151018868 | 0.100000000000000 | 0.100000000000000 |
| 55 | 0.017601909597820 | 0.082398090402180 | 0.100000000000000 | 0.100000000000000 |
| 56 | 0.018553566780084 | 0.081446433219916 | 0.100000000000000 | 0.100000000000000 |
| 57 | 0.019603677755387 | 0.080396322244613 | 0.100000000000000 | 0.100000000000000 |
| 58 | 0.020763743220215 | 0.079236256779785 | 0.100000000000000 | 0.100000000000000 |
| 59 | 0.022039555076787 | 0.077960444923213 | 0.100000000000000 | 0.100000000000000 |
| 60 | 0.023444294789462 | 0.076555705210537 | 0.100000000000000 | 0.100000000000000 |
| 61 | 0.024999474689780 | 0.075000525310220 | 0.100000000000000 | 0.100000000000000 |
| 62 | 0.028911597431727 | 0.071088402568273 | 0.100000000000000 | 0.100000000000000 |
| 63 | 0.033805146771139 | 0.066194853228861 | 0.100000000000000 | 0.100000000000000 |
| 64 | 0.039874361828110 | 0.060125638171890 | 0.100000000000000 | 0.100000000000000 |
| 65 | 0.047298426634372 | 0.052701573365628 | 0.100000000000000 | 0.100000000000000 |
| 66 | 0.056022871299171 | 0.043977128700828 | 0.100000000000000 | 0.100000000000000 |

|    | c_K(t) +          | c_{SK}(t) =       | c_K(t)+c_{SK}(t)   | c_K(0)            |
|----|-------------------|-------------------|--------------------|-------------------|
| 67 | 0.065595562861107 | 0.034404437138893 | 0.100000000000000  | 0.100000000000000 |
| 68 | 0.073511122002664 | 0.026488877997336 | 0.100000000000000  | 0.100000000000000 |
| 69 | 0.080746807667357 | 0.019253192332643 | 0.100000000000000  | 0.100000000000000 |
| 70 | 0.085608616303471 | 0.014391383696529 | 0.100000000000000  | 0.100000000000000 |
| 71 | 0.089600233184702 | 0.010399766815298 | 0.100000000000000  | 0.100000000000000 |
| 72 | 0.092717138429277 | 0.007282861570723 | 0.100000000000000  | 0.100000000000000 |
| 73 | 0.095037153610585 | 0.004962846389415 | 0.100000000000000  | 0.100000000000000 |
| 74 | 0.096692862597670 | 0.003307137402330 | 0.100000000000000  | 0.100000000000000 |
| 75 | 0.097836560950040 | 0.002163439049960 | 0.100000000000000  | 0.100000000000000 |
| 76 | 0.098659818335973 | 0.001340181664027 | 0.100000000000000  | 0.100000000000000 |
| 77 | 0.099183275117585 | 0.000816724882415 | 0.100000000000000  | 0.100000000000000 |
| 78 | 0.099457399957585 | 0.000542600042415 | 0.100000000000000  | 0.100000000000000 |

```
table(Ct(:,1),Ct(:,3),Ct(:,4),Ct(:,1)+Ct(:,3)+Ct(:,4),repelem(Ct(1,1),lCt,1),'V
ariableNames',{'c_S(t) +','c_{SK}(t) +','c_P(t)
=','c_S(t)+c_{SK}(t)+c_P(t)','c_S(0)'})
```

ans = 78×5 table

|   | c_S(t) +          | c_{SK}(t) +       | c_P(t) =          | c_S(t)+c_{SK}(t)+c_P(t) | ... |
|---|-------------------|-------------------|-------------------|-------------------------|-----|
| 1 | 1                 | 0                 | 0                 | 1                       |     |
| 2 | 0.999981031475416 | 0.000018968069065 | 0.000000000455519 | 1                       |     |
| 3 | 0.999962066829422 | 0.000037931733012 | 0.000000001437566 | 1                       |     |
| 4 | 0.999946298943917 | 0.000053698418636 | 0.000000002637447 | 1                       |     |
| 5 | 0.999930533817956 | 0.000069461970081 | 0.000000004211963 | 1                       |     |
| 6 | 0.999914771441127 | 0.000085222399163 | 0.000000006159710 | 1                       |     |

| | c_S(t) + | c_{SK}(t) + | c_P(t) = | c_S(t)+c_{SK}(t)+c_P(t) | ... |
|---|---|---|---|---|---|
| 7  | 0.999813859449419 | 0.000186113524681 | 0.000000027025900 | 1 | |
| 8  | 0.999713058361731 | 0.000286878686503 | 0.000000062951766 | 1 | |
| 9  | 0.999612369183457 | 0.000387516740283 | 0.000000114076260 | 1 | |
| 10 | 0.999511792168843 | 0.000488027394696 | 0.000000180436461 | 1 | |
| 11 | 0.999254233505318 | 0.000745346203057 | 0.000000420291625 | 1 | |
| 12 | 0.998997410032895 | 0.001001829895218 | 0.000000760071887 | 1 | |
| 13 | 0.998741319130854 | 0.001257481406345 | 0.000001199462801 | 1 | |
| 14 | 0.998485958100514 | 0.001512303759417 | 0.000001738140069 | 1 | |
| 15 | 0.997310524679806 | 0.002683947592789 | 0.000005527727405 | 1 | |
| 16 | 0.996150485092044 | 0.003838098534819 | 0.000011416373137 | 1 | |
| 17 | 0.995005579308064 | 0.004975048031623 | 0.000019372660313 | 1 | |
| 18 | 0.993875552989212 | 0.006095081253014 | 0.000029365757774 | 1 | |
| 19 | 0.992760157176902 | 0.007198477434746 | 0.000041365388353 | 1 | |
| 20 | 0.987778542612394 | 0.012099205557115 | 0.000122251830492 | 1 | |
| 21 | 0.983082654126592 | 0.016674621770877 | 0.000242724102530 | 0.999999999999999 | |
| 22 | 0.978651504712362 | 0.020948330158646 | 0.000400165128988 | 0.999999999999995 | |
| 23 | 0.974465860448249 | 0.024941994110668 | 0.000592145441073 | 0.999999999999990 | |
| 24 | 0.970508066515016 | 0.028675525488639 | 0.000816407996331 | 0.999999999999986 | |
| 25 | 0.966761906034554 | 0.032167238488059 | 0.001070855477371 | 0.999999999999984 | |
| 26 | 0.956257785282224 | 0.041702190985579 | 0.002040023732183 | 0.999999999999986 | |
| 27 | 0.947294995102439 | 0.049470794127623 | 0.003234210769449 | 0.999999999999511 | |
| 28 | 0.939573033039863 | 0.055814861542914 | 0.004612105415500 | 0.999999999998276 | |
| 29 | 0.932854341655054 | 0.061005362400691 | 0.006140295940940 | 0.999999999996684 | |

| | c_S(t) + | c_{SK}(t) + | c_P(t) = | c_S(t)+c_{SK}(t)+c_P(t) | ... |
|---|---|---|---|---|---|
| 30 | 0.926948697652883 | 0.065259745097148 | 0.007791557245351 | 0.999999999995383 | |
| 31 | 0.921703790847992 | 0.068752493927735 | 0.009543715218896 | 0.999999999994623 | |
| 32 | 0.914243355362366 | 0.073178980851669 | 0.012577663780362 | 0.999999999994397 | |
| 33 | 0.907827243159659 | 0.076399598814472 | 0.015773158017879 | 0.999999999992010 | |
| 34 | 0.902171842309864 | 0.078741394928906 | 0.019086762750543 | 0.999999999989313 | |
| 35 | 0.897070709782259 | 0.080442869989026 | 0.022486420215240 | 0.999999999986525 | |
| 36 | 0.892372317685017 | 0.081679237307914 | 0.025948444991024 | 0.999999999983955 | |
| 37 | 0.887967404498956 | 0.082576970830713 | 0.029455624652440 | 0.999999999982109 | |
| 38 | 0.882355864636325 | 0.083409757192954 | 0.034234378151820 | 0.999999999981099 | |
| 39 | 0.877007287559026 | 0.083940943897759 | 0.039051768524322 | 0.999999999981107 | |
| 40 | 0.871836773155247 | 0.084269360233169 | 0.043893866592793 | 0.999999999981210 | |
| 41 | 0.866786183286304 | 0.084462808625131 | 0.048751008068612 | 0.999999999980047 | |
| 42 | 0.861815202051116 | 0.084568173962134 | 0.053616623965006 | 0.999999999978256 | |
| 43 | 0.856897593513953 | 0.084615925329709 | 0.058486481133371 | 0.999999999977034 | |
| 44 | 0.848870072123484 | 0.084618874022541 | 0.066511053830094 | 0.999999999976119 | |
| 45 | 0.840903708501659 | 0.084563477850486 | 0.074532813624041 | 0.999999999976186 | |
| 46 | 0.832978366883572 | 0.084473606135804 | 0.082548026956812 | 0.999999999976188 | |
| 47 | 0.825081008611200 | 0.084364891727045 | 0.090554099636462 | 0.999999999974707 | |
| 48 | 0.817202453315287 | 0.084248111331572 | 0.098549435326058 | 0.999999999972916 | |
| 49 | 0.809338837765879 | 0.084127782664930 | 0.106533379541456 | 0.999999999972265 | |
| 50 | 0.792936138666642 | 0.083868791071771 | 0.123195070233292 | 0.999999999971705 | |
| 51 | 0.776601931654413 | 0.083594124242097 | 0.139803944074817 | 0.999999999971327 | |
| 52 | 0.760337081873154 | 0.083304997316167 | 0.156357920780559 | 0.999999999969880 | |

|    | c_S(t) + | c_{SK}(t) + | c_P(t) = | c_S(t)+c_{SK}(t)+c_P(t) | ... |
|----|----------|-------------|----------|-------------------------|-----|
| 53 | 0.744136700924349 | 0.083009049655737 | 0.172854249388331 | 0.999999999968416 | |
| 54 | 0.728000807460025 | 0.082708151018868 | 0.189291041489223 | 0.999999999968116 | |
| 55 | 0.711934908464484 | 0.082398090402180 | 0.205667001101894 | 0.999999999968558 | |
| 56 | 0.665985547208777 | 0.081446433219916 | 0.252568019537290 | 0.999999999965983 | |
| 57 | 0.620706798087609 | 0.080396322244613 | 0.298896879635522 | 0.999999999967743 | |
| 58 | 0.576168068633679 | 0.079236256779785 | 0.344595674555765 | 0.999999999969229 | |
| 59 | 0.532440678911392 | 0.077960444923213 | 0.389598876134322 | 0.999999999968927 | |
| 60 | 0.489609123638135 | 0.076555705210537 | 0.433835171120711 | 0.999999999969384 | |
| 61 | 0.447774966376944 | 0.075000525310220 | 0.477224508284228 | 0.999999999971392 | |
| 62 | 0.362680494094298 | 0.071088402568273 | 0.566231103300636 | 0.999999999963207 | |
| 63 | 0.283903574320504 | 0.066194853228861 | 0.649901572392553 | 0.999999999941918 | |
| 64 | 0.212951899368885 | 0.060125638171890 | 0.726922462384838 | 0.999999999925613 | |
| 65 | 0.151549291723079 | 0.052701573365628 | 0.795749134828036 | 0.999999999916743 | |
| 66 | 0.101271817417802 | 0.043977128700828 | 0.854751053795383 | 0.999999999914013 | |
| 67 | 0.063021096055700 | 0.034404437138893 | 0.902574466719493 | 0.999999999914086 | |
| 68 | 0.040313161066519 | 0.026488877997336 | 0.933197960850366 | 0.999999999914220 | |
| 69 | 0.024624845796406 | 0.019253192332643 | 0.956121961785052 | 0.999999999914102 | |
| 70 | 0.016230720564216 | 0.014391383696529 | 0.969377895653293 | 0.999999999914038 | |
| 71 | 0.010479024425279 | 0.010399766815298 | 0.979121208673467 | 0.999999999914044 | |
| 72 | 0.006659708696054 | 0.007282861570723 | 0.986057429647291 | 0.999999999914067 | |
| 73 | 0.004183919685525 | 0.004962846389415 | 0.990853233839135 | 0.999999999914075 | |
| 74 | 0.002605656949010 | 0.003307137402330 | 0.994087205562730 | 0.999999999914070 | |
| 75 | 0.001611952264519 | 0.002163439049960 | 0.996224608599586 | 0.999999999914064 | |

|    | c_S(t) + | c_{SK}(t) + | c_P(t) = | c_S(t)+c_{SK}(t)+c_P(t) | ... |
|----|----------|-------------|----------|--------------------------|-----|
| 76 | 0.000951005447611 | 0.001340181664027 | 0.997708812802425 | 0.999999999914064 | |
| 77 | 0.000559027357348 | 0.000816724882415 | 0.998624247674303 | 0.999999999914066 | |
| 78 | 0.000364726144740 | 0.000542600042415 | 0.999092673726911 | 0.999999999914067 | |

```
format default
```

The simple Matlab command `Aest = Dm'/Cest'` can be used for the unique determination for noisy data?

```
sd =0;
Dm=D+sd*randn(size(D));
% Aest = (Ct\Dm)';
Aest = Dm'/Ct';
DfA = Aest-A;
figure,
plot3(xa,ya,Dm); hold on
plot3(xb,yb,Dm); hold off
xlabel('time'); ylabel('wavelength'); zlabel('abs., a.u.');
title('Noisy Michaelis-Menten kinetic data matrix, Dm');
```

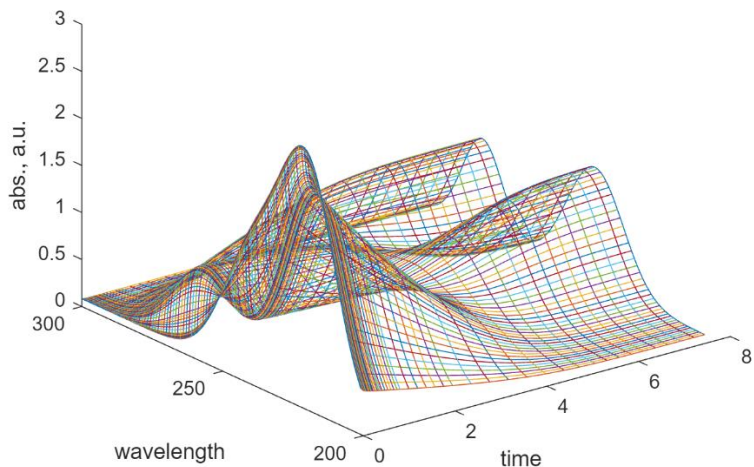

```
figure,
subplot(2,1,1);
colororder({'#0072BD','#D95319','#EDB120','#7E2F8E'})
```

```matlab
plot3(xc,yc,Aest,'--'); hold on
colororder({'#0072BD','#D95319','#EDB120','#7E2F8E'})
plot3(xc,yc,A,'-'); hold off
legend('A_{est}:S','A_{est}:K','A_{est}:SK','A_{est}:P','A:S','A:K','A:SK','A:P')
title('A_{est} and A');
xlabel('wavelength'); ylabel('component'); zlabel('abs., a.u.');
yticks([1 2 3 4]); yticklabels({'S','K','SK','P'});
subplot(2,1,2);
plot3(xc,yc,DfA);
title('Difference');
xlabel('wavelength'); ylabel('component'); zlabel('A_{est}-A');
yticks([1 2 3 4]); yticklabels({'S','K','SK','P'});
```

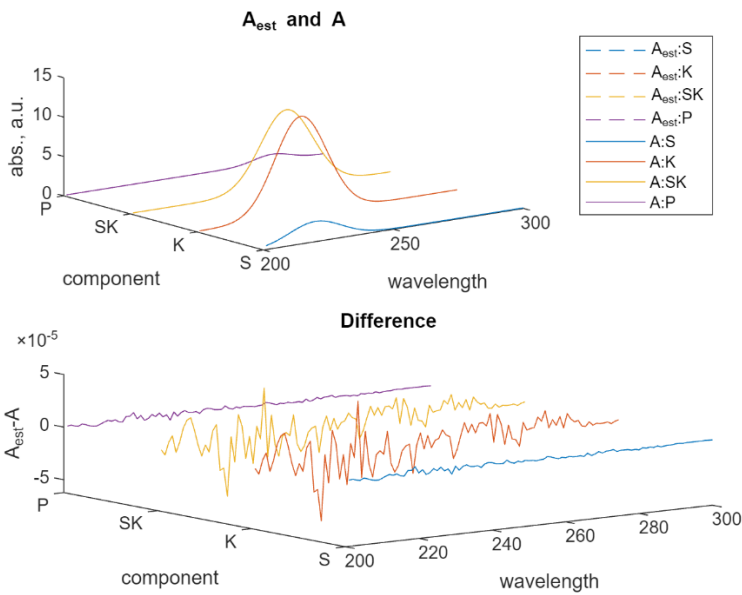

Proper ode solver in Matlab:

```matlab
% % ode89
k1=p1.Value; k1r=p1r.Value; k2=p2.Value;
% y0 = [s1.Value s2.Value s3.Value s4.Value];
y0 = [S_conc0 K_conc0 SK_conc0 P_conc0];
tspan = [0 csObj.Stoptime];
[tw,Cw] = ode89(@(t,y) odefn_MichMent(t,y,k1,k1r,k2), tspan, y0);
lCtw=length(tw);

figure, plot(tw,Cw,'LineWidth',2); hold on
plot(t,Ct,'k.','MarkerSize', 12); hold off
legend([species; {SolvTyp}]);
xlabel('Time');
```

```
ylabel('Species Amount');
title('ode89 - solving nonstiff differential equations — high order method');
```

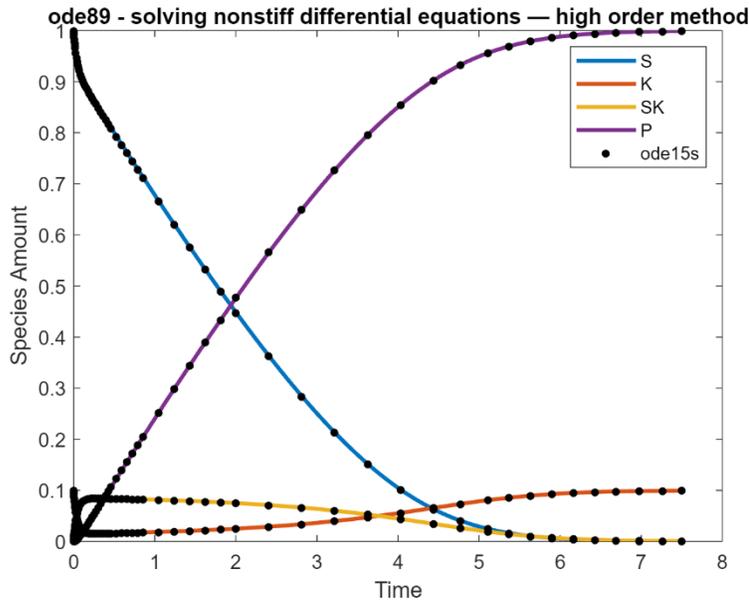

Singular values of matrix C by SVD:

```
format long
sv_Cw = svd(Cw); display(sv_Cw);
```

sv_Cw = 4×1

```
  10.487255413923357
   7.918856903253778
   0.656617933338373
   0.000000000000000
```

```
format default
```

Singular values of matrix D by SVD:

```
Dw = Cw*A';
format long
sv_Dw = svd(Dw); display(sv_Dw);
```

sv_Dw = 100×1

$10^2 \times$

```
   1.338873482091462
   0.424617307961927
   0.236383352548320
   0.000000000000000
   0.000000000000000
   0.000000000000000
```

```
             0.000000000000000
             0.000000000000000
             0.000000000000000
             0.000000000000000
                   ⋮
```

```
format default

format long
table(Cw(:,2),Cw(:,3),Cw(:,2)+Cw(:,3),repelem(Cw(1,2),lCtw,1),'VariableNames',{
'c_K(t) +','c_{SK}(t) =','c_K(t)+c_{SK}(t)','c_K(0)'})
```

ans = 241×4 table

|    | c_K(t) +          | c_{SK}(t) =       | c_K(t)+c_{SK}(t)  | c_K(0)            |
|----|-------------------|-------------------|-------------------|-------------------|
| 1  | 0.100000000000000 | 0                 | 0.100000000000000 | 0.100000000000000 |
| 2  | 0.099953597627809 | 0.000046402372191 | 0.100000000000000 | 0.100000000000000 |
| 3  | 0.099907222276180 | 0.000092777723820 | 0.100000000000000 | 0.100000000000000 |
| 4  | 0.099860873927535 | 0.000139126072465 | 0.100000000000000 | 0.100000000000000 |
| 5  | 0.099814552564306 | 0.000185447435694 | 0.100000000000000 | 0.100000000000000 |
| 6  | 0.099768258168939 | 0.000231741831061 | 0.100000000000000 | 0.100000000000000 |
| 7  | 0.099721990723896 | 0.000278009276104 | 0.100000000000000 | 0.100000000000000 |
| 8  | 0.099675750211650 | 0.000324249788350 | 0.100000000000000 | 0.100000000000000 |
| 9  | 0.099629536614687 | 0.000370463385313 | 0.100000000000000 | 0.100000000000000 |
| 10 | 0.099398871747103 | 0.000601128252897 | 0.100000000000000 | 0.100000000000000 |
| 11 | 0.099168877142316 | 0.000831122857684 | 0.100000000000000 | 0.100000000000000 |
| 12 | 0.098939550626167 | 0.001060449373833 | 0.100000000000000 | 0.100000000000000 |
| 13 | 0.098710890032764 | 0.001289109967236 | 0.100000000000000 | 0.100000000000000 |
| 14 | 0.098482893204448 | 0.001517106795552 | 0.100000000000000 | 0.100000000000000 |
| 15 | 0.098255557991748 | 0.001744442008252 | 0.100000000000000 | 0.100000000000000 |
| 16 | 0.098028882253353 | 0.001971117746647 | 0.100000000000000 | 0.100000000000000 |
| 17 | 0.097802863856070 | 0.002197136143930 | 0.100000000000000 | 0.100000000000000 |

|    | c_K(t) +         | c_{SK}(t) =      | c_K(t)+c_{SK}(t)  | c_K(0)            |
|----|------------------|------------------|-------------------|-------------------|
| 18 | 0.096682557888051 | 0.003317442111949 | 0.100000000000000 | 0.100000000000000 |
| 19 | 0.095578370071779 | 0.004421629928221 | 0.100000000000000 | 0.100000000000000 |
| 20 | 0.094490042671066 | 0.005509957328934 | 0.100000000000000 | 0.100000000000000 |
| 21 | 0.093417322753307 | 0.006582677246693 | 0.100000000000000 | 0.100000000000000 |
| 22 | 0.092359962081911 | 0.007640037918089 | 0.100000000000000 | 0.100000000000000 |
| 23 | 0.091317717011609 | 0.008682282988391 | 0.100000000000000 | 0.100000000000000 |
| 24 | 0.090290348386582 | 0.009709651613418 | 0.100000000000000 | 0.100000000000000 |
| 25 | 0.089277621441283 | 0.010722378558717 | 0.100000000000000 | 0.100000000000000 |
| 26 | 0.084425690477672 | 0.015574309522328 | 0.100000000000000 | 0.100000000000000 |
| 27 | 0.079906949089213 | 0.020093050910787 | 0.100000000000000 | 0.100000000000000 |
| 28 | 0.075696455206483 | 0.024303544793517 | 0.100000000000000 | 0.100000000000000 |
| 29 | 0.071771386099778 | 0.028228613900222 | 0.100000000000000 | 0.100000000000000 |
| 30 | 0.068110827811141 | 0.031889172188859 | 0.100000000000000 | 0.100000000000000 |
| 31 | 0.064695589184798 | 0.035304410815202 | 0.100000000000000 | 0.100000000000000 |
| 32 | 0.061508037061442 | 0.038491962938558 | 0.100000000000000 | 0.100000000000000 |
| 33 | 0.058531949993379 | 0.041468050006621 | 0.100000000000000 | 0.100000000000000 |
| 34 | 0.050878989356642 | 0.049121010643359 | 0.100000000000000 | 0.100000000000000 |
| 35 | 0.044601714157693 | 0.055398285842307 | 0.100000000000000 | 0.100000000000000 |
| 36 | 0.039442402030795 | 0.060557597969205 | 0.100000000000000 | 0.100000000000000 |
| 37 | 0.035194639184386 | 0.064805360815614 | 0.100000000000000 | 0.100000000000000 |
| 38 | 0.031692435763988 | 0.068307564236012 | 0.100000000000000 | 0.100000000000000 |
| 39 | 0.028801919015099 | 0.071198080984901 | 0.100000000000000 | 0.100000000000000 |
| 40 | 0.026414257560847 | 0.073585742439153 | 0.100000000000000 | 0.100000000000000 |

| | c_K(t) + | c_{SK}(t) = | c_K(t)+c_{SK}(t) | c_K(0) |
|---|---|---|---|---|
| 41 | 0.024440446601026 | 0.075559553398974 | 0.100000000000000 | 0.100000000000000 |
| 42 | 0.022384841135156 | 0.077615158864844 | 0.100000000000000 | 0.100000000000000 |
| 43 | 0.020776855704413 | 0.079223144295587 | 0.100000000000000 | 0.100000000000000 |
| 44 | 0.019519225289761 | 0.080480774710239 | 0.100000000000000 | 0.100000000000000 |
| 45 | 0.018536344327955 | 0.081463655672045 | 0.100000000000000 | 0.100000000000000 |
| 46 | 0.017769247934712 | 0.082230752065288 | 0.100000000000000 | 0.100000000000000 |
| 47 | 0.017171808417260 | 0.082828191582740 | 0.100000000000000 | 0.100000000000000 |
| 48 | 0.016707936404723 | 0.083292063595277 | 0.100000000000000 | 0.100000000000000 |
| 49 | 0.016349384221102 | 0.083650615778898 | 0.100000000000000 | 0.100000000000000 |
| 50 | 0.015969284768495 | 0.084030715231505 | 0.100000000000000 | 0.100000000000000 |
| 51 | 0.015715731523063 | 0.084284268476937 | 0.100000000000000 | 0.100000000000000 |
| 52 | 0.015550978025704 | 0.084449021974296 | 0.100000000000000 | 0.100000000000000 |
| 53 | 0.015451243229542 | 0.084548756770458 | 0.100000000000000 | 0.100000000000000 |
| 54 | 0.015398614787886 | 0.084601385212114 | 0.100000000000000 | 0.100000000000000 |
| 55 | 0.015377605396501 | 0.084622394603499 | 0.100000000000000 | 0.100000000000000 |
| 56 | 0.015379069384416 | 0.084620930615584 | 0.100000000000000 | 0.100000000000000 |
| 57 | 0.015399314862249 | 0.084600685137751 | 0.100000000000000 | 0.100000000000000 |
| 58 | 0.015423866490562 | 0.084576133509438 | 0.100000000000000 | 0.100000000000000 |
| 59 | 0.015454771907284 | 0.084545228092716 | 0.100000000000000 | 0.100000000000000 |
| 60 | 0.015490566116299 | 0.084509433883701 | 0.100000000000000 | 0.100000000000000 |
| 61 | 0.015530156805355 | 0.084469843194645 | 0.100000000000000 | 0.100000000000000 |
| 62 | 0.015572713647624 | 0.084427286352376 | 0.100000000000000 | 0.100000000000000 |
| 63 | 0.015617592308256 | 0.084382407691744 | 0.100000000000000 | 0.100000000000000 |

| | c_K(t) + | c_{SK}(t) = | c_K(t)+c_{SK}(t) | c_K(0) |
|---|---|---|---|---|
| 64 | 0.015664312261928 | 0.084335687738072 | 0.100000000000000 | 0.100000000000000 |
| 65 | 0.015712524689123 | 0.084287475310877 | 0.100000000000000 | 0.100000000000000 |
| 66 | 0.015788449568746 | 0.084211550431254 | 0.100000000000000 | 0.100000000000000 |
| 67 | 0.015866379657562 | 0.084133620342438 | 0.100000000000000 | 0.100000000000000 |
| 68 | 0.015945912366136 | 0.084054087633864 | 0.100000000000000 | 0.100000000000000 |
| 69 | 0.016027007065395 | 0.083972992934605 | 0.100000000000000 | 0.100000000000000 |
| 70 | 0.016109425015513 | 0.083890574984487 | 0.100000000000000 | 0.100000000000000 |
| 71 | 0.016192751271291 | 0.083807248728709 | 0.100000000000000 | 0.100000000000000 |
| 72 | 0.016276957262022 | 0.083723042737978 | 0.100000000000000 | 0.100000000000000 |
| 73 | 0.016362256302927 | 0.083637743697073 | 0.100000000000000 | 0.100000000000000 |
| 74 | 0.016449422268771 | 0.083550577731228 | 0.100000000000000 | 0.100000000000000 |
| 75 | 0.016537435588452 | 0.083462564411548 | 0.100000000000000 | 0.100000000000000 |
| 76 | 0.016626293381961 | 0.083373706618039 | 0.100000000000000 | 0.100000000000000 |
| 77 | 0.016716117837831 | 0.083283882162169 | 0.100000000000000 | 0.100000000000000 |
| 78 | 0.016806898781865 | 0.083193101218135 | 0.100000000000000 | 0.100000000000000 |
| 79 | 0.016898499251601 | 0.083101500748399 | 0.100000000000000 | 0.100000000000000 |
| 80 | 0.016990946471292 | 0.083009053528708 | 0.100000000000000 | 0.100000000000000 |
| 81 | 0.017084384798506 | 0.082915615201494 | 0.100000000000000 | 0.100000000000000 |
| 82 | 0.017189139979753 | 0.082810860020247 | 0.100000000000000 | 0.100000000000000 |
| 83 | 0.017294885732210 | 0.082705114267790 | 0.100000000000000 | 0.100000000000000 |
| 84 | 0.017401740212417 | 0.082598259787583 | 0.100000000000000 | 0.100000000000000 |
| 85 | 0.017509973518827 | 0.082490026481173 | 0.100000000000000 | 0.100000000000000 |
| 86 | 0.017619531042344 | 0.082380468957656 | 0.100000000000000 | 0.100000000000000 |

|     | c_K(t) +          | c_{SK}(t) =       | c_K(t)+c_{SK}(t)  | c_K(0)            |
| --- | ----------------- | ----------------- | ----------------- | ----------------- |
| 87  | 0.017730148577856 | 0.082269851422144 | 0.100000000000000 | 0.100000000000000 |
| 88  | 0.017841906218684 | 0.082158093781316 | 0.100000000000000 | 0.100000000000000 |
| 89  | 0.017955057667605 | 0.082044942332395 | 0.100000000000000 | 0.100000000000000 |
| 90  | 0.018071540550951 | 0.081928459449049 | 0.100000000000000 | 0.100000000000000 |
| 91  | 0.018189236664676 | 0.081810763335324 | 0.100000000000000 | 0.100000000000000 |
| 92  | 0.018308216459931 | 0.081691783540069 | 0.100000000000000 | 0.100000000000000 |
| 93  | 0.018428741510104 | 0.081571258489896 | 0.100000000000000 | 0.100000000000000 |
| 94  | 0.018550789080146 | 0.081449210919854 | 0.100000000000000 | 0.100000000000000 |
| 95  | 0.018674109840605 | 0.081325890159395 | 0.100000000000000 | 0.100000000000000 |
| 96  | 0.018798772205302 | 0.081201227794698 | 0.100000000000000 | 0.100000000000000 |
| 97  | 0.018925041132226 | 0.081074958867774 | 0.100000000000000 | 0.100000000000000 |
| 98  | 0.019057712344682 | 0.080942287655318 | 0.100000000000000 | 0.100000000000000 |
| 99  | 0.019191874451134 | 0.080808125548866 | 0.100000000000000 | 0.100000000000000 |
| 100 | 0.019327591145975 | 0.080672408854025 | 0.100000000000000 | 0.100000000000000 |

⋮

```
table(Cw(:,1),Cw(:,3),Cw(:,4),Cw(:,1)+Cw(:,3)+Cw(:,4),repelem(Cw(1,1),lCtw,1),'
VariableNames',{'c_S(t) +','c_{SK}(t) +','c_P(t)
=','c_S(t)+c_{SK}(t)+c_P(t)','c_S(0)'})
```

ans = 241×5 table

|   | c_S(t) +          | c_{SK}(t) +       | c_P(t) =          | c_S(t)+c_{SK}(t)+c_P(t) | ... |
| - | ----------------- | ----------------- | ----------------- | ----------------------- | --- |
| 1 | 1                 | 0                 | 0                 | 1                       |     |
| 2 | 0.999953596012296 | 0.000046402372191 | 0.000000001615512 | 1                       |     |
| 3 | 0.999907215815386 | 0.000092777723820 | 0.000000006460795 | 1                       |     |

|    | c_S(t) +          | c_{SK}(t) +       | c_P(t) =          | c_S(t)+c_{SK}(t)+c_P(t) | ... |
|----|-------------------|-------------------|-------------------|-------------------------|-----|
| 4  | 0.999860859393568 | 0.000139126072465 | 0.000000014533967 | 1 | |
| 5  | 0.999814526731156 | 0.000185447435694 | 0.000000025833149 | 1 | |
| 6  | 0.999768217812476 | 0.000231741831061 | 0.000000040356464 | 1 | |
| 7  | 0.999721932621863 | 0.000278009276104 | 0.000000058102033 | 1 | |
| 8  | 0.999675671143669 | 0.000324249788350 | 0.000000079067981 | 1 | |
| 9  | 0.999629433362253 | 0.000370463385313 | 0.000000103252434 | 1 | |
| 10 | 0.999398599360359 | 0.000601128252897 | 0.000000272386744 | 1 | |
| 11 | 0.999168355439131 | 0.000831122857684 | 0.000000521703185 | 1 | |
| 12 | 0.998938699657362 | 0.001060449373833 | 0.000000850968805 | 1 | |
| 13 | 0.998709630081357 | 0.001289109967236 | 0.000001259951407 | 1 | |
| 14 | 0.998481144784901 | 0.001517106795552 | 0.000001748419546 | 1 | |
| 15 | 0.998253241849220 | 0.001744442008252 | 0.000002316142528 | 1 | |
| 16 | 0.998025919362950 | 0.001971117746647 | 0.000002962890403 | 1 | |
| 17 | 0.997799175422102 | 0.002197136143930 | 0.000003688433969 | 1 | |
| 18 | 0.996674067771324 | 0.003317442111949 | 0.000008490116728 | 1 | |
| 19 | 0.995563142337748 | 0.004421629928221 | 0.000015227734032 | 1 | |
| 20 | 0.994466169215361 | 0.005509957328934 | 0.000023873455705 | 1 | |
| 21 | 0.993382922857316 | 0.006582677246693 | 0.000034399895991 | 1 | |
| 22 | 0.992313181976628 | 0.007640037918089 | 0.000046780105284 | 1 | |
| 23 | 0.991256729449565 | 0.008682282988391 | 0.000060987562044 | 1 | |
| 24 | 0.990213352221679 | 0.009709651613418 | 0.000076996164903 | 1 | |
| 25 | 0.989182841216358 | 0.010722378558717 | 0.000094780224926 | 1 | |
| 26 | 0.984216229275982 | 0.015574309522328 | 0.000209461201690 | 1 | |

|    | c_S(t) +          | c_{SK}(t) +       | c_P(t) =          | c_S(t)+c_{SK}(t)+c_P(t) | ... |
|----|-------------------|-------------------|-------------------|-------------------------|-----|
| 27 | 0.979542049439804 | 0.020093050910787 | 0.000364899649410 | 1 | |
| 28 | 0.975138149220705 | 0.024303544793517 | 0.000558305985778 | 1 | |
| 29 | 0.970984287794363 | 0.028228613900222 | 0.000787098305416 | 1 | |
| 30 | 0.967061942930999 | 0.031889172188859 | 0.001048884880142 | 1 | |
| 31 | 0.963354140805382 | 0.035304410815202 | 0.001341448379416 | 1 | |
| 32 | 0.959845305463844 | 0.038491962938558 | 0.001662731597598 | 1 | |
| 33 | 0.956521125454134 | 0.041468050006621 | 0.002010824539245 | 1 | |
| 34 | 0.947707516004977 | 0.049121010643359 | 0.003171473351665 | 1 | |
| 35 | 0.940092233171122 | 0.055398285842307 | 0.004509480986571 | 1 | |
| 36 | 0.933449261327363 | 0.060557597969205 | 0.005993140703432 | 1 | |
| 37 | 0.927598044688995 | 0.064805360815614 | 0.007596594495391 | 1 | |
| 38 | 0.922393707822055 | 0.068307564236012 | 0.009298727941933 | 1 | |
| 39 | 0.917719629667258 | 0.071198080984901 | 0.011082289347840 | 1 | |
| 40 | 0.913481164787030 | 0.073585742439153 | 0.012933092773817 | 1 | |
| 41 | 0.909601025940534 | 0.075559553398974 | 0.014839420660492 | 1 | |
| 42 | 0.905012046883498 | 0.077615158864844 | 0.017372794251658 | 1 | |
| 43 | 0.900810429968219 | 0.079223144295587 | 0.019966425736194 | 1 | |
| 44 | 0.896912038010450 | 0.080480774710239 | 0.022607187279312 | 1 | |
| 45 | 0.893251542772944 | 0.081463655672045 | 0.025284801555010 | 1 | |
| 46 | 0.889778044816617 | 0.082230752065288 | 0.027991203118095 | 1 | |
| 47 | 0.886451761252781 | 0.082828191582740 | 0.030720047164480 | 1 | |
| 48 | 0.883241591700588 | 0.083292063595277 | 0.033466344704136 | 1 | |
| 49 | 0.880123211403305 | 0.083650615778898 | 0.036226172817797 | 1 | |

|    | c_S(t) +          | c_{SK}(t) +       | c_P(t) =          | c_S(t)+c_{SK}(t)+c_P(t) | … |
|----|-------------------|-------------------|-------------------|-------------------------|---|
| 50 | 0.875682488997012 | 0.084030715231505 | 0.040286795771483 | 1 | |
| 51 | 0.871353180221567 | 0.084284268476937 | 0.044362551301497 | 1 | |
| 52 | 0.867102874534059 | 0.084449021974296 | 0.048448103491646 | 1 | |
| 53 | 0.862911162876980 | 0.084548756770458 | 0.052540080352563 | 1 | |
| 54 | 0.858762686373025 | 0.084601385212114 | 0.056635928414861 | 1 | |
| 55 | 0.854644190357788 | 0.084622394603499 | 0.060733415038713 | 1 | |
| 56 | 0.850547857172960 | 0.084620930615584 | 0.064831212211457 | 1 | |
| 57 | 0.846470535734081 | 0.084600685137751 | 0.068928779128168 | 1 | |
| 58 | 0.843116769816562 | 0.084576133509438 | 0.072307096674000 | 1 | |
| 59 | 0.839770470453095 | 0.084545228092716 | 0.075684301454189 | 1 | |
| 60 | 0.836430397182912 | 0.084509433883701 | 0.079060168933387 | 1 | |
| 61 | 0.833095628673353 | 0.084469843194645 | 0.082434528132002 | 1 | |
| 62 | 0.829765467953463 | 0.084427286352376 | 0.085807245694160 | 1 | |
| 63 | 0.826439377365567 | 0.084382407691744 | 0.089178214942689 | 1 | |
| 64 | 0.823116959353438 | 0.084335687738072 | 0.092547352908491 | 1 | |
| 65 | 0.819797928925137 | 0.084287475310877 | 0.095914595763986 | 1 | |
| 66 | 0.814732897253611 | 0.084211550431254 | 0.101055552315135 | 1 | |
| 67 | 0.809674586994719 | 0.084133620342438 | 0.106191792662843 | 1 | |
| 68 | 0.804622703028488 | 0.084054087633864 | 0.111323209337648 | 1 | |
| 69 | 0.799577257479306 | 0.083972992934605 | 0.116449749586090 | 1 | |
| 70 | 0.794538093945541 | 0.083890574984487 | 0.121571331069972 | 1 | |
| 71 | 0.789504908674037 | 0.083807248728709 | 0.126687842597254 | 1 | |
| 72 | 0.784477725586853 | 0.083723042737978 | 0.131799231675169 | 1 | |

| | $c_S(t)$ + | $c_{SK}(t)$ + | $c_P(t)$ = | $c_S(t)+c_{SK}(t)+c_P(t)$ | ... |
|---|---|---|---|---|---|
| 73 | 0.779456773056219 | 0.083637743697073 | 0.136905483246707 | 1 | |
| 74 | 0.774371549067328 | 0.083550577731228 | 0.142077873201444 | 1 | |
| 75 | 0.769292600237217 | 0.083462564411548 | 0.147244835351234 | 1 | |
| 76 | 0.764219977822748 | 0.083373706618039 | 0.152406315559213 | 1 | |
| 77 | 0.759153838454379 | 0.083283882162169 | 0.157562279383452 | 1 | |
| 78 | 0.754094227389351 | 0.083193101218135 | 0.162712671392514 | 1 | |
| 79 | 0.749041084610459 | 0.083101500748399 | 0.167857414641142 | 1 | |
| 80 | 0.743994488604757 | 0.083009053528708 | 0.172996457866535 | 1 | |
| 81 | 0.738954616413115 | 0.082915615201494 | 0.178129768385392 | 1 | |
| 82 | 0.733363805693593 | 0.082810860020247 | 0.183825334286160 | 1 | |
| 83 | 0.727781244910167 | 0.082705114267790 | 0.189513640822043 | 1 | |
| 84 | 0.722207115389749 | 0.082598259787583 | 0.195194624822668 | 1 | |
| 85 | 0.716641725489730 | 0.082490026481173 | 0.200868248029097 | 1 | |
| 86 | 0.711085111534975 | 0.082380468957656 | 0.206534419507370 | 1 | |
| 87 | 0.705537137622517 | 0.082269851422144 | 0.212193010955339 | 1 | |
| 88 | 0.699997955653124 | 0.082158093781316 | 0.217843950565559 | 1 | |
| 89 | 0.694467862104165 | 0.082044942332395 | 0.223487195563439 | 1 | |
| 90 | 0.688836315026808 | 0.081928459449049 | 0.229235225524143 | 1 | |
| 91 | 0.683214210733192 | 0.081810763335324 | 0.234975025931484 | 1 | |
| 92 | 0.677601705626162 | 0.081691783540069 | 0.240706510833768 | 1 | |
| 93 | 0.671999114156070 | 0.081571258489896 | 0.246429627354034 | 1 | |
| 94 | 0.666406515344167 | 0.081449210919854 | 0.252144273735980 | 1 | |
| 95 | 0.660823804220900 | 0.081325890159395 | 0.257850305619705 | 1 | |

| | c_S(t) + | c_{SK}(t) + | c_P(t) = | c_S(t)+c_{SK}(t)+c_P(t) | ... |
|---|---|---|---|---|---|
| 96 | 0.655251138925126 | 0.081201227794698 | 0.263547633280176 | 1 | |
| 97 | 0.649688839716120 | 0.081074958867774 | 0.269236201416106 | 1 | |
| 98 | 0.643914749381654 | 0.080942287655318 | 0.275142962963028 | 1 | |
| 99 | 0.638151899628861 | 0.080808125548866 | 0.281039974822273 | 1 | |
| 100 | 0.632400467043700 | 0.080672408854025 | 0.286927124102275 | 1 | |

⋮

```
table(1/Cw(1,1)*Cw(:,1)-1/Cw(1,2)*Cw(:,2)+(1/Cw(1,1)-
1/Cw(1,2))*Cw(:,3)+1/Cw(1,1)*Cw(:,4),'VariableNames',{'cS(t)/cS0-
cK(t)/cK0+(1/cS0-1/cK0)*cSK(t)+cP(t)/cS0'})
```

ans = 241×1 table

| | cS(t)/cS0-cK(t)/cK0+(1/cS0-1/cK0)*cSK(t)+cP(t)/cS0 |
|---|---|
| 1 | 0 |
| 2 | -9.168737916056226e-17 |
| 3 | 5.806185860301614e-18 |
| 4 | -1.147689756880440e-16 |
| 5 | -8.295927704809058e-17 |
| 6 | -4.804430433830495e-17 |
| 7 | -6.051696374829803e-17 |
| 8 | -1.083245686999597e-16 |
| 9 | -4.190859619174181e-17 |
| 10 | 2.957516120500876e-17 |
| 11 | -5.731850778246169e-17 |
| 12 | -2.432233932217042e-17 |

| | cS(t)/cS0-cK(t)/cK0+(1/cS0-1/cK0)*cSK(t)+cP(t)/cS0 |
|---|---:|
| 13 | 6.566813506002096e-17 |
| 14 | 9.287292750169776e-17 |
| 15 | 1.735528157276699e-17 |
| 16 | -2.033048479999772e-17 |
| 17 | 1.155357175219602e-16 |
| 18 | 1.360978658330320e-16 |
| 19 | 5.826909050751783e-17 |
| 20 | 6.035617968955242e-17 |
| 21 | -1.759118024857731e-17 |
| 22 | 7.064254780100865e-17 |
| 23 | 4.967001202699217e-17 |
| 24 | 1.457032194548957e-16 |
| 25 | 1.974603206639225e-17 |
| 26 | -4.385597787703865e-17 |
| 27 | 7.318364664277155e-18 |
| 28 | 6.136584296267955e-17 |
| 29 | 1.155759515869548e-16 |
| 30 | 8.153200337090993e-17 |
| 31 | 3.057450126409123e-17 |
| 32 | 1.277190159187924e-16 |
| 33 | -4.336808689942018e-19 |
| 34 | 4.727121472036799e-17 |
| 35 | 9.801187639268960e-17 |

| | cS(t)/cS0-cK(t)/cK0+(1/cS0-1/cK0)*cSK(t)+cP(t)/cS0 |
|---|---|
| 36 | -4.510281037539698e-17 |
| 37 | -8.673617379884035e-18 |
| 38 | -2.081668171172169e-17 |
| 39 | 1.665334536937735e-16 |
| 40 | 3.295974604355933e-17 |
| 41 | 1.040834085586084e-17 |
| 42 | 6.938893903907228e-18 |
| 43 | 1.804112415015879e-16 |
| 44 | 8.326672684688674e-17 |
| 45 | -1.734723475976807e-17 |
| 46 | 9.020562075079397e-17 |
| 47 | 1.040834085586084e-16 |
| 48 | 1.734723475976807e-16 |
| 49 | 1.665334536937735e-16 |
| 50 | 1.249000902703301e-16 |
| 51 | 1.665334536937735e-16 |
| 52 | 5.551115123125783e-17 |
| 53 | 8.326672684688674e-17 |
| 54 | 1.387778780781446e-16 |
| 55 | 1.110223024625157e-16 |
| 56 | 1.110223024625157e-16 |
| 57 | 1.249000902703301e-16 |
| 58 | 1.942890293094024e-16 |

| | cS(t)/cS0-cK(t)/cK0+(1/cS0-1/cK0)*cSK(t)+cP(t)/cS0 |
|---|---:|
| 59 | 1.249000902703301e-16 |
| 60 | 3.053113317719180e-16 |
| 61 | 1.387778780781446e-16 |
| 62 | 8.326672684688674e-17 |
| 63 | 1.665334536937735e-16 |
| 64 | -2.775557561562891e-17 |
| 65 | 2.498001805406602e-16 |
| 66 | 6.938893903907228e-17 |
| 67 | 2.914335439641036e-16 |
| 68 | 2.220446049250313e-16 |
| 69 | 2.914335439641036e-16 |
| 70 | 2.220446049250313e-16 |
| 71 | 3.053113317719180e-16 |
| 72 | 1.665334536937735e-16 |
| 73 | 1.942890293094024e-16 |
| 74 | 2.775557561562891e-16 |
| 75 | 2.220446049250313e-16 |
| 76 | 2.498001805406602e-16 |
| 77 | 2.220446049250313e-16 |
| 78 | 3.053113317719180e-16 |
| 79 | 2.220446049250313e-16 |
| 80 | 1.665334536937735e-16 |
| 81 | 2.498001805406602e-16 |

|    | cS(t)/cS0-cK(t)/cK0+(1/cS0-1/cK0)*cSK(t)+cP(t)/cS0 |
|----|---:|
| 82 | 2.498001805406602e-16 |
| 83 | 2.498001805406602e-16 |
| 84 | 2.775557561562891e-16 |
| 85 | 3.330669073875470e-16 |
| 86 | 3.330669073875470e-16 |
| 87 | 3.608224830031759e-16 |
| 88 | 2.498001805406602e-16 |
| 89 | 3.330669073875470e-16 |
| 90 | 3.330669073875470e-16 |
| 91 | 2.498001805406602e-16 |
| 92 | 2.498001805406602e-16 |
| 93 | 3.885780586188048e-16 |
| 94 | 2.220446049250313e-16 |
| 95 | 2.220446049250313e-16 |
| 96 | 3.885780586188048e-16 |
| 97 | 2.775557561562891e-16 |
| 98 | 3.885780586188048e-16 |
| 99 | 3.330669073875470e-16 |
| 100 | 2.220446049250313e-16 |

⋮

```
format default
```

The simple Matlab command `Aest = Dm'/Cest'` can be used for the unique determination for noisy data?

```
sd =0;
Dm=Dw+sd*randn(size(Dw));
% Aest = (Cw\Dm)';
Aest = Dm'/Cw';
```

Warning: Rank deficient, rank = 3, tol = 5.057951e-13.

```
DfA = Aest-A;
[xa, ya] = meshgrid(tw,x);
[yb, xb] = meshgrid(x,tw);
figure,
plot3(xa,ya,Dm); hold on
plot3(xb,yb,Dm); hold off
xlabel('time'); ylabel('wavelength'); zlabel('abs., a.u.');
title('Noisy Michaelis-Menten kinetic data matrix, Dm');
```

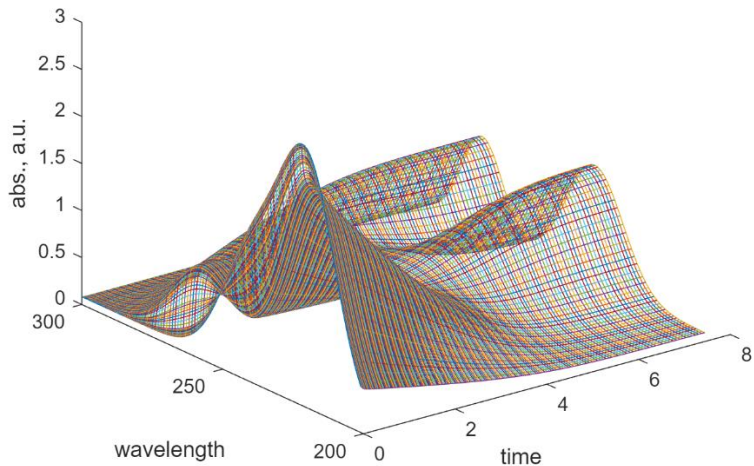

```
figure,
subplot(2,1,1);
colororder({'#0072BD','#D95319','#EDB120','#7E2F8E'})
plot3(xc,yc,Aest,'--'); hold on
colororder({'#0072BD','#D95319','#EDB120','#7E2F8E'})
plot3(xc,yc,A,'-'); hold off
legend('A_{est}:S','A_{est}:K','A_{est}:SK','A_{est}:P','A:S','A:K','A:SK','A:P')
title('A_{est} and A');
xlabel('wavelength'); ylabel('component'); zlabel('abs., a.u.');
yticks([1 2 3 4]); yticklabels({'S','K','SK','P'});
subplot(2,1,2);
```

```
plot3(xc,yc,DfA);
title('Difference');
xlabel('wavelength'); ylabel('component'); zlabel('A_{est}-A');
yticks([1 2 3 4]); yticklabels({'S','K','SK','P'});
```

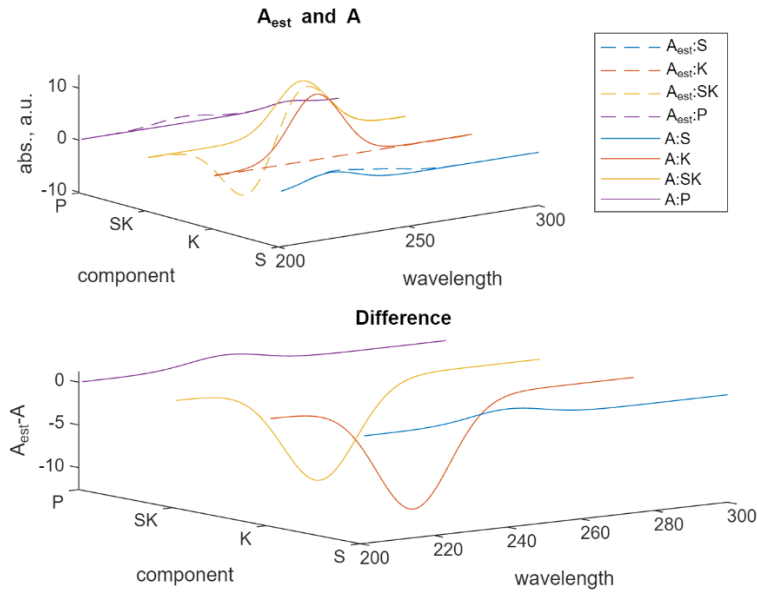

Continuous dosing for tackling rank deficiency:

```
m = sbiomodel('m');
SolvTypd = "ode45";
EnzymeDose = 0.0005;
configsetObj = getconfigset(m); set(configsetObj, 'SolverType', SolvTypd);
comp = addcompartment(m,'comp');
s1 = addspecies(m,'S' ,'InitialAmount',S_conc0,'InitialAmountUnits','mole');
s2 = addspecies(m,'K' ,'InitialAmount',K_conc0-
EnzymeDose,'InitialAmountUnits','mole');
s3 = addspecies(m,'SK','InitialAmount',SK_conc0,'InitialAmountUnits','mole');
s4 = addspecies(m,'P' ,'InitialAmount',P_conc0,'InitialAmountUnits','mole');
r1 = addreaction(m,'S + K -> SK');
lk1 = addkineticlaw(r1,'MassAction');
p1 = addparameter(lk1,'k1','Value',20,'ValueUnits','1/(mole*second)');
lk1.ParameterVariableNames = 'k1';
r1r = addreaction(m,'SK -> S + K');
lk1r = addkineticlaw(r1r,'MassAction');
p1r = addparameter(lk1r,'k1r','Value',0.1,'ValueUnits','1/second');
lk1r.ParameterVariableNames = 'k1r';
r2 = addreaction(m,'SK -> K + P');
lk2 = addkineticlaw(r2,'MassAction');
p2 = addparameter(lk2,'k2','Value',3,'ValueUnits','1/second');
```

```matlab
lk2.ParameterVariableNames = 'k2';

xmi=0.0005; xma=0.1; ymi=7.5; yma=15;
st_time = (yma-ymi)/(xma-xmi)*(EnzymeDose-xmi)+ymi; %0.0005 - 7.5, 0.1 - 15;
d1 = adddose(m,'dose');
d1.Amount = EnzymeDose;
d1.TargetName = 'K';
d1.Rate = EnzymeDose/st_time;
d1.RateUnits = 'mole/second';
d1.Active = true;

csObj = getconfigset(m,'active');
set(csObj,'Stoptime',st_time);
sb = sbiosimulate(m,csObj);
% [td,Ctd,species] = sbiosimulate(m,csObj);
[td,Ctd,species] = getdata(sb);
% sb = sbiosimulate(m,d1);
% [t,Ct,species] = sbiosimulate(m,d1);
lCt = length(Ct);
%sbioplot(sb);
figure, plot(td,Ctd,'LineWidth',5); hold on
plot(t,Ct,'k--','LineWidth',1.5); plot(t,Ct,'k-','LineWidth',1.5);
plot(t,Ct,'w--','LineWidth',1.5); hold off
legend([species; {"orig ("+SolvTyp+")"}]);
xlabel('Time');
ylabel('Species Amount');
title('Continuous dosing for breaking rank deficiency');
```

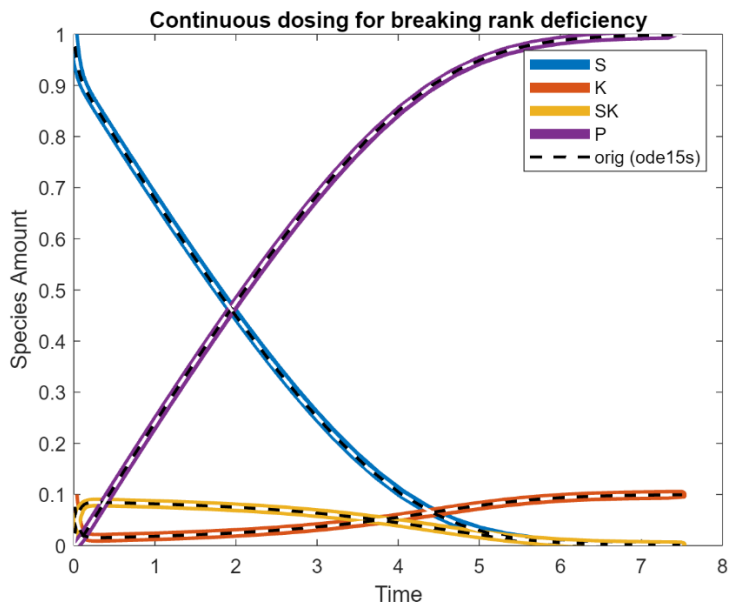

The figure of the Michaelis-Menten simulated spectrokinetic data matrix with doses:

```matlab
Dd = Ctd*A';
sd =0.0001;
Dmd=Dd+sd*randn(size(Dd));
% Aest = (Ctd\Dmd)';
Aest = Dmd'/Ctd';
DfA = Aest-A;
[xa, ya] = meshgrid(td,x);
[yb, xb] = meshgrid(x,td);
figure,
plot3(xa,ya,Dmd); hold on
plot3(xb,yb,Dmd); hold off
xlabel('time'); ylabel('wavelength'); zlabel('abs., a.u.');
title('Noisy Michaelis-Menten kinetic data matrix with doses, Dmd');
```

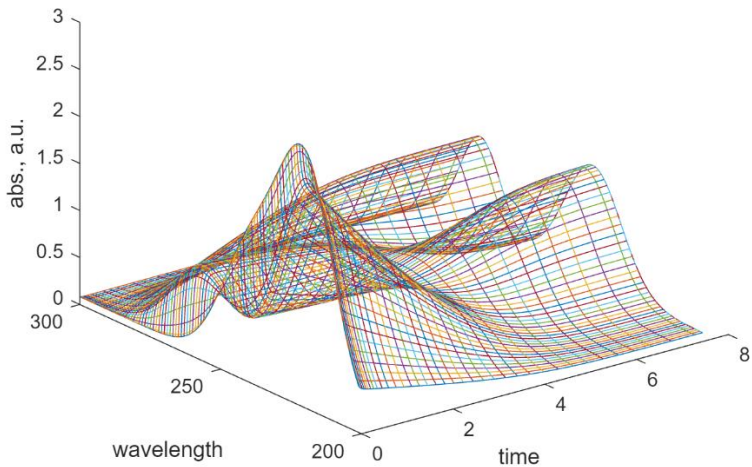

```matlab
figure,
subplot(2,1,1);
colororder({'#0072BD','#D95319','#EDB120','#7E2F8E'})
plot3(xc,yc,Aest,'--'); hold on
colororder({'#0072BD','#D95319','#EDB120','#7E2F8E'})
plot3(xc,yc,A,'-'); hold off
legend('A_{est}:S','A_{est}:K','A_{est}:SK','A_{est}:P','A:S','A:K','A:SK','A:P')
title('A_{est} and A');
xlabel('wavelength'); ylabel('component'); zlabel('abs., a.u.');
yticks([1 2 3 4]); yticklabels({'S','K','SK','P'});
subplot(2,1,2);
plot3(xc,yc,DfA);
title('Difference');
```

```
xlabel('wavelength'); ylabel('component'); zlabel('A_{est}-A');
yticks([1 2 3 4]); yticklabels({'S','K','SK','P'});
```

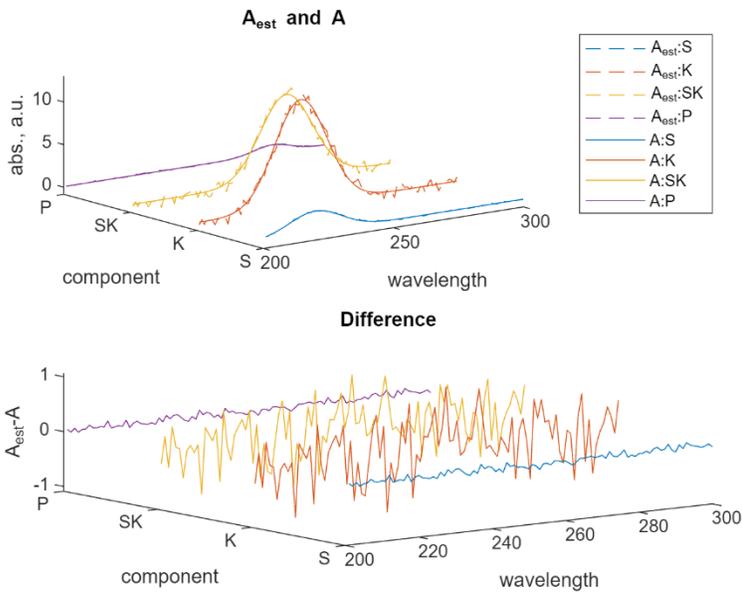

Singular values of matrix C with doses by SVD:

```
format long
sv_Cd = svd(Ctd); display(sv_Cd);
```

sv_Cd = 4×1

```
   4.381204273923026
   3.479031483896810
   0.261733260800627
   0.000176374752617
```

```
format default
```

Singular values of matrix Dmd with doses by SVD:

```
format long
sv_Dmd = svd(Dmd); display(sv_Dmd);
```

sv_Dmd = 42×1

```
  55.583717176021899
  18.624075230010302
   9.488837220098089
   0.005689597142808
   0.001584780793803
   0.001443108078605
   0.001405546721179
   0.001388975928315
   0.001335702575579
```

```
        0.001274277036946
           ⋮
```

```matlab
format default
```

Add discrete dose to the reaction (spiked reaction) for tackling rank deficiency

```matlab
m = sbiomodel('m');
SolvTypd = "ode45";
EnzymeDose = 0.0005;
configsetObj = getconfigset(m); set(configsetObj, 'SolverType', SolvTypd);
comp = addcompartment(m,'comp');
s1 = addspecies(m,'S' ,'InitialAmount',S_conc0,'InitialAmountUnits','mole');
s2 = addspecies(m,'K' ,'InitialAmount',K_conc0-EnzymeDose,'InitialAmountUnits','mole');
s3 = addspecies(m,'SK','InitialAmount',SK_conc0,'InitialAmountUnits','mole');
s4 = addspecies(m,'P' ,'InitialAmount',P_conc0,'InitialAmountUnits','mole');
r1 = addreaction(m,'S + K -> SK');
lk1 = addkineticlaw(r1,'MassAction');
p1 = addparameter(lk1,'k1','Value',20,'ValueUnits','1/(mole*second)');
lk1.ParameterVariableNames = 'k1';
r1r = addreaction(m,'SK -> S + K');
lk1r = addkineticlaw(r1r,'MassAction');
p1r = addparameter(lk1r,'k1r','Value',0.1,'ValueUnits','1/second');
lk1r.ParameterVariableNames = 'k1r';
r2 = addreaction(m,'SK -> K + P');
lk2 = addkineticlaw(r2,'MassAction');
p2 = addparameter(lk2,'k2','Value',3,'ValueUnits','1/second');
lk2.ParameterVariableNames = 'k2';

xmi=0.0005; xma=0.09; ymi=7.5; yma=12;
st_time = (yma-ymi)/(xma-xmi)*(EnzymeDose-xmi)+ymi; %0.0005 - 7.5, 0.09 - 12;
objd1 = adddose(m,'d1','schedule');
objd1.Amount = EnzymeDose;
objd1.AmountUnits = 'mole';
objd1.TimeUnits = 'second';
objd1.Time = 3;
objd1.TargetName = 'K';

csObj = getconfigset(m,'active');
set(csObj,'Stoptime',st_time);
sb = sbiosimulate(m,csObj,objd1);
% [td,Ctd,species] = sbiosimulate(m,csObj,objd1);
[td,Ctd,species] = getdata(sb);
lCt = length(Ct);
figure, plot(td,Ctd,'LineWidth',5); hold on
```

```
plot(t,Ct,'k--','LineWidth',1.5); plot(t,Ct,'k-','LineWidth',1.5);
plot(t,Ct,'w--','LineWidth',1.5); hold off
legend([species; {"orig ("+SolvTyp+")"}]);
xlabel('Time');
ylabel('Species Amount');
title('Discrete dose for breaking rank deficiency');
```

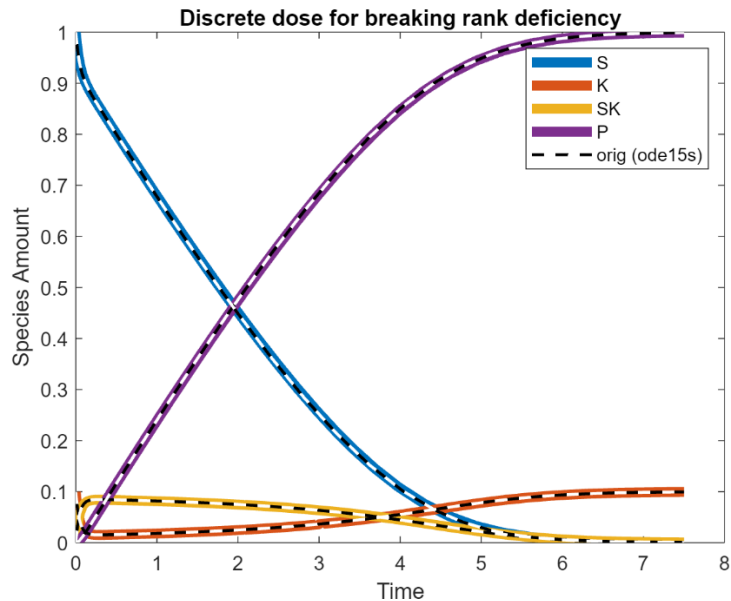

```
Dd = Ctd*A';
sd =0.001;
Dmd=Dd+sd*randn(size(Dd));
% Aest = (Ctd\Dmd)';
Aest = Dmd'/Ctd';
DfA = Aest-A;
[xa, ya] = meshgrid(td,x);
[yb, xb] = meshgrid(x,td);
figure,
subplot(2,1,1);
colororder({'#0072BD','#D95319','#EDB120','#7E2F8E'})
plot3(xc,yc,Aest,'--'); hold on
colororder({'#0072BD','#D95319','#EDB120','#7E2F8E'})
plot3(xc,yc,A,'-'); hold off
legend('A_{est}:S','A_{est}:K','A_{est}:SK','A_{est}:P','A:S','A:K','A:SK','A:P
')
title('A_{est} and A');
xlabel('wavelength'); ylabel('component'); zlabel('abs., a.u.');
yticks([1 2 3 4]); yticklabels({'S','K','SK','P'});
subplot(2,1,2);
```

```
plot3(xc,yc,DfA);
title('Difference');
xlabel('wavelength'); ylabel('component'); zlabel('A_{est}-A');
yticks([1 2 3 4]); yticklabels({'S','K','SK','P'});
```

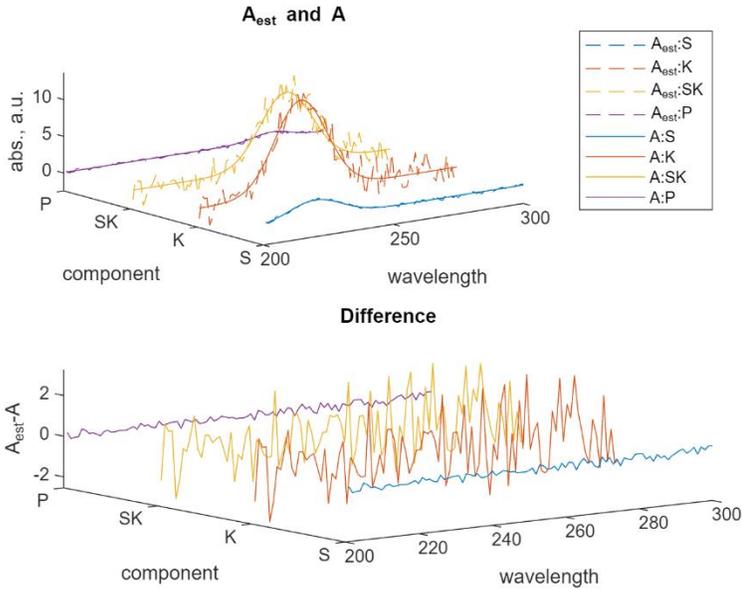

Singular values of matrix C with discrete dose by SVD:

```
format long
sv_Cd = svd(Ctd); display(sv_Cd);
```

sv_Cd = 4×1

    4.366731985527608
    3.348556356261246
    0.241674779524008
    0.000556057067659

```
format default
```

Singular values of matrix Dmd with discrete dose by SVD:

```
format long
sv_Dmd = svd(Dmd); display(sv_Dmd);
```

sv_Dmd = 42×1

   54.429845283684223
   18.148646659966406
    8.810401498197697
    0.021040533085921
    0.015749612598964
    0.014865817757359

```
    0.014372182007768
    0.013941050875394
    0.013531529282057
    0.013086961405522
         ⋮
```

```
format default
```

Only two differential equations for the Michaelis-Menten kinetics:

$$\frac{d}{dt}c_S(t) = -k_1\Big(c_S(t) + c_P(t) + c_{K,0} - c_{S,0}\Big)c_S(t) + k_{-1}\Big(c_{S,0} - c_S(t) - c_P(t)\Big)$$

$$\frac{d}{dt}c_P(t) = k_2\Big(c_{S,0} - c_S(t) - c_P(t)\Big)$$

```
% % ode89
% k1=p1.Value; k1r=p1r.Value; k2=p2.Value;
% y0 = [s1.Value s2.Value s3.Value s4.Value];
% tspan = [0 csObj.Stoptime];
tspan = t;
opts = odeset('Reltol',1e-4,'AbsTol',1e-7);
[t2d,C2d(:,[1 4])] = ode89(@(t,y) ode2fn_MichMent(t,y,k1,k1r,k2,y0), tspan, y0([1 4]),opts);
lCt2d=length(t2d);
C2d(:,3) = y0(1)-C2d(:,1)-C2d(:,4);
C2d(:,2) = y0(2)-C2d(:,3);
figure, plot(t2d,C2d,'LineWidth',5); hold on
plot(t,Ct,'k--','LineWidth',1.5); plot(t,Ct,'k-','LineWidth',1.5);
plot(t,Ct,'w--','LineWidth',1.5); hold off
legend([species; {"orig ("+SolvTyp+")"}]);
xlabel('Time');
ylabel('Species Amount');
title('Only two differential equations for the Michaelis-Menten kinetics');
```

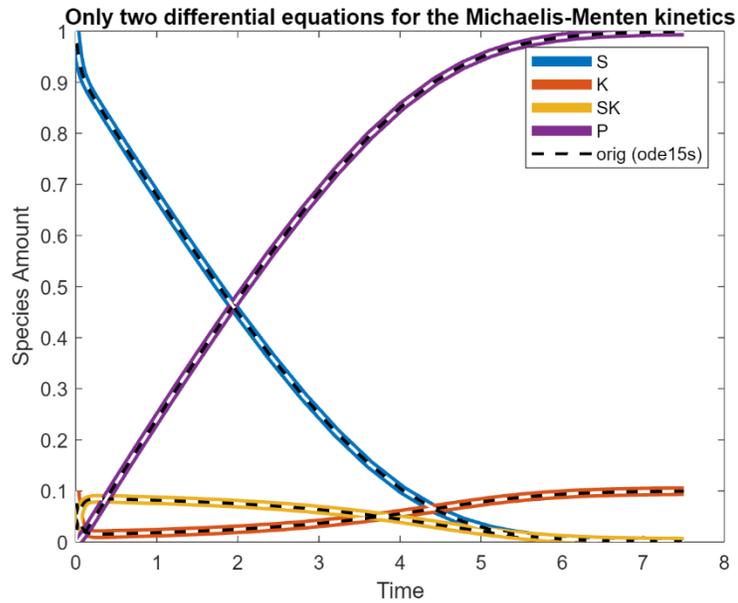

```
D2d = C2d*A';
format long
sv_C2d = svd(C2d); display(sv_C2d);
```

sv_C2d = 4×1

```
   7.054540593911640
   3.848460906924819
   0.434945450721970
   0.000000000000000
```

```
sv_D2d = svd(D2d); display(sv_D2d);
```

sv_D2d = 78×1

```
  93.841268056280867
  20.309619194402760
  15.269049424593062
   0.000000000000055
   0.000000000000044
   0.000000000000043
   0.000000000000041
   0.000000000000034
   0.000000000000032
   0.000000000000030
        ⋮
```

```
format default

sd =0;
Dm2d=D2d+sd*randn(size(D2d));
% Aest2d = (Ct\Dm2d)';
```

```
Aest2d = Dm2d'/Ct';
DfA2d = Aest2d-A;
figure, plot(xc, Aest2d);
title('Estimated spectra based on dc_S(t)/dt and dc_P(t)/dt');
xlabel('wavelength');
ylabel('absorbance, a.u.');
legend('S','K','SK','P');
```

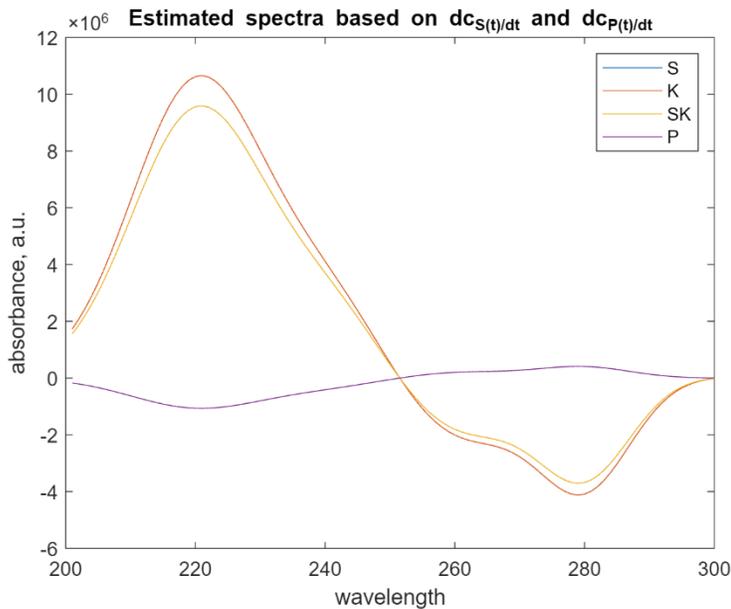

Only two differential equations for the Michaelis-Menten kinetics:

$$\frac{d}{dt} c_S(t) = -k_1 \, c_S(t) \, c_K(t) + k_{-1}\left(c_{K,0} - c_K(t)\right)$$

$$\frac{d}{dt} c_K(t) = -k_1 \, c_S(t) \, c_K(t) + \left(k_{-1} + k_2\right)\left(c_{K,0} - c_K(t)\right)$$

```
% % ode89
% k1=p1.Value; k1r=p1r.Value; k2=p2.Value;
% y0 = [s1.Value s2.Value s3.Value s4.Value];
% tspan = [0 csObj.Stoptime];
tspan = t;
opts = odeset('Reltol',1e-5,'AbsTol',1e-8);
[t2bd,C2bd(:,[1 2])] = ode15s(@(t,y) ode2bfn_MichMent(t,y,k1,k1r,k2,y0), tspan, y0([1 2]),opts);
lCt2d=length(t2bd);
C2bd(:,3) = y0(2)-C2bd(:,2);
```

```
C2bd(:,4) = y0(1)-C2bd(:,3)-C2bd(:,1);
figure, plot(t2d,C2bd,'LineWidth',5); hold on
plot(t,Ct,'k--','LineWidth',1.5); plot(t,Ct,'k-','LineWidth',1.5);
plot(t,Ct,'w--','LineWidth',1.5); hold off
legend([species; {"orig ("+SolvTyp+")"}]);
xlabel('Time');
ylabel('Species Amount');
title('Only two differential equations for the Michaelis-Menten kinetics');
```

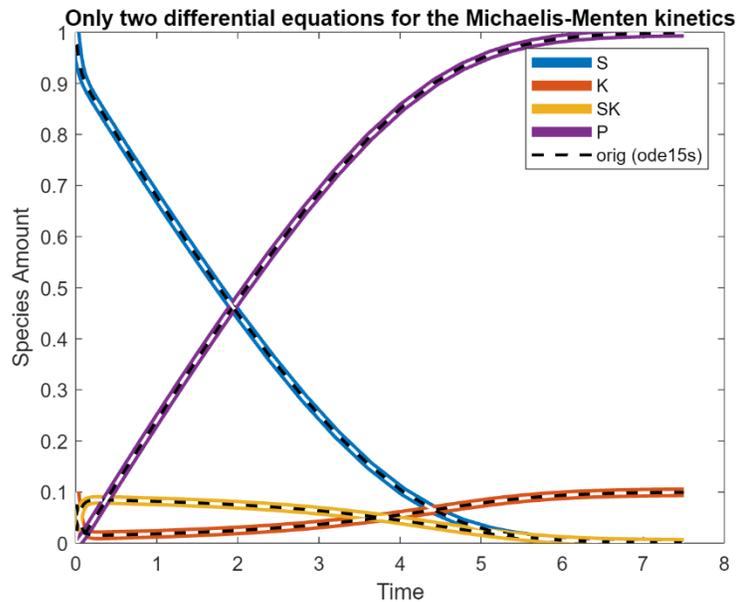

```
D2bd = C2bd*A';
format long
sv_C2bd = svd(C2bd); display(sv_C2bd);
```

sv_C2bd = 4×1

```
   7.054540505168170
   3.848461742679671
   0.434945904629261
   0.000000000000000
```

```
sv_D2bd = svd(D2bd); display(sv_D2bd);
```

sv_D2bd = 78×1

```
  93.841264611670468
  20.309619537765801
  15.269068785486118
   0.000000000000074
   0.000000000000055
   0.000000000000049
   0.000000000000046
```

```
    0.000000000000043
    0.000000000000039
    0.000000000000037
       ⋮
```

```matlab
format default

sd =0;
Dm2bd=D2bd+sd*randn(size(D2bd));
% Aest2bd = (Ct\Dm2bd)';
Aest2bd = Dm2bd'/Ct';
DfA2bd = Aest2bd-A;
figure, plot(xc, Aest2bd);
title('Estimated spectra based on dc_S(t)/dt and dc_K(t)/dt');
xlabel('wavelength');
ylabel('absorbance, a.u.');
legend('S','K','SK','P');
```

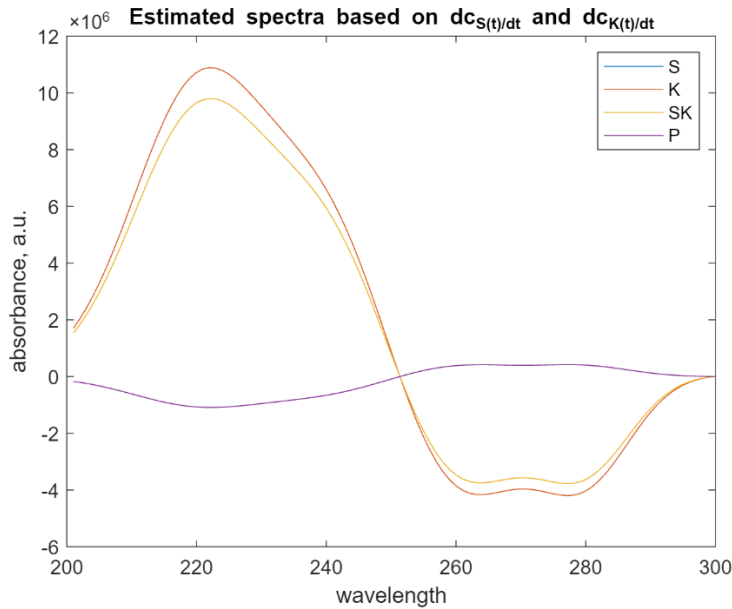

Only one equation is enough for the Michaelis-Menten kinetics (closed form solution, but it uses quasi-steady-state approximation):

$$c_S(t) = K_M W\left(\frac{c_{S,0}}{K_M}\exp\left(\frac{-v_{max}\,t + c_{S,0}}{K_M}\right)\right)$$

$$c_{SK}(t) = \frac{c_{S,0}\,c_S(t)}{K_M + c_S(t)}\left(1 - \exp(-(K_M + c_S(t))\,k_1\,t)\right)$$

$$c_E(t) = c_{E,0} - c_{SK}(t)$$

$$c_P(t) = c_{S,0} - c_S(t) - c_{SK}(t)$$

```matlab
K_M = (k1r+k2)/k1;
vmax = k2*y0(2);
t_C = 1/(k1*(K_M+y0(1)));
t_S = (K_M+y0(1))/vmax;
Ccf(:,1) = K_M*lambertw(exp((-vmax*t+y0(1))/K_M)*y0(1)/K_M);
Ccf(:,3) = (y0(2)*Ccf(:,1)./(K_M+Ccf(:,1))).*(1-exp(-(K_M+Ccf(:,1)).*t*k1));
% FastTransReg = t<t_C;
% Ccf(FastTransReg,3) = (y0(2)*y0(1)./(K_M+y0(1))).*(1-exp(-(K_M+y0(1)).*t(FastTransReg)*k1));
Ccf(:,2) = y0(2)-Ccf(:,3);
Ccf(:,4) = y0(1)-Ccf(:,1);

Dcf = Ccf*A';
format long
sv_Ccf = svd(Ccf); display(sv_Ccf);
```

```
sv_Ccf = 4×1

   7.365605158315339
   3.895934628211725
   0.445320880848155
   0.000000000000000
```

```matlab
sv_Dcf = svd(Dcf); display(sv_Dcf);
```

```
sv_Dcf = 78×1

  96.837522843633522
  20.341961350939190
  15.741537329903462
   0.000000000000062
   0.000000000000057
   0.000000000000052
   0.000000000000046
   0.000000000000042
   0.000000000000038
   0.000000000000034
       ⋮
```

```matlab
format default

figure, plot(t,Ccf,'LineWidth',5); hold on
plot(t,Ct,'k--','LineWidth',1.5); plot(t,Ct,'k-','LineWidth',1.5);
plot(t,Ct,'w--','LineWidth',1.5); hold off
legend([species; {"orig ("+SolvTyp+")"}]);
xlabel('Time');
ylabel('Species Amount');
title('Closed form solution for the Michaelis-Menten kinetics');
```

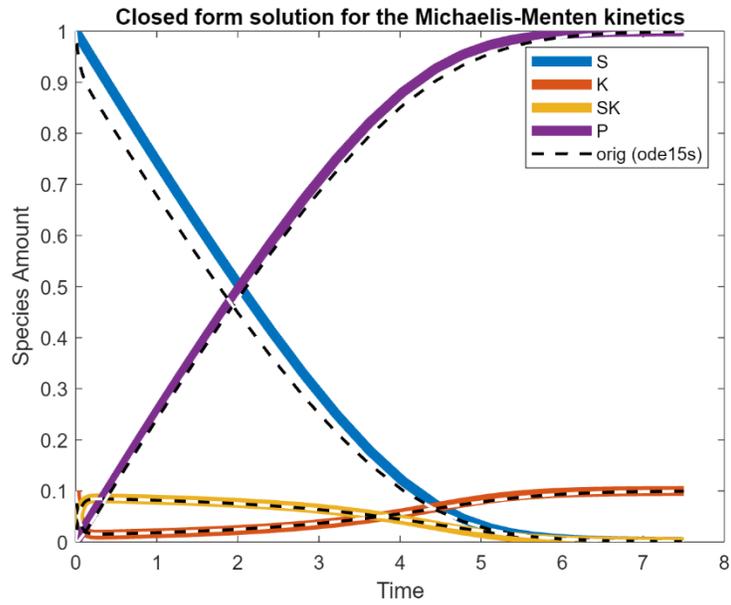

```
sd =0;
Dmcf=Dcf+sd*randn(size(Dcf));
% Aestcf = (Ct\Dmcf)';
Aestcf = Dmcf'/Ct';
DfAcf = Aestcf-A;
figure, plot(xc, Aestcf);
title('Estimated spectra based on the closed form solution');
xlabel('wavelength');
ylabel('absorbance, a.u.');
legend('S','K','SK','P');
```

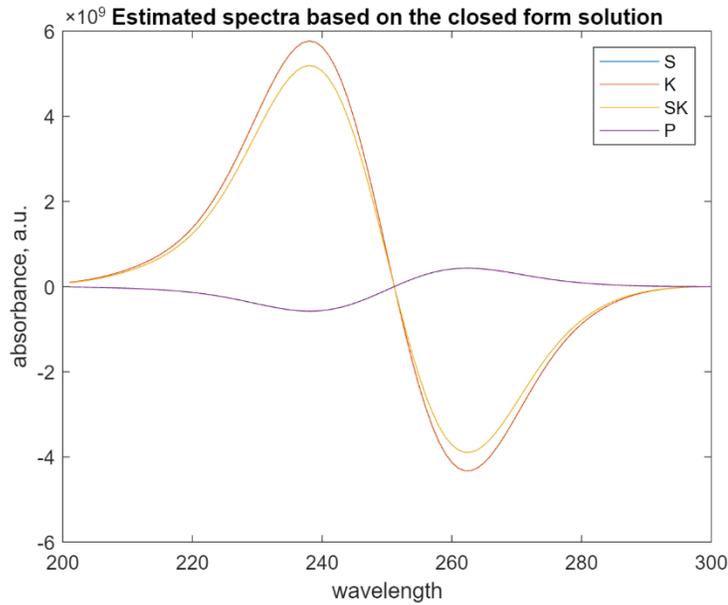

The spectra of the substrate and enzyme are known, or they can be measured before the enzyme reaction.

```
sd =0.1;
Dmcw=Dw+sd*randn(size(Dw));
Aest_kn_un = NaN(size(A));
Aest_kn_un_jsubs = Aest_kn_un;
Aest_kn_un_jenz = Aest_kn_un;
Aest_kn_un_ztsp = Aest_kn_un;

% Dw = Cw*A' = C_kn*A_kn + C_un*A_un ==> Aest = pinv(Cw_un)*(Dw-Cw_kn*A_kn')
% Aest_kn_un(:,3:4) = (Cw(:,3:4)\(Dmcw-Cw(:,1:2)*A(:,1:2)'))';
Aest_kn_un(:,3:4) = (Dmcw-Cw(:,1:2)*A(:,1:2)')'/Cw(:,3:4)';
figure, plot(xc, Aest_kn_un,'LineWidth',5); hold on, plot(xc, A,'-.k','LineWidth',1); hold off
title('Estimated spectra based on the known spectra of substrate and enzyme');
xlabel('wavelength');
ylabel('absorbance, a.u.');
legend('S','K','SK','P');
```

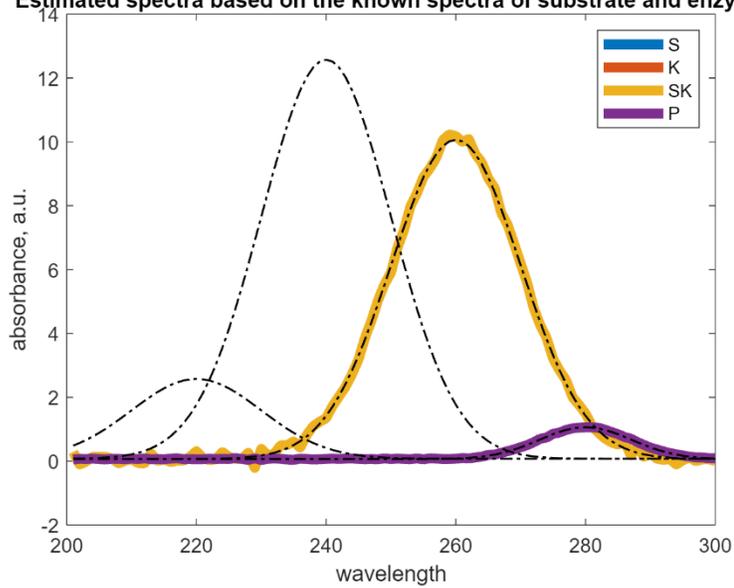

```
% only the spectrum of the substrate is known
% Aest_kn_un_jsubs(:,2:4) = (Cw(:,2:4)\(Dmcw-Cw(:,1)*A(:,1)'))';
Aest_kn_un_jsubs(:,2:4) = (Dmcw-Cw(:,1)*A(:,1)')'/Cw(:,2:4)';
figure, plot(xc, Aest_kn_un_jsubs,'LineWidth',5); hold on, plot(xc, A,'-.k','LineWidth',1); hold off
title('Estimated spectra based on the known spectrum of substrate');
xlabel('wavelength');
ylabel('absorbance, a.u.');
legend('S','K','SK','P');
```

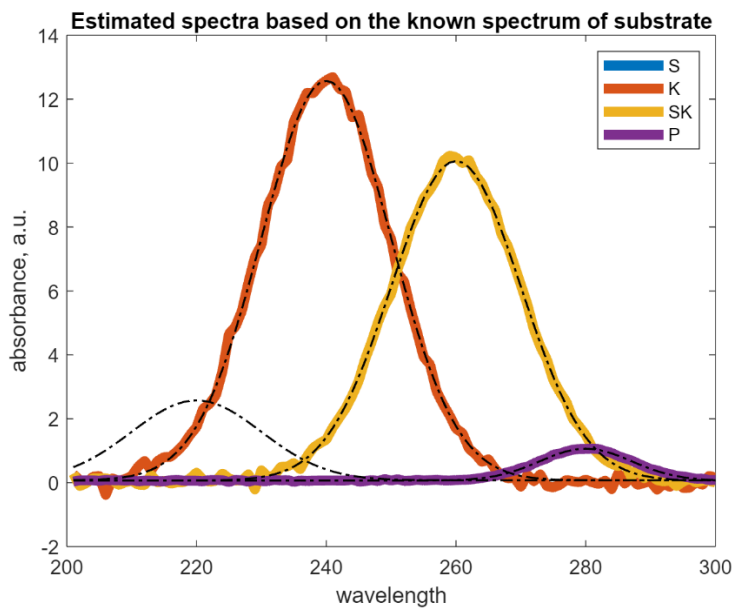

```
% only the spectrum of the enzyme is known
% Aest_kn_un_jenz(:,[1 3 4]) = (Cw(:,[1 3 4])\(Dmcw-Cw(:,2)*A(:,2)'))';
Aest_kn_un_jenz(:,[1 3 4]) = (Dmcw-Cw(:,2)*A(:,2)')'/Cw(:,[1 3 4])';
figure, plot(xc, Aest_kn_un_jenz,'LineWidth',5); hold on, plot(xc, A,'-.k','LineWidth',1); hold off
title('Estimated spectra based on the known spectrum of enzyme');
xlabel('wavelength');
ylabel('absorbance, a.u.');
legend('S','K','SK','P');
```

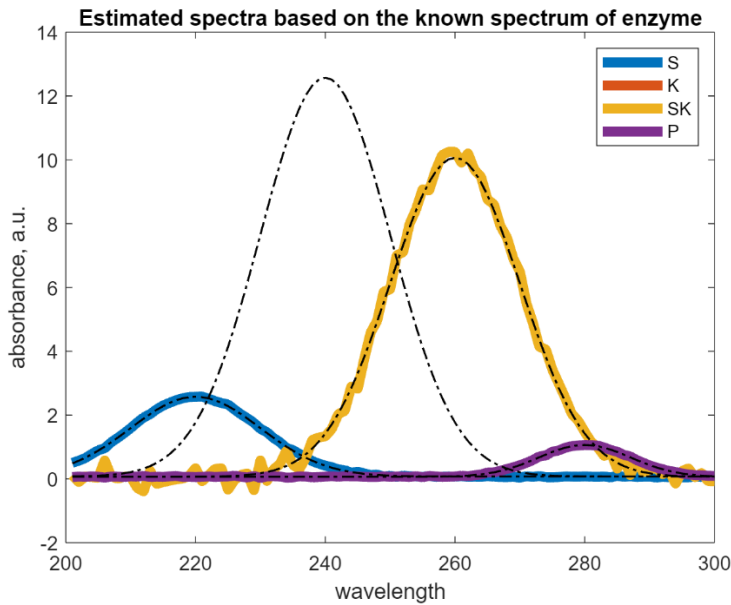

The spectrum of the substrate before the reaction (t<0) is definitely known (becuase it can be measured), after that the spectrum of the initial mixture (t=0) of the substrate and the enzyme (dropped to the solution of substrate) is known as well.

```
A_Sm = A(:,1)*S_conc0+sd*randn(size(A(:,1)));

% only the spectrum of the substrate is known (t<0)
% Aest_kn_un_jsubs(:,2:4) = (Cw(:,2:4)\(Dmcw-Cw(:,1)*A(:,1)'))';
Aest_kn_un_jsubs_fromD = [A_Sm (Dmcw-Cw(:,1)*A_Sm(:,1)')'/Cw(:,2:4)'];
figure, plot(xc, Aest_kn_un_jsubs_fromD,'LineWidth',5); hold on, plot(xc, A,'-.k','LineWidth',1); hold off
title('Estimated spectra based on the known spectrum of substrate (t<0)');
xlabel('wavelength');
ylabel('absorbance, a.u.');
legend('S','K','SK','P');
```

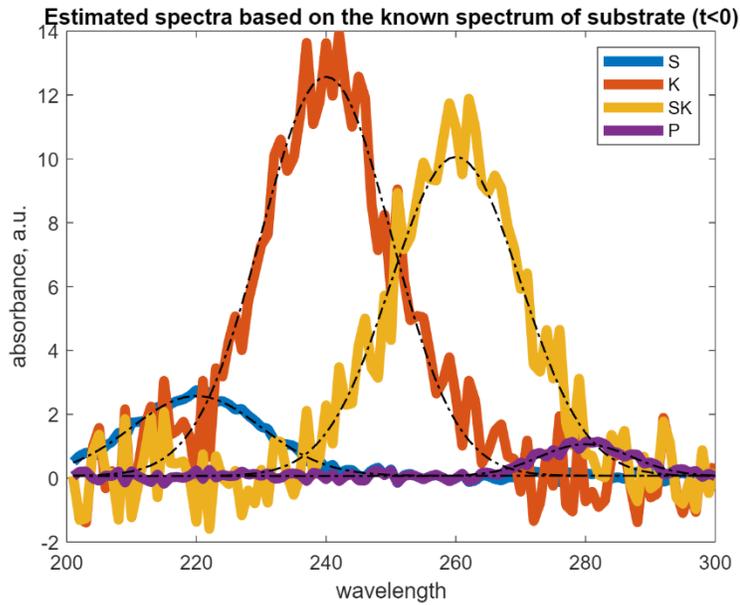

```matlab
% spectrum of the mixture of substrate and enzyme is known (t<=0)
% A_Kest = (Dmcw(1,:)'-A_Sm)/K_conc0;
[A_S_Kest] = [A_Sm Dmcw(1,:)']/[S_conc0 S_conc0; 0 K_conc0];
A_Sm = A_S_Kest(:,1);
A_Kest = A_S_Kest(:,2);
% Dw = Cw*A' = C_kn*A_kn + C_un*A_un ==> Aest = pinv(Cw_un)*(Dw-Cw_kn*A_kn')
% Aest_kn_un(:,3:4) = (Cw(:,3:4)\(Dmcw-Cw(:,1:2)*A(:,1:2)'))';
Aest_kn_un_fromD = [A_Sm A_Kest ((Dmcw-Cw(:,1:2)*[A_Sm A_Kest]'))'/Cw(:,3:4)'];
figure, plot(xc, Aest_kn_un_fromD,'LineWidth',5); hold on, plot(xc, A,'-.k','LineWidth',1); hold off
title('Using spectrum of the mixture of substrate and enzyme as well (t<=0)');
xlabel('wavelength');
ylabel('absorbance, a.u.');
legend('S','K','SK','P');
```

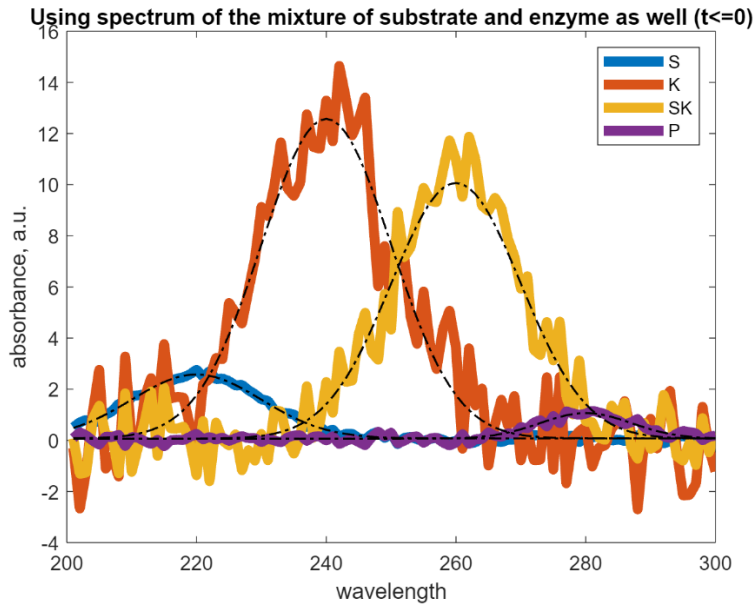

Using spectrum of the mixture of substrate and enzyme as well (t<=0)

Practical-like simulation, i.e., even six order of magnitude may occur in concentration ratio of substrate and enzyme.

```matlab
m = sbiomodel('m');
SolvTypp = "ode45";
S_conc0 = 1;
rat_conc0_S_K = 1000;
K_conc0 = S_conc0/rat_conc0_S_K;
SK_conc0 =0;
P_conc0 =  0;
% k1 = 20; k1r = 0.1; k2 = 3;
k1 = 20;
k1r = 0.118;
k2 = 3;
configsetObj = getconfigset(m); set(configsetObj, 'SolverType', SolvTypp);
comp = addcompartment(m,'comp');
s1 = addspecies(m,'S' ,'InitialAmount',S_conc0,'InitialAmountUnits','mole');
s2 = addspecies(m,'K' ,'InitialAmount',K_conc0,'InitialAmountUnits','mole');
s3 = addspecies(m,'SK','InitialAmount',SK_conc0,'InitialAmountUnits','mole');
s4 = addspecies(m,'P' ,'InitialAmount',P_conc0,'InitialAmountUnits','mole');
r1 = addreaction(m,'S + K -> SK');
lk1 = addkineticlaw(r1,'MassAction');
p1 = addparameter(lk1,'k1','Value',k1,'ValueUnits','1/(mole*second)');
lk1.ParameterVariableNames = 'k1';
r1r = addreaction(m,'SK -> S + K');
lk1r = addkineticlaw(r1r,'MassAction');
p1r = addparameter(lk1r,'k1r','Value',k1r,'ValueUnits','1/second');
lk1r.ParameterVariableNames = 'k1r';
```

```matlab
r2 = addreaction(m,'SK -> K + P');
lk2 = addkineticlaw(r2,'MassAction');
p2 = addparameter(lk2,'k2','Value',k2,'ValueUnits','1/second');
lk2.ParameterVariableNames = 'k2';

st_time = rat_conc0_S_K;

csObj = getconfigset(m,'active');
set(csObj,'Stoptime',st_time);
sb = sbiosimulate(m,csObj);
% [td,Ctp,species] = sbiosimulate(m,csObj);
[td,Ctp,species] = getdata(sb);
lCt = length(Ct);
figure, plot(td,Ctp,'LineWidth',5);
legend(species);
xlabel('Time');
ylabel('Species Amount');
title({"Practical-like simulation (ODE solver: "+SolvTypp+")"});
```

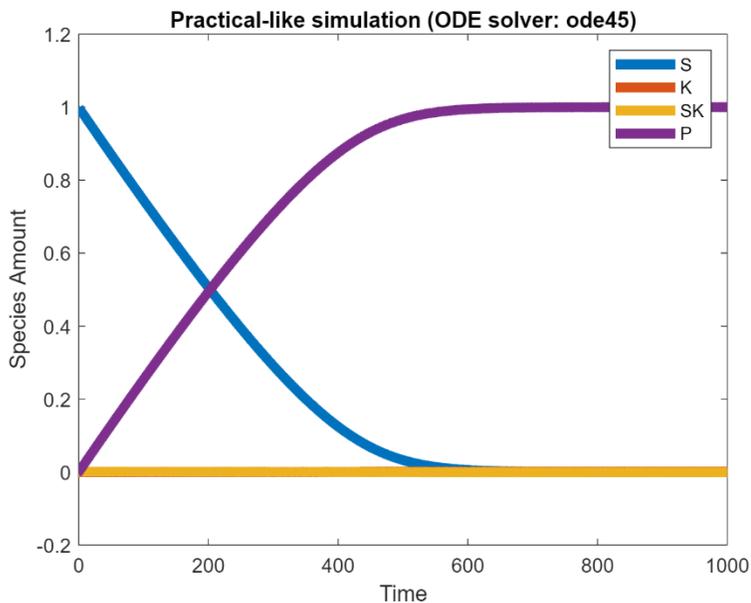

```matlab
x = (1:100);
A(:,1) = 2.5*exp(-(x-20).^2./200) + 0.075;
A(:,2) = 12.5*exp(-(x-40).^2./200) + 0.075;
A(:,3) = 10*exp(-(x-60).^2./200) + 0.065;
A(:,4) = exp(-(x-80).^2./100) + 0.065;

Dp = Ctp*A';
sd =0.003;
```

```matlab
Dmp=Dp+sd*randn(size(Dp));
format long
sv_Ctp = svd(Ctp); display(sv_Ctp);
```

sv_Ctp = *4×1*

     33.835880907017476
     22.880312619089516
      0.009704932944434
      0.000000000000000

```matlab
sv_Dmp = svd(Dmp); display(sv_Dmp);
```

sv_Dmp = *100×1*

$10^2 \times$

     3.085801990220252
     1.065885298536425
     0.003745157882322
     0.001712290785874
     0.001700929472219
     0.001693939050571
     0.001677304825045
     0.001671313064059
     0.001655661717768
     0.001650479588727
           ⋮

```matlab
format default

A_Sm = A(:,1)*S_conc0+sd*randn(size(A(:,1)));

% only the spectrum of the substrate is known (t<0)
% Aest_kn_un_jsubs(:,2:4) = (Cw(:,2:4)\(Dmcw-Cw(:,1)*A(:,1)'))';
Aest_kn_un_jsubs_fromDmp = [A_Sm (Dmp-Ctp(:,1)*A_Sm(:,1)')'/Ctp(:,2:4)'];
figure, plot(xc, Aest_kn_un_jsubs_fromDmp,'LineWidth',5); hold on, plot(xc, A,'-.k','LineWidth',1); hold off
title('Est. spectra based on the known spectrum of substrate (t<0)');
xlabel('wavelength');
ylabel('absorbance, a.u.');
legend('S','K','SK','P');
```

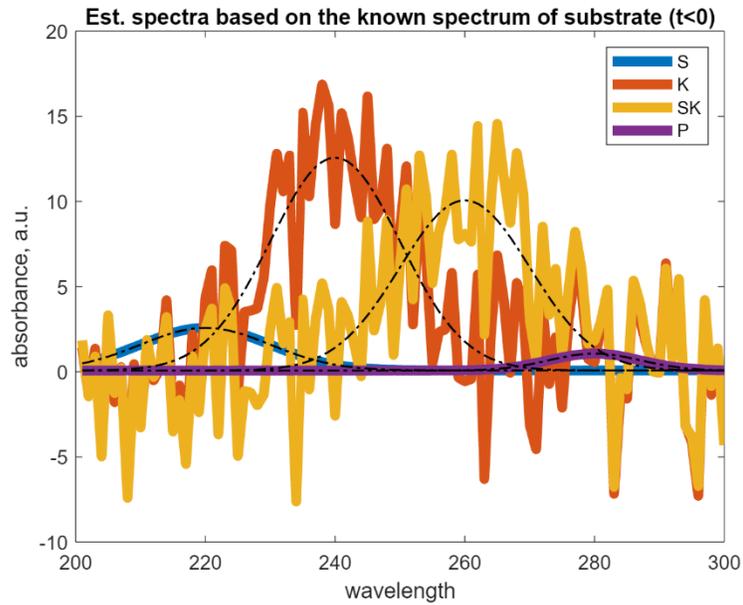

```matlab
% spectrum of the mixture of substrate and enzyme is known (t<=0)
% A_Kest = (Dmcw(1,:)'-A_Sm)/K_conc0;
[A_S_Kest] = [A_Sm Dmp(1,:)']/[S_conc0 S_conc0; 0 K_conc0];
A_Sm = A_S_Kest(:,1);
A_Kest = A_S_Kest(:,2);
% Dw = Cw*A' = C_kn*A_kn + C_un*A_un ==> Aest = pinv(Cw_un)*(Dw-Cw_kn*A_kn')
% Aest_kn_un(:,3:4) = (Cw(:,3:4)\(Dmcw-Cw(:,1:2)*A(:,1:2)'))';
Aest_kn_un_fromD = [A_Sm A_Kest ((Dmp-Ctp(:,1:2)*[A_Sm 
A_Kest]'))'/Ctp(:,3:4)'];
 figure, plot(xc, Aest_kn_un_fromD,'LineWidth',5); hold on, plot(xc, A,'-
.k','LineWidth',1); hold off
 title('Using spectrum of the mixture of substrate and enzyme as well (t<=0)');
 xlabel('wavelength');
 ylabel('absorbance, a.u.');
 legend('S','K','SK','P');
```

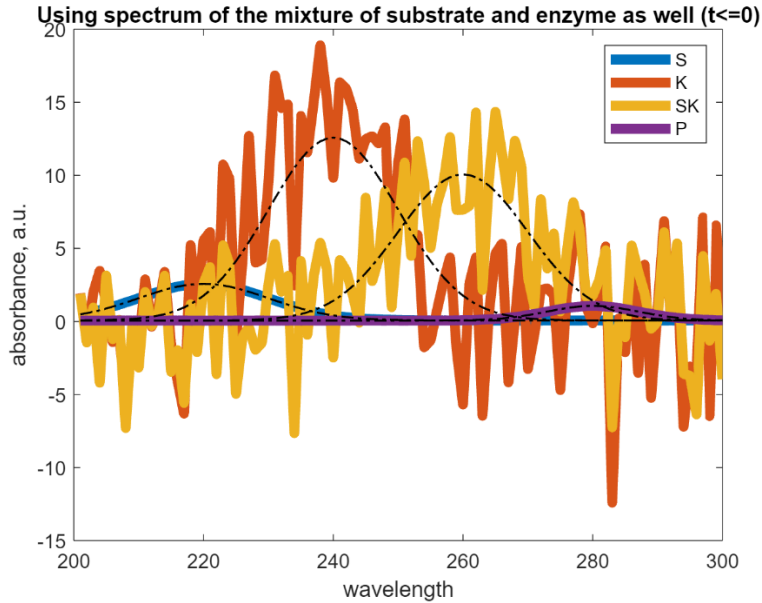

# Multiple equilibria titration of monoprotic dyes

This is an acid-base titration of a mixture of three monoprotic dyes with NaOH, the three indicators are Phenol red (P), Methyl orange (M) and Bromocresol green (B). There are three equilibria among the different forms of the dyes:

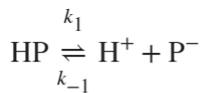

$$HP \underset{k_{-1}}{\overset{k_1}{\rightleftharpoons}} H^+ + P^-$$

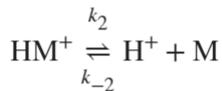

$$HM^+ \underset{k_{-2}}{\overset{k_2}{\rightleftharpoons}} H^+ + M$$

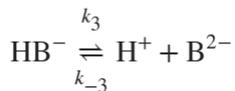

$$HB^- \underset{k_{-3}}{\overset{k_3}{\rightleftharpoons}} H^+ + B^{2-}$$

The initial solution was acidified with 0.001 M HCl, thus this task is equivalent to titrating strong acid with strong base. The titration was carried out with 0:11 mL of 0.005 M NaOH getting 60 titration points (i.e. 60 is the number of solutions as well).

```
clearvars
spec_names = {'P^-'  'M'   'B^{2-}' 'H^+'   'HP'  'HM^+' 'HB^-'  'OH^-'};
Model     = [  1     0      0        0       1     0      0       0;   ... % P-
               0     1      0        0       0     1      0       0;   ... % M
               0     0      1        0       0     0      1       0;   ... % B2-
               0     0      0        1       1     1      1      -1];  ... % H
% instead of L, mL can be used
% NaOH titrant with indicator(s)
L_mL = 1000; % L:1, mL:1000
inds_in_titrant = 1e-10; % indicators conc. in titrant
```

```matlab
    x_val_ind = 1; % independent variable for figure
    log_beta = log10(L_mL)+ [0      0      0      0         7.66   3.43   4.62   -14];
    beta =10.^log_beta;
    c_0 =             L_mL*[3e-5   3e-5   2e-5    0.001]; % tot conc in initial solution P,M,B,H
    c_added =         L_mL*[0      0      0      -0.005]; % tot conc titration solution P,M,B,H
    ncomp = length(c_0);                % number of components
    v_0 = L_mL*0.05;                    % initial volume (50mL)
    v_added =1e-3*L_mL*[0:0.6:9 9.05:0.05:9.7 9.725:0.025:9.9 9.95:0.05:11]';   % added solution (0-11mL)
    %           16            14              8               22
    v_tot =v_0+v_added;                 % total volume
    nvol =length(v_added);              % number of solutions
    c_tot=NaN(nvol,ncomp);
    C=NaN(nvol,size(Model,2));
    for j=1:ncomp
        ind_range = 1:16;
        c_tot(ind_range,j)=(v_0*c_0(j)+v_added(ind_range)*c_added(j))./v_tot(ind_range);
        ind_range = 17:30;
        c_added = L_mL*[inds_in_titrant 0 0 -0.005];
        c_tot(ind_range,j)=(v_0*c_0(j)+v_added(ind_range)*c_added(j))./v_tot(ind_range);
        ind_range = 31:38;
        c_added = L_mL*[0 inds_in_titrant 0 -0.005];
        c_tot(ind_range,j)=(v_0*c_0(j)+v_added(ind_range)*c_added(j))./v_tot(ind_range);
        ind_range = 39:60;
        c_added = L_mL*[inds_in_titrant inds_in_titrant inds_in_titrant -0.005];
        c_tot(ind_range,j)=(v_0*c_0(j)+v_added(ind_range)*c_added(j))./v_tot(ind_range);
    end

    c_comp_guess = c_0; %[1e-10 1e-10 1e-10 1e-10];       % initial guess for [P-],[M],[B2-],[H]
    for i=1:nvol
        C(i,:)=NewtonRaphson(Model,beta,c_tot(i,:),c_comp_guess,i);
        c_comp_guess=C(i,1:ncomp);
    end
    switch x_val_ind
        case 1
            x_val = v_added;
        case 2
            x_val = C(:,4);
        case 3
```

```matlab
        x_val = -log10(C(:,4)/L_mL);
end
figure
pl2 = plot(x_val,C(:,[1:3 5:7]),'LineWidth',2);
pl2(1).LineStyle = '--';
pl2(2).LineStyle = '--';
pl2(3).LineStyle = '--';
pl2(4).LineStyle = '-';
pl2(5).LineStyle = '-';
pl2(6).LineStyle = '-';

pl2(1).Color = "#FF0000"; %red
pl2(2).Color = "#FF8800"; %orange
pl2(3).Color = "#77AC30"; %green
pl2(4).Color = "#FF0000"; %red
pl2(5).Color = "#FF8800"; %orange
pl2(6).Color = "#77AC30"; %green

legend(spec_names([1:3 5:7]),'Location','bestoutside');
x_lab = '';
y_lab = '';
switch x_val_ind
    case 1
        if L_mL>1
            x_lab = 'V_{NaOH} (mL)';
        else
            x_lab = 'V_{NaOH} (L)';
        end
    case 2
        if L_mL>1
            x_lab = '[H^+] [mol/mL]';
        else
            x_lab = '[H^+] [mol/L]';
        end
    case 3
        x_lab = 'pH';
end
if L_mL>1
    y_lab = 'conc. [mol/mL]';
else
    y_lab = 'conc. [mol/L]';
end
xlabel(x_lab);
ylabel(y_lab);
title('Acid-base titration of a mixture of three monoprotic dyes with NaOH');
```

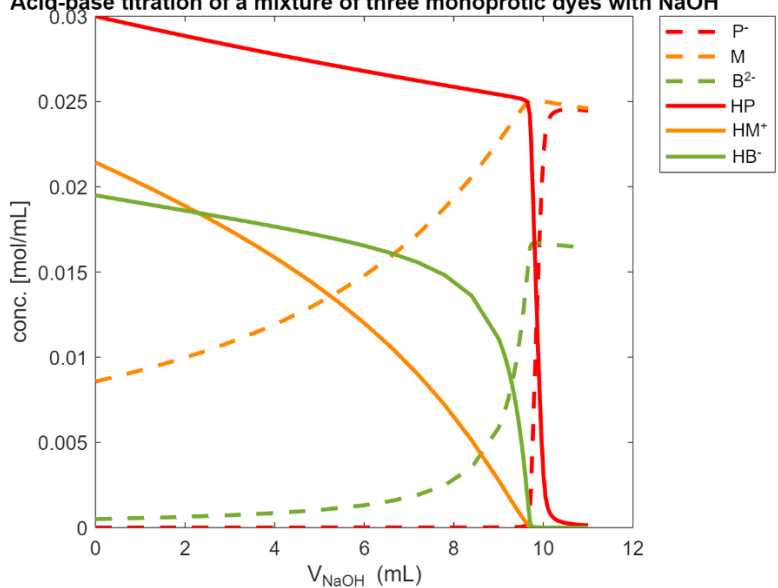

```
drawnow;

format long
if inds_in_titrant==0
    disp('rank deficient concentration matrix');
else
    disp('full rank concentration matrix');
end
```

full rank concentration matrix

```
sv_C6 = svd(C(:,[1:3 5:7]))
```

sv_C6 = 6×1

      0.244565342039891
      0.149238105847751
      0.057345153259102
      0.018292524950773
      0.000000039070176
      0.000000012964462

```
format default

figure
pl2 = plot(x_val,C(:,1:3)+C(:,5:7),'LineWidth',2);
pl2(1).Color = "#FF0000"; %red
pl2(2).Color = "#FF8800"; %orange
pl2(3).Color = "#77AC30"; %green
xlabel(x_lab);
ylabel(y_lab);
```

```
line(x_val([1 end]),C([1 end],1)+C([1 end],5),'Color','black','LineStyle','-
.');
line(x_val([1 end]),C([1 end],2)+C([1 end],6),'Color','black','LineStyle','-
.');
line(x_val([1 end]),C([1 end],3)+C([1 end],7),'Color','black','LineStyle','-
.');
legend(strcat({'['},spec_names(1:3)',{'] + 
'},{'['},spec_names(5:7)',{']'}),'Location','bestoutside');
title('The titration dilution is not straight line');
```

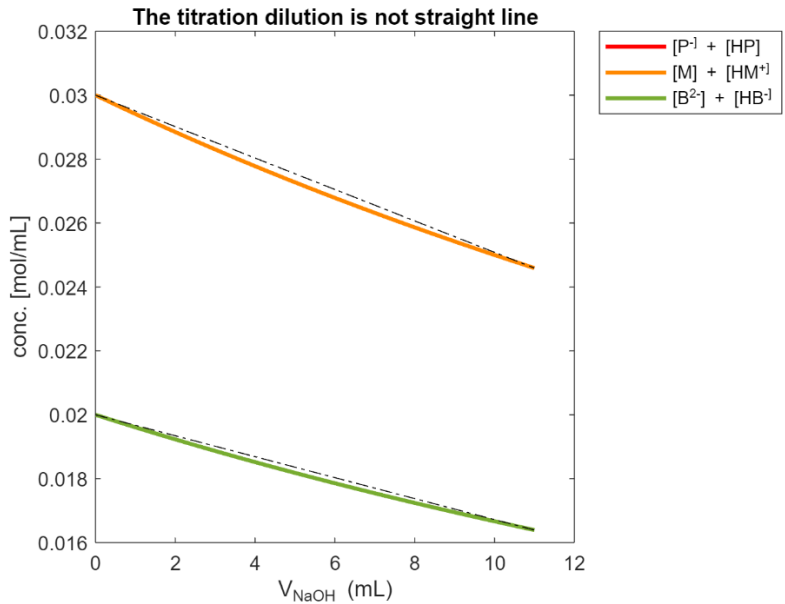

```
drawnow;
```

# Internal and external normalizations

```
clearvars
% data from the paper of Henry & Kim, ChemoLab, 8 (1990) 205-216
% https://doi.org/10.1016/0169-7439(90)80136-T
    %Source compositions
        %Marine    Udust     Auto
    C = [[ 0.40000  0.01250  0.00000]; ...    %Na
         [ 0.00000  0.08840  0.01100]; ...    %Al
         [ 0.00000  0.22300  0.00820]; ...    %Si
         [ 0.40000  0.00000  0.03000]; ...    %Cl
         [ 0.01400  0.01030  0.00072]; ...    %K
         [ 0.01400  0.02440  0.01250]; ...    %Ca
         [ 0.00000  0.00640  0.00000]; ...    %Ti
         [ 0.00000  0.06000  0.02100]; ...    %Fe
         [ 0.00200  0.00020  0.05000]; ...    %Br
```

```
            [ 0.00000  0.00370  0.20000]];        %Pb

     %Source apportionment ug/m3
           %Marine    Udust     Auto
     A = [[    3         8        19]; ...
          [    3         8        16]; ...
          [    5         7        12]; ...
          [    6        49        13]; ...
          [    6        39         7]; ...
          [    6         9        12]; ...
          [    6        17        18]; ...
          [    2         6         6]; ...
          [    4        42         7]; ...
          [    4        49        20]; ...
          [    4        29        14]; ...
          [    5        44         9]; ...
          [    5        32        12]; ...
          [    5         9        11]; ...
          [    6        26         8]; ...
          [    3         8        14]; ...
          [    2        39        11]; ...
          [    3         6        20]; ...
          [    4        48        15]; ...
          [    5        37        14]];

     ElemsTxt = {'Na'; 'Al'; 'Si'; 'Cl'; 'K'; 'Ca'; 'Ti'; 'Fe'; 'Br'; 'Pb'};
     ApporTxt = {'Marine';'Udust';'Auto'};
     R = C*A'; R_dual = R';
     [I, J] = size(R);
     [Um, Sm, Vm] = svds(R,3);
     loads{1}=Um*Sm; loads{2}=Vm;
     [sgnsXV,loadsXV] = sign_flip(loads,R);
     loads{1}=Vm*Sm; loads{2}=Um;
     [sgnsYU,loadsYU] = sign_flip(loads,R_dual);
     X = loadsXV{1}; V = loadsXV{2};
     Y = loadsYU{1}; U = loadsYU{2};
```

## l1-norm

External normalization

```
 R_ex_1n_r = R./sum(R,2);
 [Um, Sm, Vm] = svds(R_ex_1n_r,3);
 loads{1}=Um*Sm; loads{2}=Vm;
 [sgnsXV_ex_1n,loadsXV_ex_1n] = sign_flip(loads,R_ex_1n_r);
 X_ex_1n = loadsXV_ex_1n{1}; V_ex_1n = loadsXV_ex_1n{2};
```

```matlab
R_ex_1n_c = R_dual./sum(R_dual,2);
[Um, Sm, Vm] = svds(R_ex_1n_c,3);
loads{1}=Um*Sm; loads{2}=Vm;
[sgnsYU_ex_1n,loadsYU_ex_1n] = sign_flip(loads,R_ex_1n_c);
Y_ex_1n = loadsYU_ex_1n{1}; U_ex_1n = loadsYU_ex_1n{2};

disp('Checking the decomposition after external normalization');
```

Checking the decomposition after external normalization

### X_ex_1n*V_ex_1n'-R_ex_1n_r

ans = 10×20

$10^{-16} \times$

```
   -0.5551   -0.0694   -0.2082         0         0   -0.2082   -0.2776 ...
   -0.0694    0.2776    0.3643    0.1388    0.1388    0.4857    0.4163
   -0.8327   -0.4163   -0.3643   -0.2776   -0.1388   -0.3469   -0.3469
   -0.5551   -0.0694   -0.0694   -0.4163   -0.1388   -0.1388   -0.1388
   -0.2082    0.0694    0.0694         0         0    0.1041    0.2082
   -0.4857   -0.1041   -0.1041   -0.1388   -0.1388   -0.0694         0
   -0.6939   -0.3123   -0.2949   -0.4163   -0.4163   -0.3469   -0.3469
   -0.4857   -0.1735   -0.1388   -0.4163   -0.2776   -0.1388   -0.0694
   -0.2776         0         0   -0.2082   -0.2429   -0.2082   -0.2776
   -0.5551         0   -0.1388   -0.2082   -0.1388   -0.4163   -0.1388
```

### Y_ex_1n*U_ex_1n'-R_ex_1n_c

ans = 20×10

$10^{-15} \times$

```
    0.0278   -0.1388   -0.3886   -0.0555   -0.0139   -0.0555   -0.0087 ...
    0.0833    0.0278         0    0.0278    0.0035         0    0.0035
    0.1110    0.1527   -0.0278    0.1388    0.0017    0.0139    0.0017
    0.0139   -0.0555   -0.0555         0         0   -0.0069    0.0017
         0   -0.0278   -0.0555   -0.0278         0         0         0
    0.1110    0.1665         0    0.1110    0.0035    0.0139    0.0026
    0.0833    0.0555   -0.0278    0.0278    0.0017   -0.0069    0.0017
    0.0278    0.0278   -0.1110         0   -0.0069   -0.0139   -0.0017
    0.0278   -0.0833         0   -0.0416         0         0    0.0017
         0   -0.0833   -0.1110   -0.0694   -0.0035   -0.0208         0
      ⋮
```

```matlab
% T_A_ex_1n = (A./sum(A))'*V_ex_1n;
% T_C_ex_1n = (C./sum(C))'*U_ex_1n;
figure,
sgtitle('Depicting dimensional reduction after external normalization');
subplot(1,2,1);
```

```
plot3(X_ex_1n(:,1),X_ex_1n(:,2),X_ex_1n(:,3),'.b'), grid on,
text(X_ex_1n(:,1),X_ex_1n(:,2),X_ex_1n(:,3),ElemsTxt((1:I)));
T_A_ex_1n = (A./sum(A))'*V_ex_1n;
xlabel('X^{ex}_1'); ylabel('X^{ex}_2'); zlabel('X^{ex}_3');
hold on
plot3(T_A_ex_1n(:,1),T_A_ex_1n(:,2),T_A_ex_1n(:,3),'og'), grid on,
text(T_A_ex_1n(:,1),T_A_ex_1n(:,2),T_A_ex_1n(:,3),ApporTxt);
plot3(T_A_ex_1n([1:end 1],1),T_A_ex_1n([1:end 1],2),T_A_ex_1n([1:end 1],3),'-g');
hold off
title('V|row-space');
subplot(1,2,2);
plot3(Y_ex_1n(:,1),Y_ex_1n(:,2),Y_ex_1n(:,3),'.b'), grid on,
text(Y_ex_1n(:,1),Y_ex_1n(:,2),Y_ex_1n(:,3),num2str((1:J)'));
T_C_ex_1n = (C./sum(C))'*U_ex_1n;
xlabel('Y^{ex}_1'); ylabel('Y^{ex}_2'); zlabel('Y^{ex}_3');
hold on
plot3(T_C_ex_1n(:,1),T_C_ex_1n(:,2),T_C_ex_1n(:,3),'og'), grid on,
text(T_C_ex_1n(:,1),T_C_ex_1n(:,2),T_C_ex_1n(:,3),ApporTxt);
plot3(T_C_ex_1n([1:end 1],1),T_C_ex_1n([1:end 1],2),T_C_ex_1n([1:end 1],3),'-g');
hold off
title('U|column-space');
```

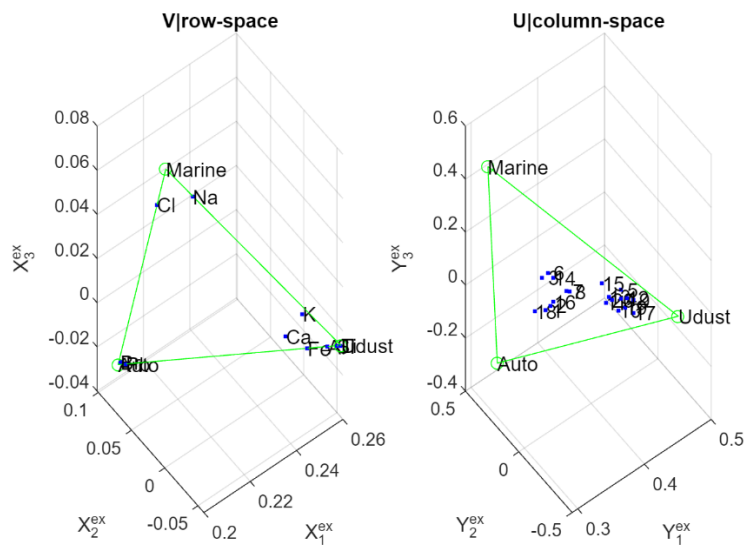

### Internal normalization

```
figure,
sgtitle('Abstract spaces before internal normalization');
```

```
 subplot(1,2,1);
 plot3(X(:,1),X(:,2),X(:,3),'.b'), grid on,
text(X(:,1),X(:,2),X(:,3),ElemsTxt((1:I)));
 T_A = A'*V;
 xlabel('X_1'); ylabel('X_2'); zlabel('X_3');
 hold on
 plot3([0 T_A(1,1)],[0 T_A(1,2)],[0 T_A(1,3)],'Color','g');
 plot3([0 T_A(2,1)],[0 T_A(2,2)],[0 T_A(2,3)],'Color','g');
 plot3([0 T_A(3,1)],[0 T_A(3,2)],[0 T_A(3,3)],'Color','g');
 text(T_A(:,1),T_A(:,2),T_A(:,3),ApporTxt);
 hold off
 title('V|row-space');
 subplot(1,2,2);
 plot3(Y(:,1),Y(:,2),Y(:,3),'.r'), grid on,
text(Y(:,1),Y(:,2),Y(:,3),num2str((1:J)'));
 xlabel('Y_1'); ylabel('Y_2'); zlabel('Y_3');
 T_C = C'*U;
 hold on
 plot3([0 T_C(1,1)],[0 T_C(1,2)],[0 T_C(1,3)],'Color','g');
 plot3([0 T_C(2,1)],[0 T_C(2,2)],[0 T_C(2,3)],'Color','g');
 plot3([0 T_C(3,1)],[0 T_C(3,2)],[0 T_C(3,3)],'Color','g');
 text(T_C(:,1),T_C(:,2),T_C(:,3),ApporTxt);
 hold off
 title('U|column-space');
```

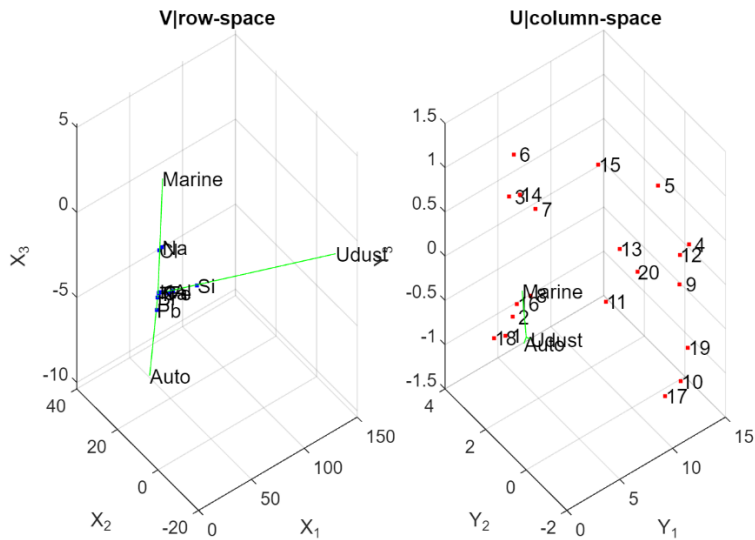

```
X_in_1n = X./sum(X,2);
Y_in_1n = Y./sum(Y,2);
```

```
 T_A_in_1n = T_A./sum(T_A,2);
 T_C_in_1n = T_C./sum(T_C,2);
 figure,
 sgtitle('Depicting dimensional reduction after internal normalization');
 subplot(1,2,1);
 plot3(X_in_1n(:,1),X_in_1n(:,2),X_in_1n(:,3),'.b'), grid on,
 text(X_in_1n(:,1),X_in_1n(:,2),X_in_1n(:,3),ElemsTxt((1:I)));
 xlabel('X^{in}_1'); ylabel('X^{in}_2'); zlabel('X^{in}_3');
 hold on
 plot3(T_A_in_1n(:,1),T_A_in_1n(:,2),T_A_in_1n(:,3),'og'), grid on,
 text(T_A_in_1n(:,1),T_A_in_1n(:,2),T_A_in_1n(:,3),ApporTxt);
 plot3(T_A_in_1n([1:end 1],1),T_A_in_1n([1:end 1],2),T_A_in_1n([1:end 1],3),'-g');
 hold off
 title('V|row-space');
 subplot(1,2,2);
 plot3(Y_in_1n(:,1),Y_in_1n(:,2),Y_in_1n(:,3),'.b'), grid on,
 text(Y_in_1n(:,1),Y_in_1n(:,2),Y_in_1n(:,3),num2str((1:J)'));
 xlabel('Y^{in}_1'); ylabel('Y^{in}_2'); zlabel('Y^{in}_3');
 hold on
 plot3(T_C_in_1n(:,1),T_C_in_1n(:,2),T_C_in_1n(:,3),'og'), grid on,
 text(T_C_in_1n(:,1),T_C_in_1n(:,2),T_C_in_1n(:,3),ApporTxt);
 plot3(T_C_in_1n([1:end 1],1),T_C_in_1n([1:end 1],2),T_C_in_1n([1:end 1],3),'-g');
 hold off
 title('U|column-space');
```

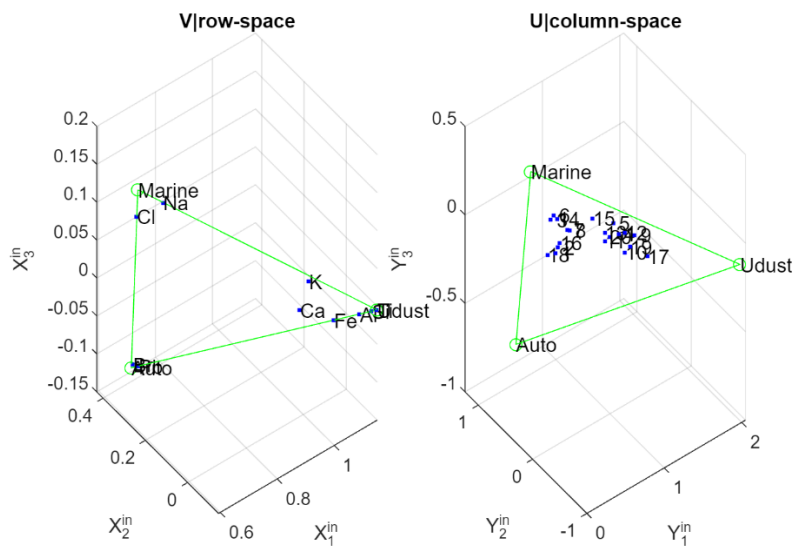

```matlab
 figure,
 sgtitle('Reduced 2D spaces with external and internal normalizations');
 subplot(2,2,1);
 plot(X_ex_1n(:,2),X_ex_1n(:,3),'.b'), grid on,
text(X_ex_1n(:,2),X_ex_1n(:,3),ElemsTxt((1:I)));
 xlabel('X^{ex}_2'); ylabel('X^{ex}_3');
 hold on
 plot(T_A_ex_1n(:,2),T_A_ex_1n(:,3),'og'), grid on,
text(T_A_ex_1n(:,2),T_A_ex_1n(:,3),ApporTxt);
 plot(T_A_ex_1n([1:end 1],2),T_A_ex_1n([1:end 1],3),'-g');
 hold off
 title('V|row-space');
 subplot(2,2,2);
 plot(Y_ex_1n(:,2),Y_ex_1n(:,3),'.b'), grid on,
text(Y_ex_1n(:,2),Y_ex_1n(:,3),num2str((1:J)'));
 xlabel('Y^{ex}_2'); ylabel('Y^{ex}_3');
 hold on
 plot(T_C_ex_1n(:,2),T_C_ex_1n(:,3),'og'), grid on,
text(T_C_ex_1n(:,2),T_C_ex_1n(:,3),ApporTxt);
 plot(T_C_ex_1n([1:end 1],2),T_C_ex_1n([1:end 1],3),'-g');
 hold off
 title('U|column-space');
 subplot(2,2,3);
 plot(X_in_1n(:,2),X_in_1n(:,3),'.b'), grid on,
text(X_in_1n(:,2),X_in_1n(:,3),ElemsTxt((1:I)));
 xlabel('X^{in}_2'); ylabel('X^{in}_3');
 hold on
 plot(T_A_in_1n(:,2),T_A_in_1n(:,3),'og'), grid on,
text(T_A_in_1n(:,2),T_A_in_1n(:,3),ApporTxt);
 plot(T_A_in_1n([1:end 1],2),T_A_in_1n([1:end 1],3),'-g');
 hold off
 subplot(2,2,4);
 plot(Y_in_1n(:,2),Y_in_1n(:,3),'.b'), grid on,
text(Y_in_1n(:,2),Y_in_1n(:,3),num2str((1:J)'));
 xlabel('Y^{in}_2'); ylabel('Y^{in}_3');
 hold on
 plot(T_C_in_1n(:,2),T_C_in_1n(:,3),'og'), grid on,
text(T_C_in_1n(:,2),T_C_in_1n(:,3),ApporTxt);
 plot(T_C_in_1n([1:end 1],2),T_C_in_1n([1:end 1],3),'-g');
 hold off
```

# Reduced 2D spaces with external and internal normalizations

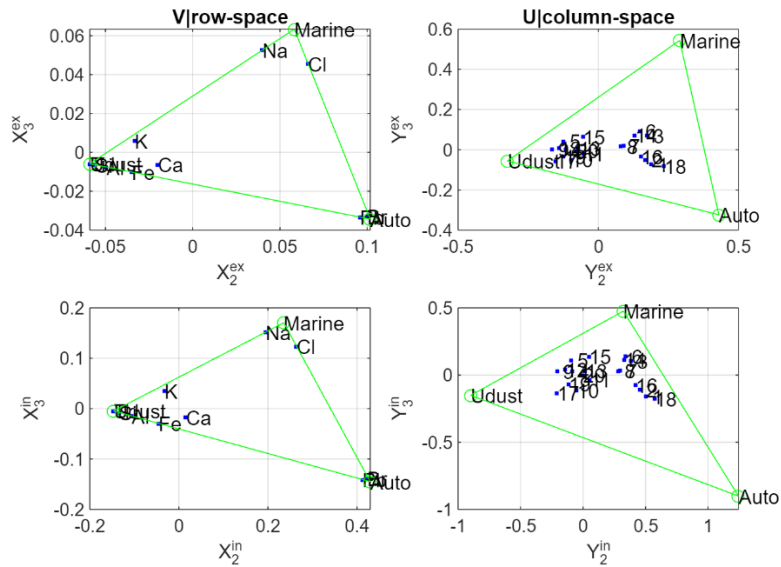

## First scores-vector-to-1 normalization (fsvt1n)

External normalization

```
epsi = 1e-15;
Xi = X;
Vi = V;
Yi = Y;
Maxcycl = 100;
icycl = 0;
while and(all(abs(Xi(:,1)-ones(I,1))>epsi),icycl<Maxcycl)
    R_ex_fsvt1n_r = inv(diag(R*Vi(:,1)))*R;
    [Ui, Si, Vi] = svds(R_ex_fsvt1n_r,3);
    if all(Ui(:,1)<0)
        Ui(:,1) = -Ui(:,1);
        Vi(:,1) = -Vi(:,1);
    end
    Xi = Ui*Si;
    icycl = icycl+1;
end
X_ex_fsvt1n = Xi;
V_ex_fsvt1n = Vi;

icycl = 0;
Ui = U;
while and(all(abs(Yi(:,1)-ones(J,1))>epsi),icycl<Maxcycl)
    R_ex_fsvt1n_c = inv(diag(R_dual*Ui(:,1)))*R_dual;
    [Vi, Si, Ui] = svds(R_ex_fsvt1n_c,3);
    if all(Vi(:,1)<0)
```

```matlab
            Vi(:,1) = -Vi(:,1);
            Ui(:,1) = -Ui(:,1);
        end
        Yi = Vi*Si;
        icycl = icycl+1;
    end
    Y_ex_fsvt1n = Yi;
    U_ex_fsvt1n = Ui;

    figure,
    sgtitle('Depicting dimensional reduction after external normalization');
    subplot(1,2,1);
    plot3(X_ex_fsvt1n(:,1),X_ex_fsvt1n(:,2),X_ex_fsvt1n(:,3),'.b'), grid on,
    text(X_ex_fsvt1n(:,1),X_ex_fsvt1n(:,2),X_ex_fsvt1n(:,3),ElemsTxt((1:I)));
    T_A_ex_fsvt1n = A'*V_ex_fsvt1n;
    T_A_ex_fsvt1n = inv(diag(T_A_ex_fsvt1n(:,1)))*T_A_ex_fsvt1n;
    xlabel('X^{ex}_1'); ylabel('X^{ex}_2'); zlabel('X^{ex}_3');
    hold on
    plot3(T_A_ex_fsvt1n(:,1),T_A_ex_fsvt1n(:,2),T_A_ex_fsvt1n(:,3),'og'), grid on,
    text(T_A_ex_fsvt1n(:,1),T_A_ex_fsvt1n(:,2),T_A_ex_fsvt1n(:,3),ApporTxt);
    plot3(T_A_ex_fsvt1n([1:end 1],1),T_A_ex_fsvt1n([1:end 1],2),T_A_ex_fsvt1n([1:end 1],3),'-g');
    hold off
    title('V|row-space');
    subplot(1,2,2);
    plot3(Y_ex_fsvt1n(:,1),Y_ex_fsvt1n(:,2),Y_ex_fsvt1n(:,3),'.b'), grid on,
    text(Y_ex_fsvt1n(:,1),Y_ex_fsvt1n(:,2),Y_ex_fsvt1n(:,3),num2str((1:J)'));
    T_C_ex_fsvt1n = C'*U_ex_fsvt1n;
    T_C_ex_fsvt1n = inv(diag(T_C_ex_fsvt1n(:,1)))*T_C_ex_fsvt1n;
    xlabel('Y^{ex}_1'); ylabel('Y^{ex}_2'); zlabel('Y^{ex}_3');
    hold on
    plot3(T_C_ex_fsvt1n(:,1),T_C_ex_fsvt1n(:,2),T_C_ex_fsvt1n(:,3),'og'), grid on,
    text(T_C_ex_fsvt1n(:,1),T_C_ex_fsvt1n(:,2),T_C_ex_fsvt1n(:,3),ApporTxt);
    plot3(T_C_ex_fsvt1n([1:end 1],1),T_C_ex_fsvt1n([1:end 1],2),T_C_ex_fsvt1n([1:end 1],3),'-g');
    hold off
    title('U|column-space');
```

## Depicting dimensional reduction after external normalization

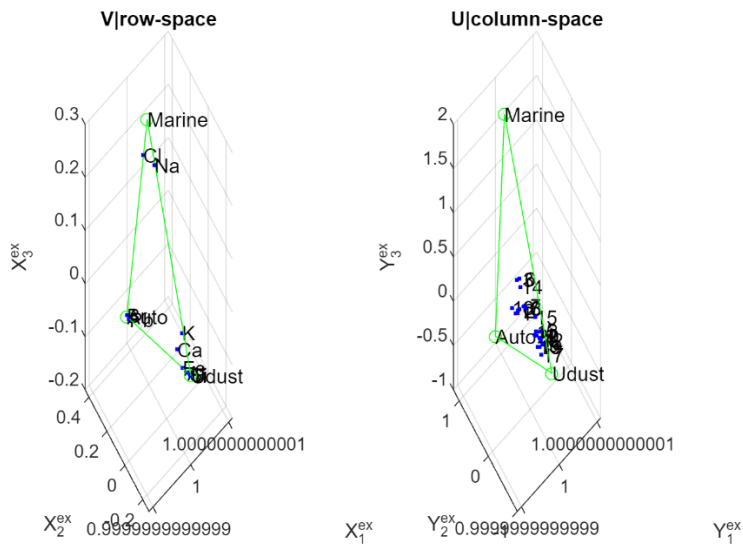

### Internal normalization

```
X_in_fsvt1n = inv(diag(X(:,1)))*X;
Y_in_fsvt1n = inv(diag(Y(:,1)))*Y;
T_A_in_fsvt1n = A'*V;
T_A_in_fsvt1n = inv(diag(T_A_in_fsvt1n(:,1)))*T_A_in_fsvt1n;
T_C_in_fsvt1n = C'*U;
T_C_in_fsvt1n = inv(diag(T_C_in_fsvt1n(:,1)))*T_C_in_fsvt1n;
figure,
sgtitle('Depicting dimensional reduction after internal normalization');
subplot(1,2,1);
plot3(X_in_fsvt1n(:,1),X_in_fsvt1n(:,2),X_in_fsvt1n(:,3),'.b'), grid on,
text(X_in_fsvt1n(:,1),X_in_fsvt1n(:,2),X_in_fsvt1n(:,3),ElemsTxt((1:I)));
xlabel('X^{in}_1'); ylabel('X^{in}_2'); zlabel('X^{in}_3');
hold on
plot3(T_A_in_fsvt1n(:,1),T_A_in_fsvt1n(:,2),T_A_in_fsvt1n(:,3),'og'), grid on,
text(T_A_in_fsvt1n(:,1),T_A_in_fsvt1n(:,2),T_A_in_fsvt1n(:,3),ApporTxt);
plot3(T_A_in_fsvt1n([1:end 1],1),T_A_in_fsvt1n([1:end 1],2),T_A_in_fsvt1n([1:end 1],3),'-g');
hold off
title('V|row-space');
subplot(1,2,2);
plot3(Y_in_fsvt1n(:,1),Y_in_fsvt1n(:,2),Y_in_fsvt1n(:,3),'.b'), grid on,
text(Y_in_fsvt1n(:,1),Y_in_fsvt1n(:,2),Y_in_fsvt1n(:,3),num2str((1:J)'));
xlabel('Y^{in}_1'); ylabel('Y^{in}_2'); zlabel('Y^{in}_3');
hold on
plot3(T_C_in_fsvt1n(:,1),T_C_in_fsvt1n(:,2),T_C_in_fsvt1n(:,3),'og'), grid on,
text(T_C_in_fsvt1n(:,1),T_C_in_fsvt1n(:,2),T_C_in_fsvt1n(:,3),ApporTxt);
```

```
plot3(T_C_in_fsvt1n([1:end 1],1),T_C_in_fsvt1n([1:end 1],2),T_C_in_fsvt1n([1:end 1],3),'-g');
hold off
title('U|column-space');
```

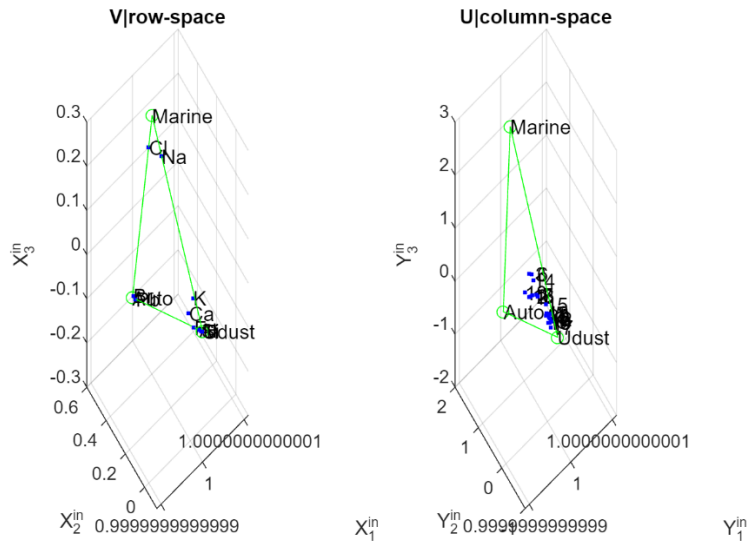

```
figure,
sgtitle('Reduced 2D spaces with external and internal normalizations');
subplot(2,2,1);
plot(X_ex_fsvt1n(:,2),X_ex_fsvt1n(:,3),'.b'), grid on,
text(X_ex_fsvt1n(:,2),X_ex_fsvt1n(:,3),ElemsTxt((1:I)));
xlabel('X^{ex}_2'); ylabel('X^{ex}_3');
hold on
plot(T_A_ex_fsvt1n(:,2),T_A_ex_fsvt1n(:,3),'og'), grid on,
text(T_A_ex_fsvt1n(:,2),T_A_ex_fsvt1n(:,3),ApporTxt);
plot(T_A_ex_fsvt1n([1:end 1],2),T_A_ex_fsvt1n([1:end 1],3),'-g');
hold off
title('V|row-space');
subplot(2,2,2);
plot(Y_ex_fsvt1n(:,2),Y_ex_fsvt1n(:,3),'.b'), grid on,
text(Y_ex_fsvt1n(:,2),Y_ex_fsvt1n(:,3),num2str((1:J)'));
xlabel('Y^{ex}_2'); ylabel('Y^{ex}_3');
hold on
plot(T_C_ex_fsvt1n(:,2),T_C_ex_fsvt1n(:,3),'og'), grid on,
text(T_C_ex_fsvt1n(:,2),T_C_ex_fsvt1n(:,3),ApporTxt);
plot(T_C_ex_fsvt1n([1:end 1],2),T_C_ex_fsvt1n([1:end 1],3),'-g');
hold off
title('U|column-space');
```

```
subplot(2,2,3);
plot(X_in_fsvt1n(:,2),X_in_fsvt1n(:,3),'.b'), grid on,
text(X_in_fsvt1n(:,2),X_in_fsvt1n(:,3),ElemsTxt((1:I)));
xlabel('X^{in}_2'); ylabel('X^{in}_3');
hold on
plot(T_A_in_fsvt1n(:,2),T_A_in_fsvt1n(:,3),'og'), grid on,
text(T_A_in_fsvt1n(:,2),T_A_in_fsvt1n(:,3),ApporTxt);
plot(T_A_in_fsvt1n([1:end 1],2),T_A_in_fsvt1n([1:end 1],3),'-g');
hold off
subplot(2,2,4);
plot(Y_in_fsvt1n(:,2),Y_in_fsvt1n(:,3),'.b'), grid on,
text(Y_in_fsvt1n(:,2),Y_in_fsvt1n(:,3),num2str((1:J)'));
xlabel('Y^{in}_2'); ylabel('Y^{in}_3');
hold on
plot(T_C_in_fsvt1n(:,2),T_C_in_fsvt1n(:,3),'og'), grid on,
text(T_C_in_fsvt1n(:,2),T_C_in_fsvt1n(:,3),ApporTxt);
plot(T_C_in_fsvt1n([1:end 1],2),T_C_in_fsvt1n([1:end 1],3),'-g');
hold off
```

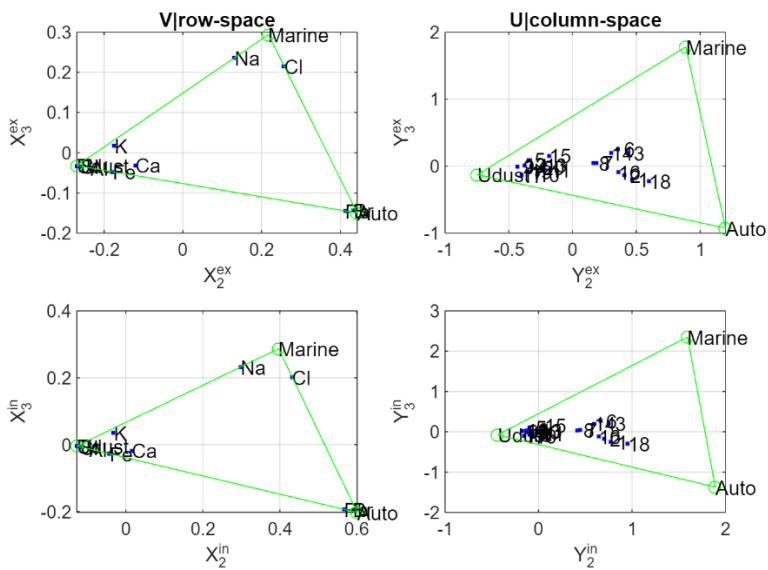

Sometimes, algorithm of the fsvt1n external normalization does not converge.

```
R = [1 0 0;2 3 4;5 6 7];
[X_100, V_100] = fsvt1n_ext_norm(R,3,epsi,100);
[X_101, V_101] = fsvt1n_ext_norm(R,3,epsi,101);
[X_102, V_102] = fsvt1n_ext_norm(R,3,epsi,102);
[X_103, V_103] = fsvt1n_ext_norm(R,3,epsi,103);
maxIts = {'';100;''};
table(maxIts,X_100,V_100)
```

`ans = 3×3 table`

|   | maxIts | X_100 | | | V_100 | | ... |
|---|---|---|---|---|---|---|---|
|   |   | 1 | 2 | 3 | 1 | 2 |   |
| 1 | '' | 1.9174 | 0.5166 | -0.0019 | 0.9656 | 0.2601 |   |
| 2 | 100 | 0.6067 | -0.8085 | -0.0285 | 0.1630 | -0.6021 |   |
| 3 | '' | 0.6896 | -0.7250 | 0.0304 | 0.2027 | -0.7549 |   |

```
maxIts = {'';102;''};
table(maxIts,X_102,V_102)
```

`ans = 3×3 table`

|   | maxIts | X_102 | | | V_102 | | ... |
|---|---|---|---|---|---|---|---|
|   |   | 1 | 2 | 3 | 1 | 2 |   |
| 1 | '' | 1.9174 | 0.5166 | -0.0019 | 0.9656 | 0.2601 |   |
| 2 | 102 | 0.6067 | -0.8085 | -0.0285 | 0.1630 | -0.6021 |   |
| 3 | '' | 0.6896 | -0.7250 | 0.0304 | 0.2027 | -0.7549 |   |

```
maxIts = {'';101;''};
table(maxIts,X_101,V_101)
```

`ans = 3×3 table`

|   | maxIts | X_101 | | | V_101 | | ... |
|---|---|---|---|---|---|---|---|
|   |   | 1 | 2 | 3 | 1 | 2 |   |
| 1 | '' | 0.5215 | 0.8947 | -0.0086 | 0.5036 | 0.8639 |   |
| 2 | 101 | 1.6482 | -0.2446 | -0.0400 | 0.5371 | -0.3055 |   |
| 3 | '' | 1.4502 | -0.0438 | 0.0486 | 0.6767 | -0.4004 |   |

```
maxIts = {'';103;''};
```

```
table(maxIts,X_103,V_103)
```

ans = 3×3 table

|   | maxIts | X_103 | | | V_103 | |  |
|---|---|---|---|---|---|---|---|
|   |   | 1 | 2 | 3 | 1 | 2 | |
| 1 | '' | 0.5215 | 0.8947 | -0.0086 | 0.5036 | 0.8639 | |
| 2 | 103 | 1.6482 | -0.2446 | -0.0400 | 0.5371 | -0.3055 | |
| 3 | '' | 1.4502 | -0.0438 | 0.0486 | 0.6767 | -0.4004 | |

## Reducibility of square matrices

```
R = [1 0 0;2 3 4;5 6 7]; % the previous matrix with failed convergence
g = digraph(logical(R));
h = transreduction(g);
[bins, binsizes] = conncomp(g);
isConnected = all(bins == 1);
R
```

R = 3×3

```
     1     0     0
     2     3     4
     5     6     7
```

```
if isConnected
    disp('This matrix is irreducible')
else
    disp('This matrix is reducible')
end
```

This matrix is reducible

```
figure,
subplot(1,2,1)
plot(g)
title('Graph with directed edges')
subplot(1,2,2)
plot(h)
title('Transitive reduction')
```

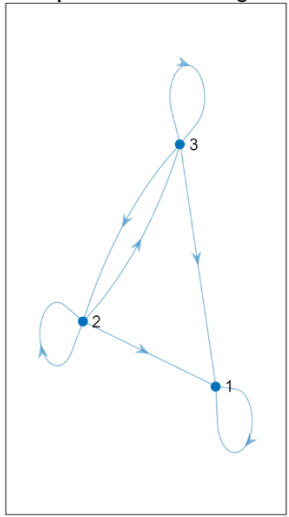
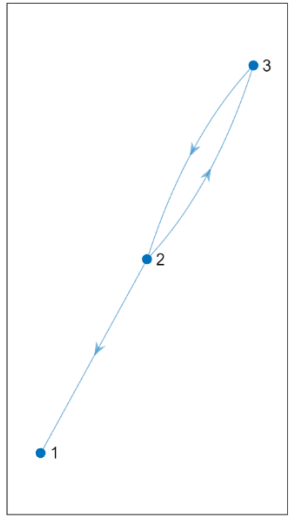

```
R = eye(3,3); % even the identity matrix is reducible!
g = digraph(logical(R));
h = transreduction(g);
[bins, binsizes] = conncomp(g);
isConnected = all(bins == 1);
R
```

R = 3×3

```
     1     0     0
     0     1     0
     0     0     1
```

```
if isConnected
    disp('This matrix is irreducible')
else
    disp('This matrix is reducible')
end
```

This matrix is reducible

```
figure,
subplot(1,2,1)
plot(g)
title('Graph with directed edges')
subplot(1,2,2)
plot(h)
title('Transitive reduction')
```

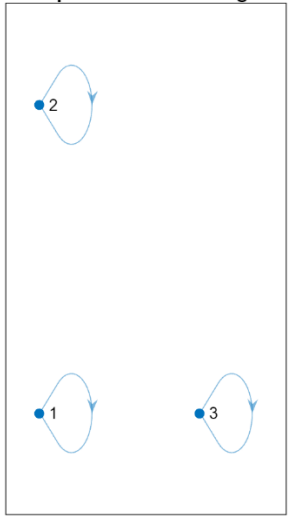
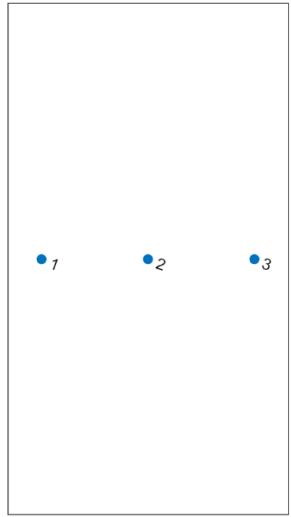

```matlab
R = [ 0 1 0; 0 0 1; 1 0 0]; %Nicolas Gillis suggested this irreducible example
g = digraph(logical(R));
h = transreduction(g);
[bins, binsizes] = conncomp(g);
isConnected = all(bins == 1);
R
```

R = 3×3

```
     0     1     0
     0     0     1
     1     0     0
```

```matlab
if isConnected
    disp('This matrix is irreducible')
else
    disp('This matrix is reducible')
end
```

This matrix is irreducible

```matlab
figure,
subplot(1,2,1)
plot(g)
title('Graph with directed edges')
subplot(1,2,2)
plot(h)
title('Transitive reduction')
```

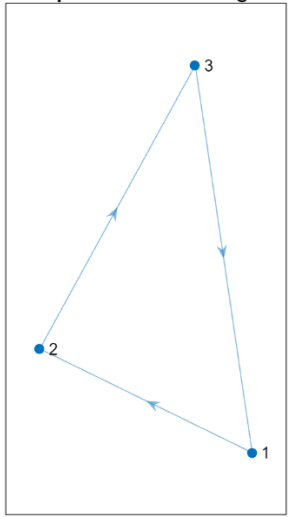
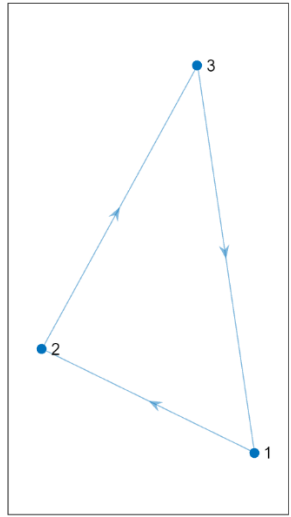

```
function dydt = odefn_MichMent(~,y,k1,k1r,k2)
    dydt = zeros(4,1);
    dydt(1)=-k1*y(1)*y(2)+k1r*y(3);
    dydt(2)=-k1*y(1)*y(2)+k1r*y(3)+k2*y(3);
    dydt(3)=k1*y(1)*y(2)-k1r*y(3)-k2*y(3);
    dydt(4)=k2*y(3);
end

function dydt = ode2fn_MichMent(~,y,k1,k1r,k2,y0)
    dydt = zeros(2,1);
    dydt(1) = -k1*(y(1)+y(2)+y0(2)-y0(1))*y(1)+k1r*(y0(1)-y(1)-y(2));
    dydt(2) = k2*(y0(1)-y(1)-y(2));
end

function dydt = ode2bfn_MichMent(~,y,k1,k1r,k2,y0)
    dydt = zeros(2,1);
    dydt(1) = -k1*y(1)*y(2)+k1r*(y0(2)-y(2));
    dydt(2) = -k1*y(1)*y(2)+(k1r+k2)*(y0(2)-y(2));
end

function c_spec=NewtonRaphson(Model, beta, c_tot, c,i)
% Copyright 2007 Marcel Maeder and Yorck-Michael Neuhold
% Practical Data Analysis in Chemistry, Elsevier, First edition 2007
% ISBN: 978-0-444-53054-7
% ISSN: 0922-3487
% MatlabFile 3-9. NewtonRaphson.m; p. 53
```

```matlab
    ncomp=length(c_tot);    % number of components
    nspec=length(beta);     % number of species
    c_tot(c_tot==0)=1e-15;  % numerical difficulties if c_tot=0
    J_s=NaN(ncomp,ncomp);   % preallocating J_s for speed
    it=0; it_endv=99;
    while it<=it_endv
        it=it+1;
        c_spec =beta.*prod(repmat(c',1,nspec).^Model,1);   %species conc
        c_tot_calc=sum(Model.*repmat(c_spec,ncomp,1),2)'; %comp ctot calc
        d =c_tot-c_tot_calc;                              % diff actual and calc
total conc
        if all(abs(d) < 1e-15)                            % return if all diff
small
            return
        end
        for j=1:ncomp              % Jacobian (J*)
            for k=j:ncomp
                J_s(j,k)=sum(Model(j,:).*Model(k,:).*c_spec);
                J_s(k,j)=J_s(j,k); % J_s is symmetric
            end
        end
        delta_c=(d/J_s)*diag(c);   % equation (2.43)
        c=c+delta_c;
        while any(c <= 0)          % take shift back if conc neg.
            delta_c=0.5*delta_c;
            c=c-delta_c;
            if all(abs(delta_c)<1e-15)
                break
            end
        end
    end
 end
 if it>it_endv; fprintf(1,'no conv. at C_spec(%i,:)\n',i); end
 end

function [X, V] = fsvt1n_ext_norm(R,ranki,epsi,max_it)
crit = true;
it = 0;
while crit
    [U,S,V] = svds(R,ranki);
    if max(U(:,1))<=0, U(:,1)=-U(:,1); V(:,1)=-V(:,1); end
    R = R./(R*V(:,1));
    X = U*S;
    it = it + 1;
    crit = and(max(abs(X(:,1)-1))>epsi,it<max_it);
```

```matlab
    end
end

function [sgns,loads] = sign_flip(loads,X)
% [sgns,loads] = sign_flip(loads,X)
% loads is a cell of loadings
% X     is the data array
% sgns is a MxF matrix where sgns(m,f) is the sign of loading f in mode m
%
% If using svd ([u,s,v]=svd(X)) then loads{1}=u*s; and loads{2}=v;
% If using an F-component PCA model ([t,p]=pca(X,F), then loads{1}=t; and
% loads{2}=p;
%
% Copyright 2007 R. Bro, E. Acar, T. Kolda - www.models.life.ku.dk
% for PARAFAC and two-way
% https://doi.org/10.1002/cem.1122
if isa(X,'dataset')
  inc = X.includ;
    X = X.data(inc{:});
end
order = length(size(X));
for i=1:order
  F(i) = size(loads{i},2);
end
for m = 1:order % for each mode
    for f=1:F(m) % for each component
       s=[];
       a = loads{m}(:,f);
       a = a /(a'*a);
       x = subtract_otherfactors(X, loads, m, f);
       for i=1:size(x(:,:),2) % for each column
          s(i)=(a'*x(:,i));
          s(i)=sign(s(i))*power(s(i),2);
       end
       S(m,f) =sum(s);
    end
end
sgns = sign(S);
for f=1:F(1) %each component
  for i=1:size(sgns,1) %each mode
    se = length(find(sgns(:,f)==-1));
    if (rem(se,2)==0 )
        loads{i}(:,f)=sgns(i,f)*loads{i}(:,f);
    else
```

```matlab
            disp('Odd number of negatives!')
            sgns(:,f) = handle_oddnumbers(S(:,f));
            se = length(find(sgns(:,f)==-1));
            if (rem(se,2)==0)
                loads{i}(:,f)=sgns(i,f)*loads{i}(:,f);
            else
                disp('Something Wrong!!!')
            end
        end
    end  %each mode
end %each component
end
%-------------------------------------------------------------------------
function sgns=handle_oddnumbers(Bcon)
sgns=sign(Bcon);
nb_neg=find(Bcon<0);
%[min_val, index]=min(abs(Bcon));
[~, index]=min(abs(Bcon));
if (Bcon(index)<0)
    sgns(index)=-sgns(index);
% since this function is called nb_neg should be greater than 0, anyway
elseif ((Bcon(index)>0) && (nb_neg>0))
    sgns(index)=-sgns(index);
end
end
%-------------------------------------------------------------------------
function x = subtract_otherfactors(X, loads, mode, factor)
order=length(size(X));
x = permute(X,[mode 1:mode-1 mode+1:order]);
loads = loads([mode 1:mode-1 mode+1:order]);
for m = 1: order
    loads{m}=loads{m}(:, [factor 1:factor-1 factor+1:size(loads{m},2)]);
    L{m} = loads{m}(:,2:end);
end
M = outerm(L);
x=x-M;
end
function mwa = outerm(facts,lo,vect)
if nargin < 2
  lo = 0;
end
if nargin < 3
  vect = 0;
end
order = length(facts);
if lo == 0
```

```matlab
    mwasize = zeros(1,order);
  else
    mwasize = zeros(1,order-1);
  end
  k = 0;
  for i = 1:order
    if i ~= lo
      [m,n] = size(facts{i});
      k = k + 1;
      mwasize(k) = m;
      if k > 1
      else
        nofac = n;
      end
    end
  end
  mwa = zeros(prod(mwasize),nofac);
  for j = 1:nofac
    if lo ~= 1
      mwvect = facts{1}(:,j);
      for i = 2:order
        if lo ~= i
            mwvect = mwvect*facts{i}(:,j)';
            mwvect = mwvect(:);
        end
      end
    elseif lo == 1
      mwvect = facts{2}(:,j);
       for i = 3:order
         mwvect = mwvect*facts{i}(:,j)';
         mwvect = mwvect(:);
      end
    end
    mwa(:,j) = mwvect;
  end
  % If vect isn't one, sum up the results of the factors and reshape
  if vect ~= 1
    mwa = sum(mwa,2);
    mwa = reshape(mwa,mwasize);
  end
end
```